\documentclass[11pt, a4paper, twoside, BCOR12mm, DIV10, headinclude, footinclude=false, 
bibliography=totoc, 
listof=totoc, numbers=noenddot, chapterprefix=false, appendixprefix,svgnames]{scrbook}
\pdfoutput=1

\newcommand{\mytitle}[1]{Modeling Musical Structure with Artificial Neural Networks}
\title{\mytitle}

\usepackage[english]{babel}
\usepackage{lmodern}
\usepackage[T1]{fontenc}
\usepackage[utf8]{inputenc}
\usepackage{booktabs}
\usepackage[toc,page]{appendix}
\usepackage{bibentry}
\nobibliography*

\usepackage{amsmath}
\usepackage{amssymb}  
\usepackage{graphicx}
\usepackage[hang]{caption}
\usepackage{subfig}
\usepackage{floatpag}
\usepackage{multirow}
\usepackage{natbib}
\usepackage{pdfpages}
\usepackage{nicefrac}
\usepackage[format=hang,font=small,labelfont=bf]{caption}

\usepackage{url}

\usepackage{color}

\usepackage{amssymb}
\usepackage{mathtools}
\usepackage{booktabs}

\usepackage[ruled,linesnumbered]{algorithm2e}

\DeclareMathOperator*{\argmin}{\arg\!\min}
\DeclareMathOperator*{\argmax}{\arg\!\max}
\DeclarePairedDelimiter\norm{\lVert}{\rVert}%

\usepackage{aliascnt}

\usepackage{rotating}
\usepackage{colortbl}
\usepackage{array}
\usepackage{lscape}

\usepackage{tabu}
\usepackage{multirow}
\usepackage{multicol}
\usepackage{tikz}
\usetikzlibrary{calc,tikzmark}
\usepackage[shortcuts]{extdash}

\makeatletter
\def\inmod#1{\allowbreak\mkern5mu{\operator@font mod}\,\,#1}
\makeatother
\urldef{\discovery}\url{http://www.music-ir.org/mirex/wiki/2017:Discovery_of_Repeated_Themes_%26_Sections}

\usepackage{hyperref}
\hypersetup{
	pdfstartview=FitH,
	pdffitwindow=true,
	colorlinks,
	linkcolor=black,
	anchorcolor=black,
	citecolor=black,
	urlcolor=black,
	pdftitle     = {{Modeling Musical Structure with Artificial Neural Networks}},
	pdfsubject   = {{PhD Thesis in Informatics at the Johannes Kepler University in Linz}},
	pdfauthor    = {{Stefan Lattner}},
	plainpages   = false, pdfpagelabels
}

\hyphenation{
  know-ledge
  au-to-ma-ti-cal-ly
  sa-lak-hut-di-nov
  kriz-hev-sky
  mu-sik-ähn-lich-keits-maße
}

\setcounter{tocdepth}{2}

\setcounter{secnumdepth}{3}

\newcommand{\abstract}[1]{#1}
\renewcommand{\bf}[1]{\textbf{#1}}
\renewcommand{\cite}{\citep}

\begin{document}

\includepdf[]{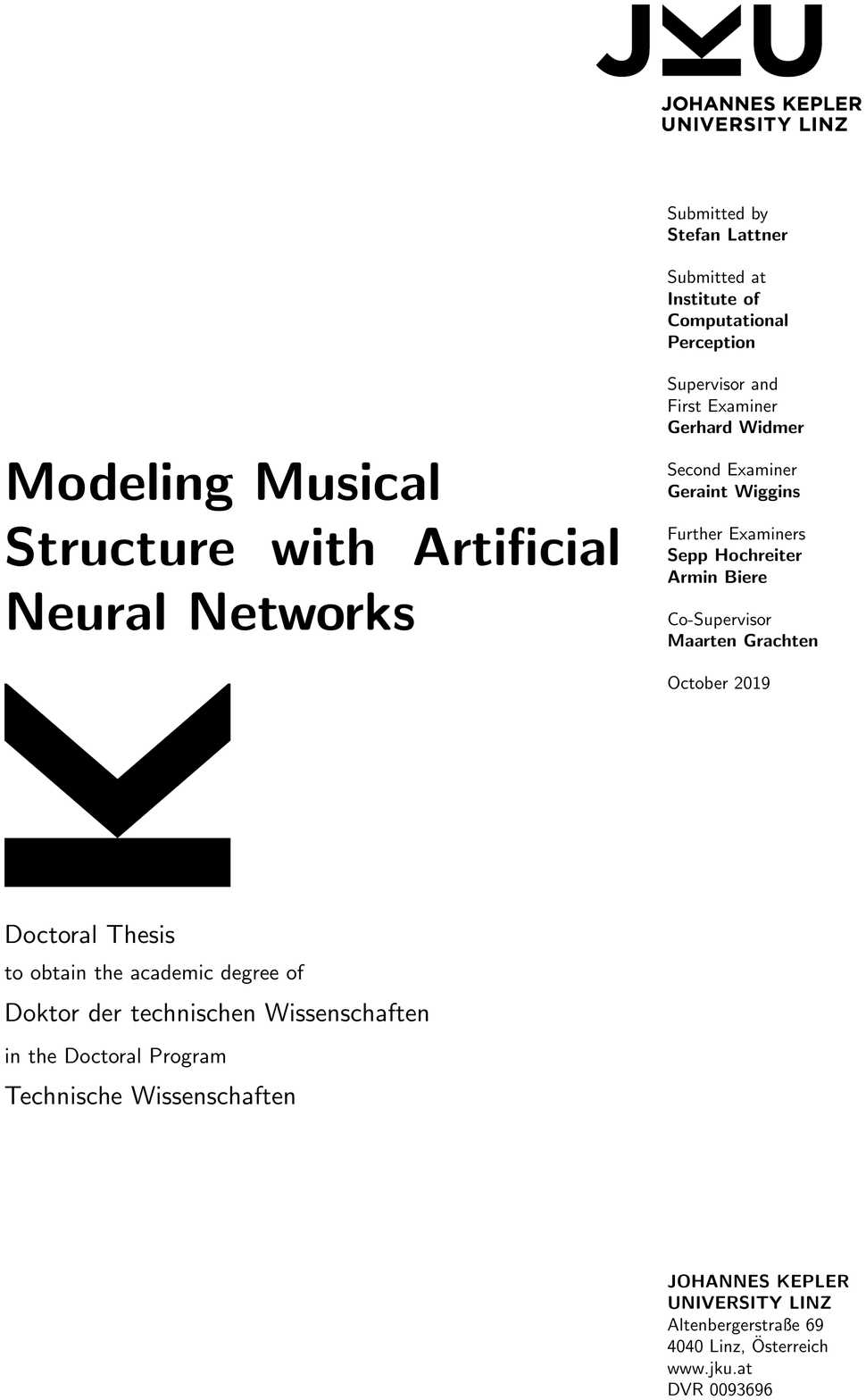}
\renewcommand{\textfraction}{0}
\sloppy




\chapter*{Eidesstattliche Erkl\"arung}
\addcontentsline{toc}{chapter}{B Eidesstattliche Erkl\"arung}
Ich erkl\"are an Eides statt, dass ich die vorliegende Dissertation selbstst\"andig und ohne fremde Hilfe verfasst, andere als die angegebenen Quellen und Hilfsmittel nicht benutzt bzw. die w\"ortlich oder sinngem\"a\ss{} entnommenen Stellen als solche kenntlich gemacht habe. Die vorliegende Dissertation ist mit dem elektronisch \"ubermittelten Textdokument identisch.

\begin{tabular}{llp{2em}l}
\\ \\ \\ \\ \\
\hspace{2cm} & \hspace{4,5cm}   && \hspace{4,5cm} \\\cline{2-4}
& Ort, Datum     && Unterschrift
\end{tabular}

\pagenumbering{Roman}
\setcounter{page}{1}
\pagestyle{plain}
\chapter*{Abstract}


In recent years, artificial neural networks (ANNs) have become a universal tool for tackling real-world problems.
ANNs have also shown great success in music-related tasks including music summarization and classification, similarity estimation, computer-aided or autonomous composition, and automatic music analysis.
As \emph{structure} is a fundamental characteristic of Western music, it plays a role in all these tasks.
Some structural aspects are particularly challenging to learn with current ANN architectures.
This is especially true for mid- and high-level self-similarity, tonal and rhythmic relationships.
In this thesis, I explore the application of ANNs to different aspects of musical structure modeling, identify some challenges involved and propose strategies to address them.

First, using probability estimations of a Restricted Boltzmann Machine (RBM), a probabilistic bottom-up approach to \emph{melody segmentation} is studied.
Then, a top-down method for \emph{imposing a high-level structural template} in music generation is presented, which combines Gibbs sampling using a convolutional RBM with gradient-descent optimization on the intermediate solutions. Furthermore, I motivate the relevance of \emph{musical transformations} in structure modeling and show how a connectionist model, the Gated Autoencoder (GAE), can be employed to learn transformations between musical fragments.
For learning transformations in sequences, I propose a special predictive training of the GAE, which yields a representation of polyphonic music as a \emph{sequence of intervals}.
Furthermore, the applicability of these interval representations to a top-down \emph{discovery of repeated musical sections} is shown.
Finally, a recurrent variant of the GAE is proposed, and its efficacy in music prediction and modeling of \emph{low-level repetition structure} is demonstrated.

\newpage

\pagestyle{plain}
\chapter*{Kurzfassung}
In den letzten Jahren haben sich Künstliche Neuronale Netze (KNNs) zu einem universellen Werkzeug für unterschiedliche Problemstellungen entwickelt. Auch für musikalische Anwendungen erwiesen sie sich als nützlich. Sowohl für die Musikanalyse, die Klassifikation von Musikstücken, für Musikähnlichkeitsbestimmung, aber auch in der automatischen Komposition wurden sie erfolgreich eingesetzt. Da \emph{Struktur} eines der bestimmenden Merkmale von westlicher Musik ist, spielt sie auch für diese Tasks eine große Rolle. Manche Aspekte von Struktur in der Musik sind jedoch für KNNs schwer zu erlernen. Vor allem mit Aspekten von Wiederholung und Variation sowie mit tonalen und rhythmischen Zusammenhängen, die sich über längere Zeiträume erstrecken, haben KNNs oft Probleme. In dieser Dissertation beschäftige ich mich mit der Anwendung von KNNs auf unterschiedliche Aspekte von musikalischer Struktur, untersuche die damit verbundenen Probleme und diskutiere mögliche Lösungen.

Das erste Experiment behandelt das \emph{probabilistische bottom-up Segmentieren von Melodien} mithilfe einer Restricted Boltzmann Machine (RBM). Danach wird gezeigt, wie beim automatischen Generieren eines Musikstücks mit einer Convolutional RBM einer \emph{Struktur-Vorlage} gefolgt werden kann. Dafür wird beim Generieren die vorgegebene Struktur durch Optimierung der Lösung mittels des Gradientenverfahrens erzwungen. Dann erläutere ich die Relevanz vom \emph{Lernen von Transformationen} für das Modellieren von musikalischer Struktur und zeige, dass ein Modell namens Gated Autoencoder (GAE) dafür gut geeignet ist. Dann stelle ich eine Methode vor, in einem ``predictive Training'' mit dem GAE Tonhöhen-Intervalle als Transformationen zwischen Noten in polyphoner Musik zu lernen. Die dabei gelernten Intervall-Repräsentationen können für das top-down \emph{Erkennen von wiederholten musikalischen Teilen} verwendet werden. Abschließend erweitere ich den GAE mit rekurrenten Verbindungen, um ihm das Lernen von \emph{kurzen Wiederholungsstrukturen} zu ermöglichen.

\newpage

\pagestyle{plain}
\chapter*{Acknowledgements}
First and foremost, I want to thank my supervisor Gerhard Widmer for his valuable support, the opportunity to work in his welcoming team, and to obtain my Ph.D. in such a scientifically productive and professional environment.

This work would not have been possible without the great support of my co-supervisor Maarten Grachten, who has taught me the scientific tools of the trade from the first day of my Ph.D., and since that accompanied me through my scientific career.
His exceptional effectiveness, scientific precision and common sense made every collaboration with him an enlightening experience, and his unshakeable calmness when approaching deadlines kept me from panicking and helped me to keep focussed.

I also want to thank Geraint Wiggins for his effort and time to review this thesis, and for his work on computational creativity, perception and information dynamics in music, which was very inspiring for my own research.

Furthermore, I am grateful for my colleagues at the Austrian Research Institute for Artificial Intelligence in Vienna and the Johannes Kepler University in Linz. It was a pleasure to be part of the team, from both a professional and a personal point of view. Special thanks go to Carlos Eduardo Cancino-Chacon with whom I shared the office for more than three years. He was not only a delightful colleague and friend, but he also was my advisor in mathematics, helping me to structure my ideas and to turn them into equations.

Lastly, and most importantly, I would not have succeeded without my great family and friends, which helped me not to forget that there is a life outside of work. Their unconditional support and the great time we shared during the years provided the emotional basis for my professional progress.

This work has been supported by the European Commission under the
Future and Emerging Technologies (FET) programme within the Seventh Framework Programme
(FET grant number 610859, project "Learning to Create"),
and by the European Research Council (ERC) under the
European Union's Horizon 2020 Research and Innovation Programme
(grant agreement 670035, project "Con Espressione").

\vspace{0.2cm}
\begin{tabular}{p{8cm}l}
&\includegraphics[width=0.35\textwidth, trim=0 160 0 160, clip]{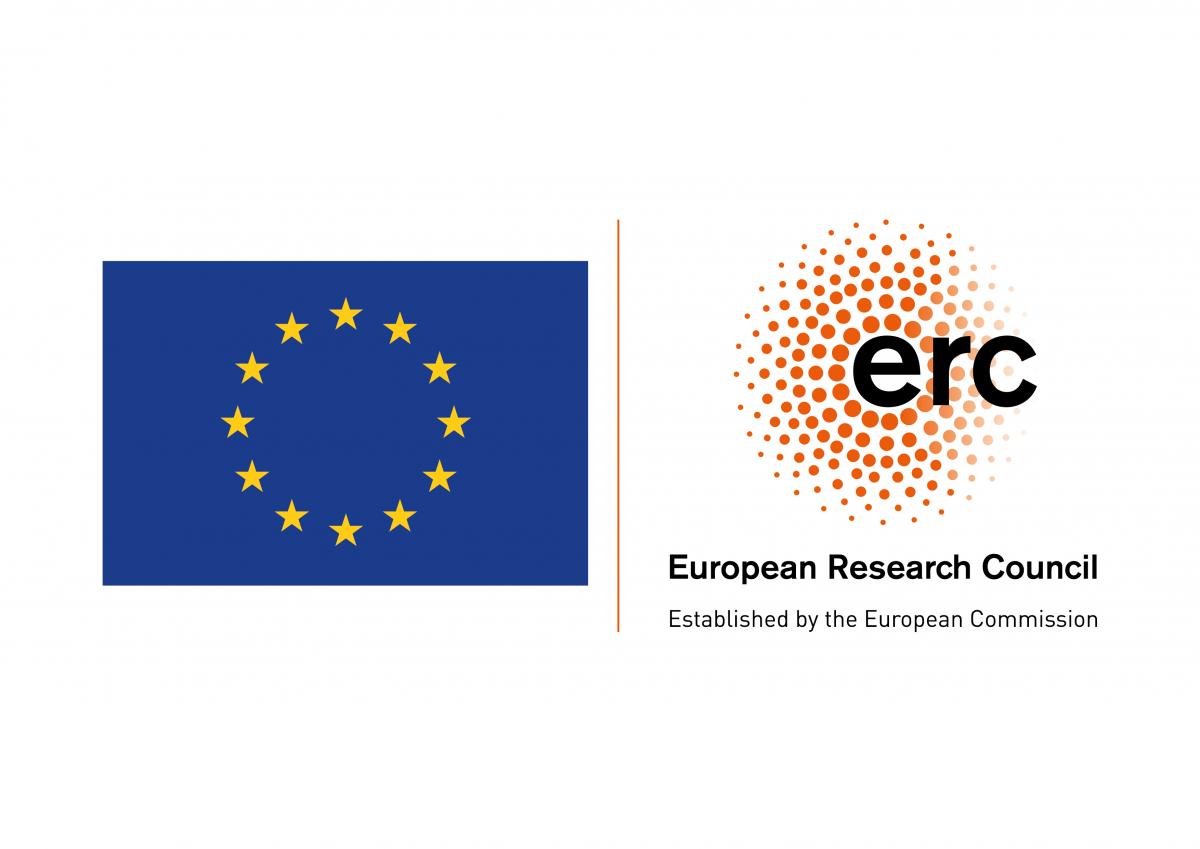}
\end{tabular}

\tableofcontents
\cleardoublepage

\listoffigures
\listoftables
\cleardoublepage

\mainmatter

\captionsetup{hangindent=-5em}

\chapter{Introduction}
\section{Motivation}

\subsection{Musical Structure}
Music is a highly structured artifact. Music as we know it\footnote{In this thesis, we only consider Western tonal music.} consists of
(notated and/or sounding) events that are organized along dimensions such
as pitch, time, loudness, and timbre, via a complex network of relationships (e.g., time and duration ratios, melodic intervals, scales).
From this interplay of musical events arise higher-level structural concepts such as texture, melody, harmony, rhythm, grouping and phrase structure, repetition and variation, motivic relationships, style, or genre. In short, music is \textit{organized sound} \cite{varese1966liberation}, with organization or structure apparent at many levels.

Fundamental to this is the human perception with its abstraction capabilities.
Octave equivalence and, more generally, the ability to abstract away from absolute pitch permit us to perceive musical contents as being ``the same'' or related even when they are transposed in pitch or, in fact, modified via a number of other types of musical \textit{transformations} (such as ornamentation, diminution, inversion, etc.).
Also underlying this is our tendency to organize musical events in a network or hierarchy of relative importance, with some events being perceived as subordinate to others, at multiple levels \cite{lerd83}. This abstraction ability permits us to recognize musical passages as similar or related even in the face of surface differences.

In the literature, the term musical structure is a general term used for different relationships between musical events and features.
From an information-theoretic point of view, and in order to be able to talk about musical structure in machine learning and statistics, we can define it very broadly as a non-random organization of musical events and their properties.
This means that one can estimate the state of an event when the states of some other events are observed.
The non-randomness of music is key to a pleasant listening experience --- in fact, auditory stimuli without structure are hardly identified as being music at all.
In this work, I will tackle the problems of learning, analysis, and generation of musical textures considering the dependencies in different hierarchical levels.
In particular, I will focus on grouping and segmentation, tonal relationships, meter, self-similarity and transformation (i.e., variation).
I will discuss reasons why modeling such dependencies with machine learning (ML) methods is challenging, and I will contribute some findings to amend the situation.

\subsection{Challenges in Structure Learning}
In computer science, musical structure is approached with different intentions.
On the one hand, in music information retrieval (MIR), a typical goal is to analyze the structure of particular features of interest.
For example, harmonic analysis unveils the chord structure, and self-similarity analysis provides insight into the musical form of a song.
Such tasks focus on specific musical features and can thus be solved at a manageable cost.

On the other hand, music prediction and music generation tasks pose a much more challenging problem.
The models typically applied to this task are called \emph{sequence models} (or language models).
They are usually trained in a so-called self-supervised fashion, to predict events in the immediate future (typically the next time step or the next event) given the past.
The training objective in such models is to maximize the likelihood of the data under the model.
Thereby, the model has to pick up on the mid- and high-level structure of music, while being explicitly trained to predict elementary low-level events.
However, the capacity of current models and training methods is often too limited to learn higher-level structure by sheer exposition to elementary events.
In music generation, this poses a challenging situation, because humans have a solid expectation regarding specific norms in the musical structure and do not readily tolerate violations thereof.
Ideally, a model should, therefore, cover all the relevant structural characteristics in order to generate convincing music.


A particular case of musical structure learning with computer models is self-similarity (i.e., repetition and variation).
While higher-level features like harmony or rhythm, can be determined by local analysis of the texture, modeling self-similarity requires a comparison between at least two musical entities.
This results in two fundamental problems:
First, self-similarity can exist between any musical entities in a piece, and a pairwise comparison between all of them results in a high computational cost.
Second, it is not straight-forward to determine what similarity means in a musical context.
The perceived similarity is highly subjective and is, therefore, an insufficient measure for an objective description of similarity in musical structure.
The approach I am taking is that musical textures can be considered similar when they can be transferred into each other by \emph{transformations} which occur regularly in a corpus.
Some well-known examples thereof are retrograde, inversion, insertion and deletion, tempo change, chromatic and diatonic transposition, as well as the identity transform, denoting a  simple copy operation of a particular event or feature.
However, there may be additional, yet undefined transformations musical textures can undergo in musical variation.
Ideally, such transformations should, therefore, be learned using self-supervised training, as it is not possible or desired to pre-define them exhaustively.

In this thesis, it is shown that the architecture of Gated Autoencoders (GAEs) foster unsupervised learning of musical transformations.
By discussing the different architectures, it will also become clear that GAEs differ fundamentally from conventional prediction models.
The most striking difference of GAEs is that they represent transformations explicitly in their latent space, rather than representing the characteristics of the data itself.
The way transformation learning with GAEs is achieved (i.e., by using three-way multiplicative interactions) leads to surprisingly compact models.
This ``transformational'' (i.e., content-invariant) view on musical material, which is provided by the latent space of GAEs, has proven advantageous for several tasks.
The experiments conducted in Chapter \ref{sec:structure} show that GAEs are useful for music analysis and prediction and that the concept of learning transformations is vital for effective modeling of repetition and variation.

In some recent developments, the problem of learning musical structure is tackled with powerful sequence models, so-called \emph{Transformers} \cite{DBLP:conf/iclr/Huang19}.
(Self-) attention mechanisms \cite{DBLP:conf/nips/VaswaniSPUJGKP17} help Transformers to ``focus'' on specific subsequences or important positions in a musical piece.
The material deemed to deserve focus is then used for prediction.
The inner working of Transformers involves re-ordering of data, which relates, to some degree, to copy operations necessary for modeling repetitions.
However, these models learn to generalize to musical relations only through redundancy, rather than through learning of transformations.
For example, in the diatonic transposition operation, the input-to-output pathways of different transposition distances and different keys would be represented differently in a Transformer.
Therefore, substantial model sizes and amounts of training data are necessary to obtain some degree of mid- and high-level structure in the generated material.
Also, data augmentation is usually used during training, for example by transposing or time-stretching musical input sequences.

I believe that modeling transformations in music as a complementary approach to conventional data representation can allow for more compact models which can generalize on fewer data.
In particular repetition and variation, but also tonal and rhythmic relationships could be better learned from data when utilizing a transformational approach.

\newpage
\section{Outline}
This thesis is based upon six publications, which are grouped into three chapters of similar topics.
In Chapter \ref{sec:segment}, a connectionist n-gram model is employed to identify segment boundaries in melodies based on the information content (IC) of note events.
Chapter \ref{sec:constraints} introduces a method to impose structural constraints in music generation using a generative connectionist model and differentiable constraint functions.
Finally, in Chapter \ref{sec:structure}, I report on the successive adaption of a GAE architecture for learning musical transformations to musical sequence modeling. By learning interval representations, the GAE yields better accuracy than baseline methods in a prediction task and is also able to perform copy-and-transpose operations. Furthermore, it is shown that the interval representations themselves are useful for transposition-invariant detection of repeated sections in polyphonic music.
In summary, the contributions are structured as follows:

\begin{itemize}
    \item \textbf{Bottom-Up Structure Analysis} of monophonic melodies via probability estimation (see Chapter \ref{sec:segment}).
    \item \textbf{Top-Down Imposition of Higher-Level Structure} in polyphonic music generation, using soft-constraints (see Chapter \ref{sec:constraints}).
    \item \textbf{Learning Musical Structure} by learning transformations (see Chapter \ref{sec:structure}):
    \begin{itemize}
        \item \textbf{Learning Transformations in Music} (see Section \ref{sec:gae}).
        \item \textbf{Learning Transposition-Invariant Interval Representations} from both, symbolic music and audio, and employing these representations for top-down structure analysis (see Section \ref{sec:interval}).
        \item \textbf{Learning to Predict Musical Sequences}---with the potential to learning repetition and variation in music---using a recurrent sequence model operating on learned interval representations (see Section \ref{sec:rgae}).
    \end{itemize}
\end{itemize}

In the following, I shall briefly introduce each chapter and discuss its relevance for musical structure modeling.

\subsection*{Chapter \ref{sec:segment}: Bottom-up Structure Analysis}
The objective of Chapter \ref{sec:segment} is the unsupervised segmentation of musical sequences (i.e., estimating phrase markers as annotated in a collection of monophonic folk song melodies). 
The level of surprise when observing an event in a sequence can be quantified by the information theoretic measure \emph{information content} (IC).
Prior research shows that the IC can act as a kernel for segment boundaries \cite{Pearce:2010ig}.
This finding indicates that the conditional probabilities \emph{within} a musical segment (i.e., subsequences perceived as coherent ``chunks'' by a human listener) tend to be higher (i.e., have lower IC) than those \emph{between} segments.
Considering the evidence that IC is a useful kernel for segment boundary detection, it follows that an improved IC estimation should result in an improved segmentation performance.

The probabilistic model used so far for the IC estimation in this task is a sophisticated variable-order Markov model \cite{Pearce:2010ig}.
I show that an improvement can be reached by instead employing a \emph{connectionist n-gram model} (i.e., a Restricted Boltzmann Machine (RBM)) for the IC estimation.
Furthermore, a feed-forward neural network is trained to \emph{predict} the IC values based on the musical texture, as opposed to using the IC estimation directly.
After applying that further optimization step (denoted as ``pseudo-supervised training'', see Section \ref{sec:pseudo-superv-train}), this unsupervised method is on a par with an expert system \cite{Cambouropoulos:2001vj}, explicitly designed to detect segment boundaries.
The results strengthen the finding that IC can be used for bottom-up structure analysis and support the hypothesis of \citet{Pearce:2010ig} that there exists a relationship between expectation and grouping in auditory perception.

\subsection*{Chapter \ref{sec:constraints}: Top-Down Imposition of Higher-Level Structure}
In Chapter \ref{sec:constraints}, the focus is on the generation of structured, polyphonic music.
In music generation, it is vital that the predictions of elementary events are conditioned on representations of higher-level characteristics, in order to reproduce structural characteristics of the training data.
Such characteristics include, among others, tonality, meter, rhythmic structure, and repetition structure.
If higher-level characteristics are not adequately represented by the model, the generated output frequently drifts off into different keys, shows inconsistent harmonic successions, or has implausible and continuously changing meter.
As stated above, self-supervised sequence models have to infer such higher-level structural characteristics while being trained to predict elementary events.
In this regard, current sequence models are still limited, which becomes evident in the quality of the results in (polyphonic) music generation tasks without additional guidance (i.e., without additional inputs, like meter information or chord symbols) \footnote{Very recent models \cite{DBLP:conf/iclr/Huang19} already show a better capacity in reproducing structural characteristics in a self-supervised setting.}.
In particular, learning repetition structure is very challenging, because it is a concept which cannot be modeled by conditioning events on abstract symbols.
For example, given a specific chord symbol, it is possible to derive a distribution over note events when analyzing a corpus of musical works.
In contrast, a model has to learn a \emph{copy operation} in order to realize an exact repetition, which is fundamentally different from deriving a distribution given an abstract symbol.
Therefore, due to limited capacity, limited training data, or simply because of the delicate nature of some musical concepts, it is challenging for self-supervised sequence models to learn the different forms of higher-level structure.

In the generation of lower-level structure, however, the performance of sequence models is considerably better.
Therefore, in Chapter \ref{sec:constraints}, I restrict the learning task solely to the musical lower-level structure.
For that, I employ a Convolutional Restricted Boltzmann Machine (C-RBM) with limited (i.e., local) receptive fields to learn the local statistics of polyphonic Mozart piano sonatas.
In generation, the C-RBM is combined with differentiable functions defining higher-level structural properties (i.e., meter, tonal structure, and self-similarity structure).
By minimizing these functions during sampling, they act as external soft-constraints on the generated material.
That way, higher-level structural properties can be transferred from an existing work to a newly generated piece of music.

The idea of constrained sampling is not new (e.g., in \cite{pachet2011} hard-constraints are used, and in \cite{hadjeres2018anticipation} a Recurrent Neural Network (RNN) going backward in time ``collects'' unary constraints which are then applied in forward generation by another RNN), but it is a novel contribution to use differentiable functions as global soft-constraints and impose them in Gibbs sampling for music generation.
The method could apply to other models using Gibbs sampling for music generation (e.g., to RNNs, as used for music generation by Gibbs sampling in \cite{hadjeres2017deepbach}).

\subsection*{Chapter \ref{sec:structure}: Learning Musical Structure}
In Chapter \ref{sec:structure}, I tackle some critical problems in learning musical characteristics with sequence models.
As stated before, musical structure implies many interdependencies between elementary and abstract musical events.
Some of these dependencies can be represented as \emph{transformations} between data, which leads to more compact representations than representing the data itself.
This observation motivates the approach taken in Section \ref{sec:gae}, namely to represent musical variations as sets of \emph{transformations}.
Some experiments performed in Chapter \ref{sec:structure} show that learning transformations is challenging for common connectionist models (see Section \ref{sec:ExperimentTransform} and Section \ref{sec:experimentsGAEpredict}).

We successfully apply GAEs to learning musical transformations and use them (including some variants, specifically developed for musical problems) for several different tasks. These include learning transformations between musical n-grams (see Section \ref{sec:gae}), learning interval representations (as transformations between pitches, see Section \ref{sec:interval}), and sequence prediction based on the learned interval representations (see Section \ref{sec:rgae}). In the following, these tasks are described in more detail.

\subsubsection*{Section \ref{sec:gae}: Learning Transformations in Music}
In this Section, I test the performance of GAEs to learn transformations between n-grams of polyphonic, symbolic music in piano-roll representation.
The performed experiments constitute a proof-of-concept to learning some common musical transformations, like chromatic and diatonic transposition, retrograde, and halftime.
Pairs of n-grams are constructed which obey such transformations and the performance of two architectures in learning the transformations from these n-grams is tested.
More precisely, I test an RBM and a GAE and find that GAEs show a much better performance.
As a result, transformations are meaningfully organized in the latent space of the GAE, and the learned representations can be used to transform test data in an analogy-making task.

\subsubsection*{Section \ref{sec:interval}: Learning Transposition-Invariant Interval Representations}
In Section \ref{sec:interval}, a GAE is trained in a predictive setting (as opposed to a symmetric setting employed in Section \ref{sec:gae}).
To that end, the GAE is trained to predict a single time-step of music in piano-roll representation based on a short temporal context (similar to the setup used in Chapter \ref{sec:segment}).
By predicting the configuration of pitches based on some temporal context using transformation learning, I obtain representations which behave like \emph{interval representations}.
This is because intervals can be considered \emph{shift transformations} between single notes.
The resulting representations show musically plausible organization and transposition-invariance.
I show that such properties are relevant for tasks like transposition-invariant repeated section discovery in (polyphonic) symbolic music and audio.

\subsubsection*{Section \ref{sec:rgae}: Learning to Predict Musical Sequences}
In Section \ref{sec:rgae}, the predictive GAE proposed in Section \ref{sec:interval} is combined with an RNN, to obtain the Recurrent Gated Autoencoder (RGAE), internally operating on learned interval representations.
This internal working on interval representations resembles the relative pitch processing of humans.
The RGAE yields improved prediction accuracy in a music sequence learning task, assumingly because interval representations reduce sparsity in the data and therefore help generalization of the model.
An additional advantage of the RGAE is that it can learn copy-and-transpose operations in a self-supervised sequence prediction task (see Section \ref{sec:copy-shift}).
Compared to a baseline RNN, the RGAE performs superior in that task.
The results suggest that the RGAE is promising for learning musical repetition structure and variations, which is a challenging problem for current sequence models.

\section{List of Publications}

Chapter \ref{sec:segment}, \ref{sec:constraints} and \ref{sec:structure} of this thesis are based on the following peer-reviewed publications.

\subsection*{Chapter \ref{sec:segment}}
\begin{itemize}
\item \bibentry{lattner2015probabilistic}
\item \bibentry{lattner2015pseudo}
\end{itemize}

\subsection*{Chapter \ref{sec:constraints}}
\begin{itemize}
\item \bibentry{lattnergeneration}
\end{itemize}

\subsection*{Chapter \ref{sec:structure}}
\begin{itemize}
\item \bibentry{lattner2017relations}
\item \bibentry{lattner2018learning}
\item \bibentry{lattner2018predictive}
\end{itemize}

\subsection*{Related Publications}

The following publications are related, but not part of this thesis.

\begin{itemize}
\item \bibentry{lattner2019complex}
\item \bibentry{DBLP:conf/waspaa/Lattner19}
\item \bibentry{arzt2018alignment}
\item \bibentry{lattner2017improving}
\item \bibentry{agres2015tonality}
\item \bibentry{chacon2014developing}
\end{itemize}

\chapter{Bottom-Up Structure Analysis via Probability Estimation}\label{sec:segment}

\newcommand\Mark[1]{\textsuperscript#1}


\begin{abstract}
A salient characteristic of human perception of music is that musical events are perceived as being grouped temporally into structural units such as phrases or motifs.
Segmentation of musical sequences into structural units is a topic of ongoing research, both in cognitive psychology and music information retrieval.
Computational models of music segmentation are typically based either on explicit knowledge of music theory and human perception or on statistical and information-theoretic properties of musical data.
The former, rule-based approach has been found to better account for (human annotated) segment boundaries in music than probabilistic methods, although the statistical model proposed in~\cite{Pearce:2010ij} performs almost as well as state-of-the-art rule-based approaches.
In this chapter, we propose a new probabilistic segmentation method, based on Restricted Boltzmann Machines (RBMs).
By sampling, we determine a probability distribution over a subset of visible units in the model, conditioned on a configuration of the remaining visible units.
We apply this approach to an n-gram representation of melodies, in order to estimate the conditional probability of a note given its n-1 predecessors.
From this estimation, we calculate the information content, which is used in combination with a threshold to determine the location of segment boundaries. 
Furthermore, we demonstrate that, remarkably, a substantial increase in segmentation accuracy can be obtained by not using information content estimates directly, but rather in a bootstrapping fashion. More specifically, we use information content estimates as a target for a feed-forward neural network that is trained to estimate the information content directly from the data. We hypothesize that the improved segmentation accuracy of this bootstrapping approach may be evidence that the generative model provides noisy estimates of the information content, which are smoothed by the feed-forward neural network, yielding more accurate information content estimates.
Comparative evaluation shows that probabilistic segmentation (on a dataset of simple, monophonic melodies) is now on a par with state-of-the-art rule-based models.
\end{abstract}

\section{Introduction}\label{sec:Introduction}

Across perceptual domains, grouping and segmentation mechanisms are crucial for our disambiguation and interpretation of the world.
Both top-down, schematic processing mechanisms and bottom-up, grouping mechanisms contribute to our ability to break the world down into meaningful, coherent ``chunks''~\cite{gobet01:_chunk}.
Indeed, a salient characteristic of human perception of music is that musical sequences are not experienced as an indiscriminate stream of events, but rather as a sequence of temporally contiguous musical groups or segments.
Elements within a group are perceived to have a coherence that leads to the perception of these events as a structural unit (e.g., a musical phrase or motif).
A prominent example of chunking has been shown in the context of chess~\cite{gobet98:_exper}, where increased skill level is associated with more efficient chunking of information about board configurations.
Moreover, chunking is involved more generally in visual~\cite{McCollough30062007} and acoustic/speech processing~\cite{baddeley66:_short} tasks.
Just as in speech, perception in terms of meaningful constituents is an essential trait of music cognition.
This is immanent in the ubiquitous notion of constituent structure in music theory.

The origin and nature of this sense of musical coherence, or lack thereof, which gives rise to musical grouping and segmentation has been a topic of ongoing research.
A prominent approach from music theory and cognitive psychology has been to apply perceptual grouping mechanisms, such as those suggested by Gestalt psychology, to music perception.
\emph{Gestalt principles}, such as the laws of proximity, similarity, and closure, were first discussed in visual perception \cite{wertheimer1938laws}, and have been successfully applied to auditory scene analysis \cite{bregman90} and inspired theories of music perception \cite{meyer56,narmour90,lerd83}.
Narmour’s Implication-Realization theory \cite{narmour90}, for example, uses measures of pitch proximity and closure that offer insight into how listeners perceive the boundaries between musical phrases.
This type of theory-driven approach has given rise to various rule-based computational models of segmentation.
This class of models relies upon the specification of one or more principles according to which musical sequences are grouped.

An alternative account of grouping and segmentation is based on the intuition that the distribution, or statistical structure of the sensory information, has an essential effect on how we perceive constituent structure.
This idea has been explored for different areas, such as vision~\cite{glicksohn11:_gestal}, speech~\cite{Brent99anefficient}, and melody perception~\cite{Pearce:2010ig}. The key idea is that the sensory information that comprises a chunk is relatively constant, whereas the succession of chunks (which chunk follows which) is more variable.
In information-theoretic terms, this implies that the \emph{information content} (informally: unexpectedness) of events within a chunk is lower than that of events that mark chunk boundaries.
As a side note on vocabulary: We will use the term \emph{segment}, rather than \emph{chunk} in the rest of this chapter, to express that we take an agnostic stance toward the precise nature of constituents, and instead focus on their demarcation.

While Gestalt principles are sometimes rather abstractly defined laws, information theory has some potential to describe and quantify such perceptive phenomena formally. The Gestalt idea of grouping based on "good form" (i.e., Pr\"agnanz), for example, has an information theoretic counterpart in the work of~\citet{von2005treatise}, where human vision is assumed to resolve ambiguous perceptive stimuli by preferring the most probable interpretation. Also, it is intuitively clear that in most real-world scenarios, the uncertainty about observing specific events (i.e., the entropy) tends to increase with higher distances from already observed events in any relevant dimension. Thus, while a direct link between the two paradigms is beyond dispute, the question remains which of it is more parsimonious and might have given rise for the other to emerge as a perceptual mechanism.

Prior work has shown that the information content of music events, as estimated from a generative probabilistic model of those events, is a good indicator of segment boundaries in melodies~\cite{Pearce:2010ij}.
The statistical model proposed in~\cite{Pearce:2010ij} (IDyOM) is capable of much better segmentations than simpler statistical models based on di-gram transition probabilities and point-wise mutual information~\cite{Brent99anefficient}, but still falls slightly short of state-of-the-art rule-based models.

In this chapter, we introduce a new probabilistic segmentation method, based on a class of stochastic neural networks known as Restricted Boltzmann Machines (RBMs). 
We present a Monte-Carlo method to determine a probability distribution over a subset of visible units in the model, conditioned on a configuration of the remaining visible units.
Processing melodies as n-grams, the RBM generates the conditional probability of a note given its n-1 predecessors.
This quantity, in combination with a threshold, determines the location of segment boundaries.

Moreover, we demonstrate that a substantial increase in segmentation accuracy can be obtained by not using information content estimates directly, but rather in a bootstrapping fashion.
More specifically, we use information content estimates computed from the RBM as a target for a feed-forward neural network (FFNN) that is trained to estimate the information content directly from the data.
This method facilitates a probabilistic approach to be on par with rule-based systems.
Since the FFNN relies on computed, rather than hand-labeled targets, we call this a ``pseudo-supervised'' scenario.

In an experimental setup, we compare our approach to other methods in evaluation against human segment boundary annotations.
Moreover, we explain the improved accuracy of the pseudo-supervised approach by describing how it can be regarded as employing a form of \emph{entropy regularization}~\cite{Grandvalet:2004wl}.

In Section~\ref{sec:Related} we give a brief overview of both rule-based and statistical models for melodic segmentation, with which we compare our approach (and which were evaluated in ~\cite{Pearce:2010ij}), and discuss related work regarding the pseudo-supervised regularization scheme.
Then, we will argue that our model (explained in Section~\ref{sec:Method_seg}) has advantages over statistical models based on n-gram counting.
In Section~\ref{sec:Method_seg} we explain how we estimate the conditional probability and information content of notes using an RBM, how notes are represented as input to the model, how an FFNN is used to predict information content, and how the information content is used to predict segment boundaries.
In Section~\ref{sec:Experiment}, we reproduce a quantitative evaluation experiment by~\citet{Pearce:2010ij}. 
The results are presented and discussed in Section~\ref{sec:results}, and conclusions and future work are presented in Section~\ref{sec:concl-future-work-seg-mix}.

\section{Related Work}\label{sec:Related}

\subsection{Rule-Based Segmentation}\label{sec:rule-based-segm}

One of the first models of melodic segmentation based on Gestalt rules was proposed by~\citet{Tenney:1980wf}.
This theory quantifies some local rules to predict grouping judgments.
However, this theory does not account for vague or ambiguous grouping judgments, and the selection of their numerical weights is somewhat arbitrary \cite{Tenney:1980wf,lerd83}.
 One of the most popular music theoretical approaches is Lerdahl and Jackendoff's Generative Theory of Tonal Music (GTTM) \cite{lerd83}.
This theory pursues the formal description of musical intuitions of experienced listeners through a combination of cognitive principles and generative linguistic theory.
In GTTM,
the hierarchical segmentation of a musical piece into motifs, phrases, and sections is represented through a \emph{grouping structure}.
This structure is expressed through consecutively numbered \emph{grouping preference rules} (GPRs), which model possible structural descriptions that correspond to experienced listeners' hearing of a particular piece \cite{lerd83}.
According to GTTM, two types of evidence are involved in the determination of the grouping structure.
The first kind of evidence to perceive a phrase boundary between two melodic events is \emph{local detail}, i.e.\ relative temporal proximity like slurs and rests  (GPR 2a), inter-onset-interval (IOI) (GPR 2b) and change in register (GPR 3a), dynamics (GPR 3b), articulation (GPR 3c) or duration (GPR 3d).

The organization of \emph{larger-level} grouping involves an intensification of the effects picked out by GPRs 2 and 3 on a larger temporal scale (GPR 4), symmetry (GPR 5) and parallelism (GPR 6).
While Lerdahl and Jackendoff's work did not attempt to quantify these rules, a computational model for identification of segment boundaries that numerically quantifies the GPRs 2a, 2b, 3a and 3d was proposed by \citet{Frankland:2004ug}.
This model encodes melodic profiles using the absolute duration of the notes, and MIDI note numbers for representing absolute pitch.

A model related to the GPRs was proposed by \citet{Cambouropoulos:2001vj}.
The Local Boundary Detection Model (LBDM) consists of a \emph{change} rule and a \emph{proximity} rule, operated on melodic profiles that encode pitch, IOI and rests.
On the one hand, the change rule identifies the strength of a segment boundary in relation to the degree of change between consecutive intervals  (similar to GPR 3).
On the other hand, the proximity rule considers the size of the intervals involved (as in GPR 2).
The total boundary strength is then computed as a weighted sum of the boundaries for pitch, IOI and rests, where the weights were empirically selected.

\citet{Temperley:2001wm} introduced a similar method, called Grouper, that partitions a melody (represented by onset time, off time, chromatic pitch and a level in a metrical hierarchy) into non-overlapping groups.
 Grouper uses three \emph{phase structure preference rules} (PSPR) to assess the existence of segment boundaries.
PSPR 1 locates boundaries at large IOIs and large offset-to-onset intervals (OOIs), and is similar to GPR 2, while PSPR 3 is a rule for metrical parallelism, analogous to GPR 6.
PSPR 2 relates to the length of the phrase and was empirically determined by Temperley using the Essen Folk Song Collection (EFSC), and therefore, may not be a general rule~\cite{Pearce:2010ig}.

\subsection{Statistical and Information-Theoretic Segmentation}\label{sec:stat-inform-theor}

\citet{Pearce:2010ig} applied two information theoretic approaches, initially designed by \citet{Brent99anefficient} for word identification in unsegmented speech, to construct boundary strength profiles (BSPs) for melodic events. This method relies on the assumption that segmentation boundaries are located in places where certain information theoretic measures have a higher numerical value than in the immediately neighboring locations. The first approach constructs BSPs using transition probability (TP),  the conditional probability of an element of a sequence given the preceding element, while the second method relies on point-wise mutual information (PMI), that measures to which degree the occurrence of an event reduces the model's uncertainty about the co-occurrence of another event, to produce such BSPs

Inspired by developments in musicology, computational linguistics and machine learning, Pearce, M\"ullensiefen and Wiggins offered the IDyOM model.
IDyOM is an unsupervised, multi-layer, variable-order Markov model that computes the conditional probability and Information Content (IC) of a musical event, given the prior context.
An overview of IDyOM can be found in~\cite{Pearce:2010ij}.

\subsection{Pseudo-Supervised Training}
Although the term pseudo-supervised does not seem to have a well-established meaning, our use of the term is compatible with its use in~\cite{noklestad09:_machin_learn_approac_anaph_resol}, in the sense that a supervised approach is used to predict targets that are computed from the input data, rather than relying on hand-labeled (or otherwise authoritative) targets.
The automatically generated targets (in this case IC values) are not themselves the actual targets of interest (the boundary segments) but are instrumental to the prediction of the actual targets.

Similar methods are proposed by~\citet{Lee:2013ta} and~\citet{Hinton:2014wy}, where supervised models are used to generate targets (pseudo labels or soft-targets) from new data. But in contrast to a \emph{pseudo-supervised} approach, these methods require hand-labeled data, and are strictly taken a \emph{semi-supervised} approach, in which predictive models are trained partly in an unsupervised manner, and partly using hand-labeled data.

In general, both semi- and pseudo-supervised learning benefit from the use of unlabeled information by using Bayesian approaches, regarding the distribution of unlabeled data. From a formal standpoint, these techniques act as regularizers of the model parameters, and thus, prevent overfitting.  Approaches like \emph{entropy regularization}~\cite{Grandvalet:2004wl} use the principle of maximum entropy to select a prior distribution of the model parameters, and then optimize the model in Maximum a Posteriori (MAP) sense.

\section{Method}\label{sec:Method_seg}

The primary assumption underlying statistical models of melodic segmentation is that the perception of segment boundaries is induced by the statistical properties of the data. RBMs (Section~\ref{sec:rbm}) can be trained effectively as a generative probabilistic model of data (Section~\ref{sec:training_seg}), and are therefore a good basis for defining a segmentation method. However, in contrast to sequential models such as recurrent neural networks, RBMs are models of static data and do not model temporal dependencies. A common way to deal with this is to feed the model sub-sequences of consecutive events (n-grams) as if they were static entities, without explicitly encoding time. This n-gram approach allows the model to capture regularities among events that take place within an n-gram. With some simplification, we can state that these regularities take the form of a joint probability distribution over all events in an n-gram. With Monte-Carlo methods, we can use this joint distribution to approximate the conditional probability of some of these events, given others. This procedure is explained in Sections~\ref{sec:appr-prob-v} and~\ref{sec:Posterior}.
The representation of music events is described in Section~\ref{sec:DataRepresentation}, and Section~\ref{sec:information-content} details how the IC of music events is computed based on their estimated conditional probabilities.
In Section~\ref{sec:pseudo-superv-train}, we describe how training a supervised model using IC values as (pseudo) targets can act as a form of regularization.
Finally, Section~\ref{sec:peak-picking} describes how segment boundaries are predicted from sequences of IC values.

\newcommand{\idy}{IDyOM}

\subsection{Relation to Other Statistical Models}\label{sec:relation-idyom-model}
Although our RBM-based method works with n-gram representations just as the statistical methods discussed in Section~\ref{sec:stat-inform-theor}, the approaches are fundamentally different.
Models such as \idy{}, TP and PMI are based on n-gram counting, and as such have to deal with the trade-off between longer n-grams and sparsity of data that is inevitable when working with longer sub-sequences.
In \idy{}, this problem is countered with ``back-off'', a heuristic to dynamically decrease or increase the n-gram size as the sparsity of the data allows.
In contrast, an RBM does not assign probabilities to n-grams based directly on their frequency counts.
The non-linear connections between visible units (via a layer of hidden units) allow a much smoother probability distribution, that can also assign a non-zero probability to n-grams that were never presented as training data.
As a result, it is possible to work with a fixed, relatively large n-gram size, without the need to reduce the size in order to counter data sparsity.

Every computational model requires a set of basic features that describe musical events.
In \idy{}, these basic features are treated as statistically independent, and dependencies between features are modeled explicitly by defining combined viewpoints as cross-products of subsets of features.
An advantage of the RBM model is that dependencies between features are modeled as an integral part of learning, without the need to specify subsets of features explicitly.

Finally, the statistical methods discussed in Section~\ref{sec:Related} are fundamentally n-gram based, and it is not obvious how these methods can be adapted to work with polyphonic music rather than monophonic melodies.
Although the RBM model presented here uses an n-gram representation, it is straight-forward to adopt the same segmentation approach using a different representation of musical events, such as the note-centered representation proposed in~\cite{grachten14:_asses_learn_score_featur_model}.
This would make the RBM suitable for segmenting polyphonic music.

\subsection{Restricted Boltzmann Machines} \label{sec:rbm}
An RBM is a stochastic Neural Network with two layers, a visible layer with units $\mathbf{v} \in \{0,1\}^r$ and a hidden layer with units $\mathbf{h} \in \{0,1\}^q$ \cite{Hinton:2002ic}.
The units of both layers are fully interconnected with weights $\mathbf{W} \in \mathbb{R}^{r \times q}$, while there are no connections between the units within a layer.

In a trained RBM, the marginal probability distribution of a visible configuration $\mathbf{v}$ is given by the equation
\begin{equation}
p(\mathbf{v}) = \frac{1}{Z} \sum_{\mathbf{h}}{e^{-E(\mathbf{v}, \mathbf{h})}},
\end{equation}
where $E(\mathbf{v}, \mathbf{h})$ is an energy function.
The computation of this probability distribution is usually intractable, because it requires summing over all possible joint configurations of $\mathbf{v}$ and $\mathbf{h}$ as
\begin{equation}
Z = \sum_{\mathbf{v},\mathbf{h}}{e^{-E(\mathbf{v},\mathbf{h})}}.
\end{equation}

\subsection{Approximation of the Probability of $v$ }\label{sec:appr-prob-v}

A possibility to circumvent the intractability to compute the probability of a visible unit configuration $\mathbf{v}$ is to approximate it through Monte Carlo techniques.
To that end, for $N$ randomly initialized \emph{fantasy particles}\footnote{See \cite{tielemanPcd2008}} $\mathbf{Q}$, we execute Gibbs sampling until thermal equilibrium. 
In the visible \emph{activation vector} $\mathbf{q}_i$ of a fantasy particle $i$, element $q_{ij}$ specifies the probability that visible unit $j$ is on.
Since all visible units are independent given $\mathbf{h}$, the probability of $\mathbf{v}$ based on one fantasy particle's visible activation is computed as:

\begin{equation}
p(\mathbf{v} | \mathbf{q}_i) = \prod_j{p(v_j | q_{ij})}.
\end{equation}

As we are using binary units, such an estimate can be calculated by using a binomial distribution with one trial per unit.
We average the results over $N$ fantasy particles, leading to an increasingly close approximation of the true probability of $\mathbf{v}$ as N increases:

\begin{equation}\label{binomial}
p(\mathbf{v} | \mathbf{Q}) = \frac{1}{N}\sum_i^N{\prod_j{\binom{1}{v_j}q_{ij}^{v_j}(1-q_{ij})^{1-v_j}}}.
\end{equation}

\subsection{Posterior Probabilities of Visible Units}\label{sec:Posterior}
When the visible layer consists of many units, $N$ will need to be very large to obtain good probability estimates with the method described above.
However, for conditioning a (relatively small) subset of visible units $\mathbf{v}_y \subset \mathbf{v}$ on the remaining visible units $\mathbf{v}_x = \mathbf{v} \setminus \mathbf{v}_y$, the above method is very useful.
This can be done by Gibbs sampling after randomly initializing the units $\mathbf{v}_y$ while clamping all other units $\mathbf{v}_x$ according to their initial state in $\mathbf{v}$.
In Eq.
\ref{binomial}, all $\mathbf{v}_x$ contribute a probability of $1$, which results in the conditional probability of $\mathbf{v}_y$ given $\mathbf{v}_x$.

We use this approach to condition the units belonging to the last time step of an n-gram on the units belonging to preceding time steps. For the experiments reported in this chapter, we found that it is sufficient to use 150 fantasy particles and for each to perform 150 Gibbs sampling steps.

\subsection{Training}\label{sec:training_seg}

We train a single RBM using \emph{persistent contrastive divergence} (PCD)~\cite{tielemanPcd2008} with \emph{fast weights}~\cite{tielemanFpcd2009}, a variation of the standard \emph{contrastive divergence} (CD) algorithm~\cite{hinton06}.
PCD is more suitable for sampling than CD because it results in a better approximation of the likelihood gradient.

Based on properties of neural coding, sparsity and selectivity can be used as constraints for the optimization of the training algorithm~\cite{Goh:2010wi}.
Sparsity encourages competition between hidden units, and selectivity prevents over-dominance by any individual unit.
A parameter $\mu$ specifies the desired degree of sparsity and selectivity, whereas another parameter $\phi$ determines how strongly the sparsity/selectivity constraints are enforced. 

\subsection{Data Representation}\label{sec:DataRepresentation}

\begin{figure}[t]
\centering\footnotesize
\includegraphics[width=0.8\linewidth]{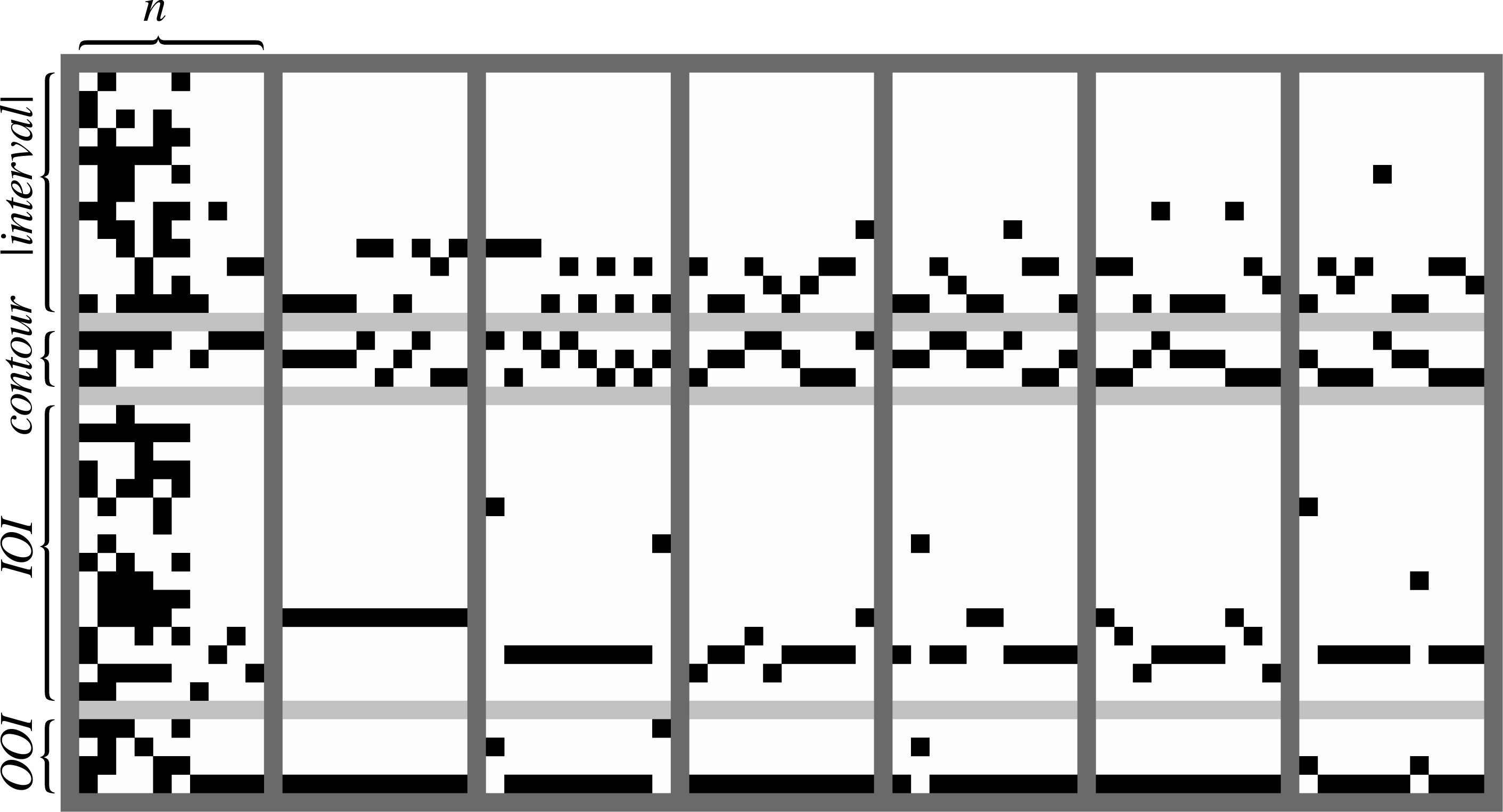}
\vspace{1mm}
\caption{Seven examples of n-gram training instances (n=10) used as input to the RBM.
Within each instance (delimited by a dark gray border), each of the ten columns represents a note.
Each column consists of four \emph{one-hot} encoded viewpoints: $|$\emph{interval}$|$, \emph{contour}, \emph{IOI} and \emph{OOI} (indicated by the braces on the left). The viewpoints are separated by horizontal light gray lines for clarity.
The first instance shows an example of noise padding (in the first six columns) to indicate the beginning of a melody.}
\label{fig:Representation}
\end{figure}

From the monophonic melodies, we construct a set of n-grams by using a sliding window of size n and a step size of 1.
For each note in the n-gram, four basic features are computed: 1) absolute values of the pitch interval between the note and its predecessor (in semitones); 2) the contour (up, down, or equal); 3) inter-onset-interval (IOI); and 4) onset-to-offset-interval (OOI). The IOI and OOI values are quantized into semiquaver and quaver, respectively.
Each of these four features is represented as a binary vector, and its respective value for any note is encoded in a one-hot representation. The first n-1 n-grams in a melody are noise-padded to account for the first n-1 prefixes of the melody. Some examples of binary representations of n-grams are given in Figure~\ref{fig:Representation}.

\subsection{Information Content}\label{sec:information-content}

\begin{figure}
\centering\footnotesize
\includegraphics[width=1.\linewidth]{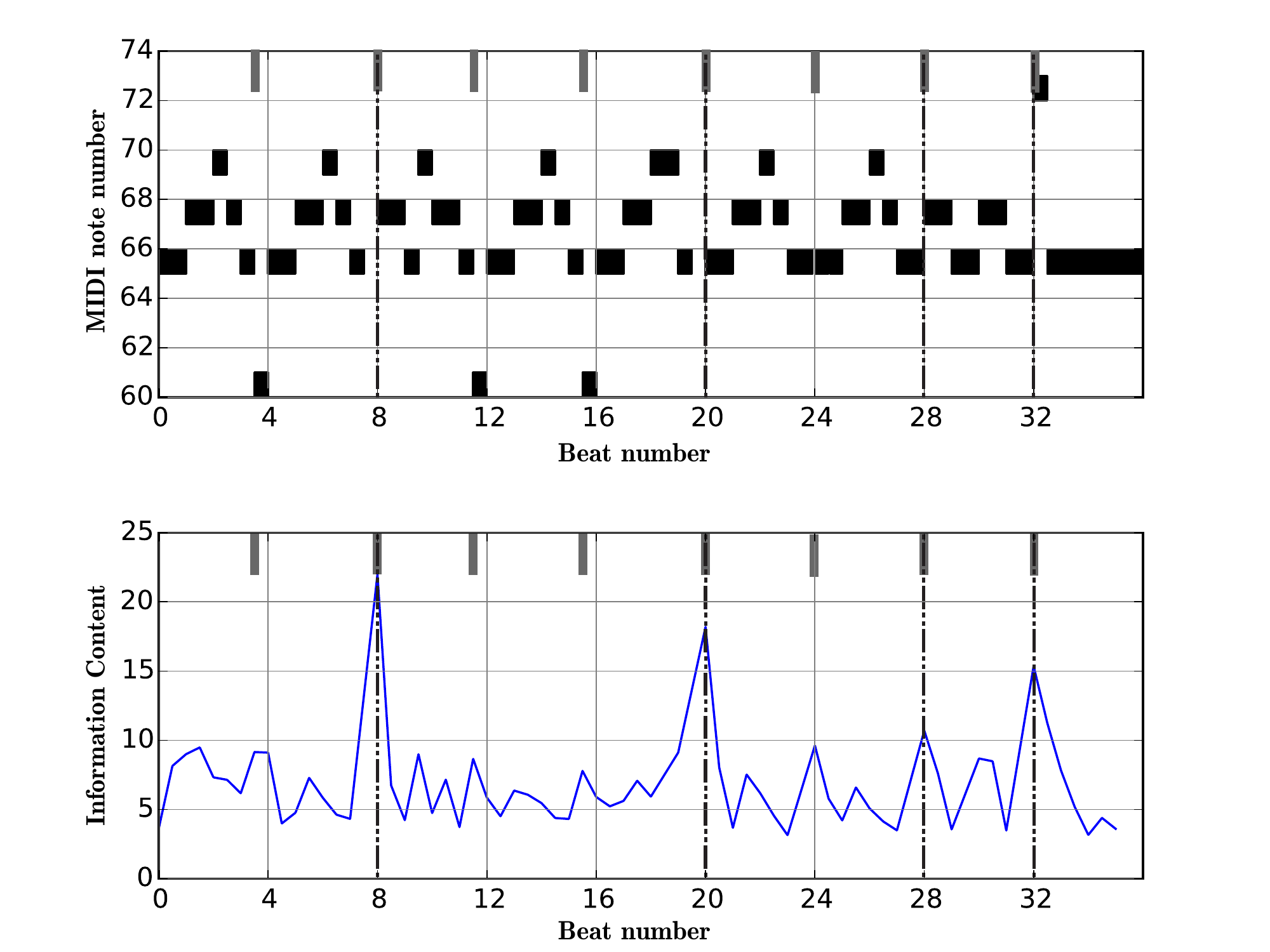}
\caption{A BSP calculated from 11-grams.
The upper figure shows the notes of 9 measures (36 beats) of a German folk song.
The lower figure shows a BSP (i.e., IC) used for segmentation.
The correct segmentation (ground truth) is depicted as vertical grey bars at the top of the figures, segment boundaries found by our model are shown as dashed vertical lines.
Note that the BSP has particularly high peaks at rests and large intervals.
However, the segment boundary found at beat $28$ does not have any of those cues and was still correctly classified.}
\label{fig:bs_curve}
\end{figure}

After training the model as described in \ref{sec:training_seg}, we estimate the probability of the last note conditioned on its preceding notes for each n-gram as introduced in \ref{sec:Posterior}.
From the probabilities $p(e_t \mid e_{t-n+1}^{t-1})$ computed thus, we calculate the IC as:

\begin{equation}\label{eq:ic}
h(e_t \mid e_{t-n+1}^{t-1}) = log_2 \frac{1}{p(e_t \mid e_{t-n+1}^{t-1})},
\end{equation}
where $e_t$ is a note event at time step t, and $e_k^l$ is a note sequence from position $k$ to $l$ of a melody.
IC is a measure of the unexpectedness of an event given its context.
According to a hypothesis of~\citet{Pearce:2010ij}, segmentation in auditory perception is determined by perceptual expectations for auditory events.
In this sense, the IC relates directly to this perceived boundary strength; thus we call the IC over a note sequence \emph{boundary strength profile}.

\subsection{Pseudo-Supervised Optimization}\label{sec:pseudo-superv-train}

\begin{algorithm}
\KwData{ Set of n-grams : $\mathbf{V}=\{ \mathbf{v}_1, \dots, \mathbf{v}_N\}$}
Train an RBM by optimizing the model parameters as
\begin{equation}
\tilde \theta = \argmax_\theta \log p(\mathbf{v} \mid \theta)
\label{eq:rbm_pretrain}
\end{equation}
 
 Compute the set of \emph{pseudo-targets} $\mathbf{T}=\{\mathbf{t}_1,\dots,\mathbf{t}_N\}$ as 
  \begin{align}
 \mathbf{t}_t(\mathbf{v}_t ; \tilde \theta) = h(e_t \mid e_{t-n+1}^{t-1}), 
 \end{align}
 where $\mathbf{v}_t$ is the encoding of the n-gram $\{e_{t-n+1}, \dots, e_{t}\} $, and $h(e_t \mid e_{t-n+1}^{t-1})$ is the IC computed as in Eq.~\eqref{eq:ic}.

Build a three layered FFNN and optimize it in a supervised way, using the set of pseudo-targets $\mathbf{T}$ as
 \begin{equation}
 \hat \theta = \argmin_\theta \sum_{i=1}^N\norm{\mathbf{t} (\mathbf{v}_t; \tilde \theta)- \mathbf{y}(\mathbf{v}_t ; \theta)}^2,
 \label{eq:ffnn_train}
 \end{equation}
 where $\mathbf{y}(\mathbf{v}_t; \theta)$ is the output of the FFNN for $\mathbf{v}_t$ given the model parameters $\theta$.
 
 \Return 
Model parameters $\hat \theta$
 \caption{Pseudo-supervised training} \label{alg:pseudo_supervised}
 \label{algo}
\end{algorithm}

\begin{figure}
\centering
\includegraphics[width=\linewidth]{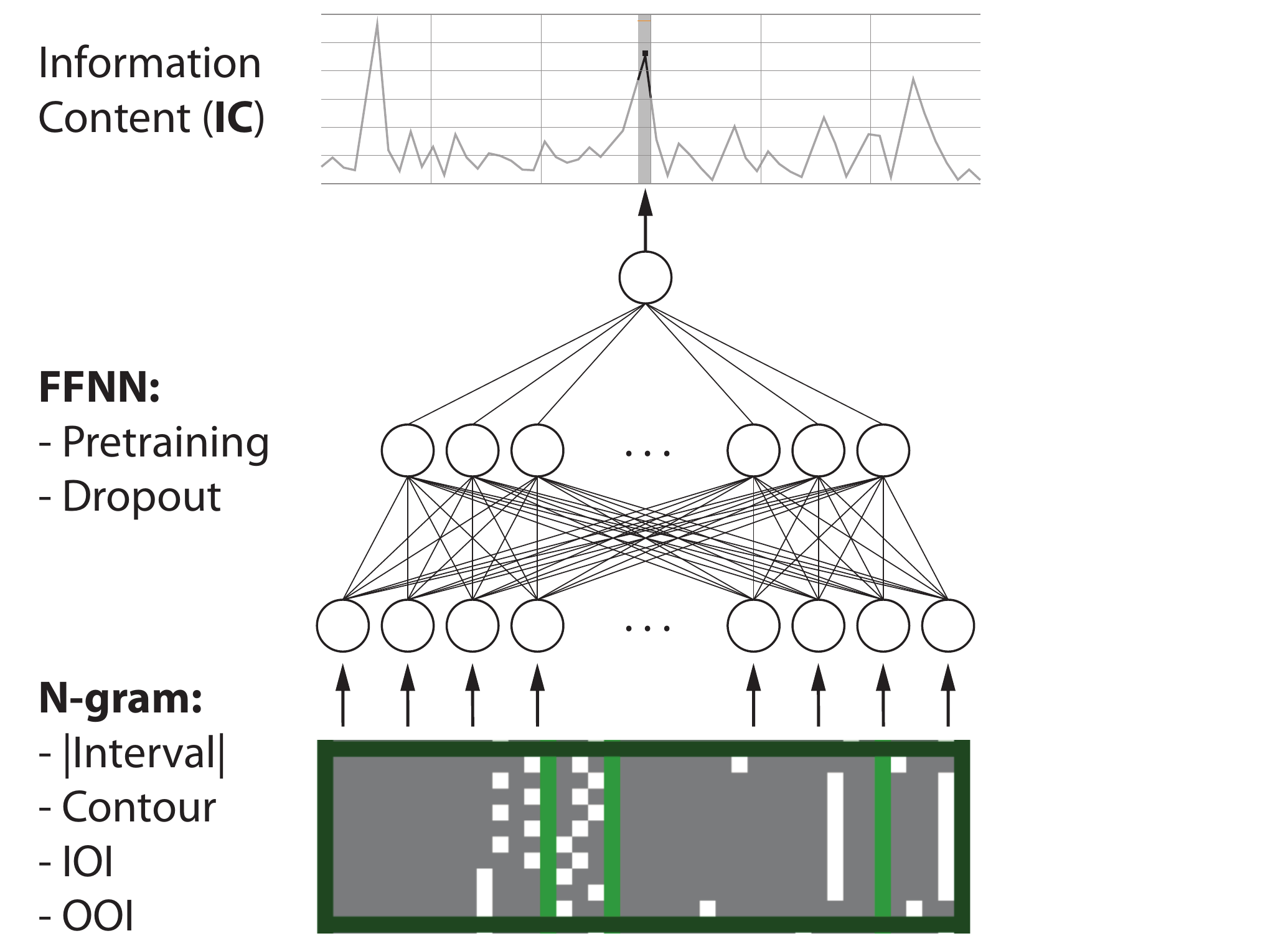}
\vspace{.3cm}
\caption{Schematic depiction of the pseudo-supervised optimization. Note that the n\=/gram is linearized before it is fed into the Feed-Forward Neural Network.}
\label{fig:pseudo}
\end{figure}

As an optimization step, we do not use the BSP estimated from the RBM for segmentation. Instead, we train an FFNN to predict the estimated BSP directly from the data in a non-probabilistic manner, and use that curve for predicting segment boundaries (by the procedure described in Section~\ref{sec:peak-picking}, see Figure \ref{fig:pseudo} for a depiction of the pseudo-supervised training). This is a way of context-sensitive smoothing, which is achieved by the generalization ability of the NN. Note that no labeled data is used at any stage of the processing pipeline. The fact that this approach still improves the segmentation results is evidence that the generative model, as described in Section \ref{sec:rbm}, provides noisy IC estimates. This is either due to poor approximations to the actual IC by the model itself, or since the data is noisy with respect to prototypical segment endings.

The proposed pseudo-supervised training method is shown in Algorithm~\ref{alg:pseudo_supervised}. Formally, this method is an approximate MAP estimation of the parameters using entropy regularization \cite{Grandvalet:2004wl}. In this method, the MAP estimate of the model parameters is computed as
\begin{equation}
\theta_{MAP} =  \argmax_\theta \log p(\mathbf{v} \mid \theta) - \lambda  H(\mathbf{t} \mid \mathbf{v} ; \theta),
\label{eq:map}
\end{equation}
where $H(\mathbf{t} \mid \mathbf{v})$ is the conditional Shannon entropy of the targets  given the inputs, and $\lambda$ is a Lagrange multiplier. In the proposed algorithm, this approximation is obtained by independently optimizing $\log p(\mathbf{v} \mid \theta)$ (see Eq.~\eqref{eq:rbm_pretrain}), and then minimizing Eq.~\eqref{eq:ffnn_train}, which is equivalent to maximizing 
\begin{equation}
p(\mathbf{t} \mid \mathbf{v}; \theta,\beta) = \mathcal{N} (\mathbf{t} \mid \mathbf{y}(\mathbf{v}, \theta), \beta^{-1}),
\end{equation}
where $\beta$ is the precision (inverse variance) of the distribution. This precision can be found by minimizing the negative log-likelihood of the above probability to give
\begin{equation}
\beta = \frac{N}{\sum_{i} \norm{\mathbf{t}_i - \mathbf{y}(\mathbf{v}_i, \theta)}^2}.
\end{equation}
The Shannon entropy for this distribution is given by
\begin{align}
H(\mathbf{t} \mid \mathbf{v} ; \theta, \beta) &=  \mathbb{E} \left\{ -  \log p(\mathbf{t} \mid \mathbf{v}) \right\} \notag \\
&= \frac{1}{2} \log \left( \frac{2\pi}{\beta}\right) + \frac{1}{2},
\end{align}
 which is minimal, since $\sum_{i} \norm{\mathbf{t}_i - \mathbf{y}(\mathbf{v}_i, \theta)}^2$ is minimal. Therefore, optimizing Eq.~\eqref{eq:ffnn_train} is equivalent to minimizing the entropy term in Eq.~\eqref{eq:map}.

We use the fact that the RBM is a generative model, and therefore, the pseudo targets $\mathbf{t}$ come from the computation of the IC from a probabilistically sound estimate of the input data. In this way, pseudo-supervised learning can be understood as a suboptimal entropy-regularized MAP model of the model parameters.

\subsubsection{Training}
To compute $\hat{\theta}$ in Equation~(\ref{eq:ffnn_train}), we use a three-layered FFNN with sigmoid hidden units and a single linear unit in the output layer. We pre-train the first hidden layer with PCD and fine-tune the whole stack with Backpropagation, by minimizing the mean square error. As targets, we use the boundary strength values, estimated by the initial model described in Section \ref{sec:Posterior}.

After training, the outputs $\mathbf{y}(\mathbf{v}_i ; \hat{\theta})$ of the FFNN are used as (improved) estimates of the information content of $e_t$, given $\{e_{i-n+1}, \dots, e_{t-1}\}$.

\subsection{Peak Picking}\label{sec:peak-picking}
Based on the BSP described in the previous section, we need to find a concrete binary segmentation vector.
For that, we adopt the peak picking method described in~\cite{Pearce:2010ij}.
This method finds all peaks in the profile and keeps those which are $k$ times the standard deviation greater than the mean boundary strength, linearly weighted from the beginning of the melody to the preceding value:

\begin{equation}
S_n > k \sqrt{\frac{\sum_{i=1}^{n-1}{(w_i S_i - \bar{S}_{w,1\dots n-1})^2}}{\sum_1^{n-1}{w_i}}} + \frac{\sum_{i=1}^{n-1}{w_i S_i}}{\sum_1^{n-1}{w_i}},
\end{equation}
where $S_m$ is the $m$-th value of the BSP, and $w_i$ are the weights which emphasize recent values over those of the beginning of the song (triangular window), and $k$ has to be found empirically.

\section{Experiment}\label{sec:Experiment}

\subsection{Training Data}\label{sec:TrainingData}

In this work, we use the Essen Folk Song Collection (EFSC) \cite{TheEssenFolksongC:1995um}.
This database is a widely used corpus in MIR for experiments on symbolic music.
This collection consists of more than 6000 transcriptions of folk songs primarily from Germany and other European regions.
The EFSC collection is commonly used for testing computational models of music segmentation, because it is annotated with phrase markers.

In accordance with~\cite{Pearce:2010ij}, we used the \emph{Erk} subset of the EFSC, which consists of 1705 German folk melodies with a total of $78,995$ note events.
Phrase boundary annotations are marked at about $12\%$ of the note events.

\subsection{Procedure}

The model is trained and tested on the data described in Section~\ref{sec:TrainingData} with various n-gram lengths between 3 and 10. For each n-gram length, we perform 5-fold cross-validation and average the results over all folds.
Similar to the approach in~\cite{Pearce:2010ij}, after computing the BSPs, we evaluate different $k$ from the set $\{0.70, 0.75, 0.80, 0.85, 0.90, 0.95, 1.00\}$ (initial IC estimation), and $\{0.24, 0.26, 0.28, 0.30, 0.32, 0.34, 0.36\}$ (after pseudo-supervised optimization), and choose the value that maximizes F1 for the respective n-gram length.
To make results comparable to those reported in~\cite{Pearce:2010ij}, the output of the model is appended with an implicit (and correct) phrase boundary at the end of each melody.

Since the hyper-parameters of the model are inter-dependent, it is infeasible to search for the optimal parameter setting exhaustively.
We have manually chosen a set of hyper-parameters that give reasonable results for the different models tested. For the initial IC estimation, we use 200 hidden units, a momentum of 0.6, and a learning rate of 0.0085 which we linearly decrease to zero during training. With increasing n-gram length we linearly adapt the batch size from 250 to 1000. Also, we use $50\%$ dropout on the hidden layer and 20\% dropout on the visible layer.

The fast weights used in the training algorithm (see Section~\ref{sec:training_seg}) help the fantasy particles mix well, even with small learning rates.
The learning rate of the fast weights is increased from 0.002 to 0.007 during training.
The training is continued until convergence of the parameters (typically between 100 and 300 epochs).
The sparsity parameters (see Section~\ref{sec:training_seg}) are set to $\mu = 0.04$, and $\phi = 0.65$, respectively.
In addition, we use a value of $0.0035$ for $L2$ weight regularization, which penalizes large weight coefficients.

For pre-training of the first layer in the FFNN, we change the learning rate to 0.005, leave the batch size constant at 250 and increase the weight regularization to 0.01. We again use dropout, for both the pre-training and the fine-tuning. 

\begin{figure}
\centering
\includegraphics[width=\linewidth]{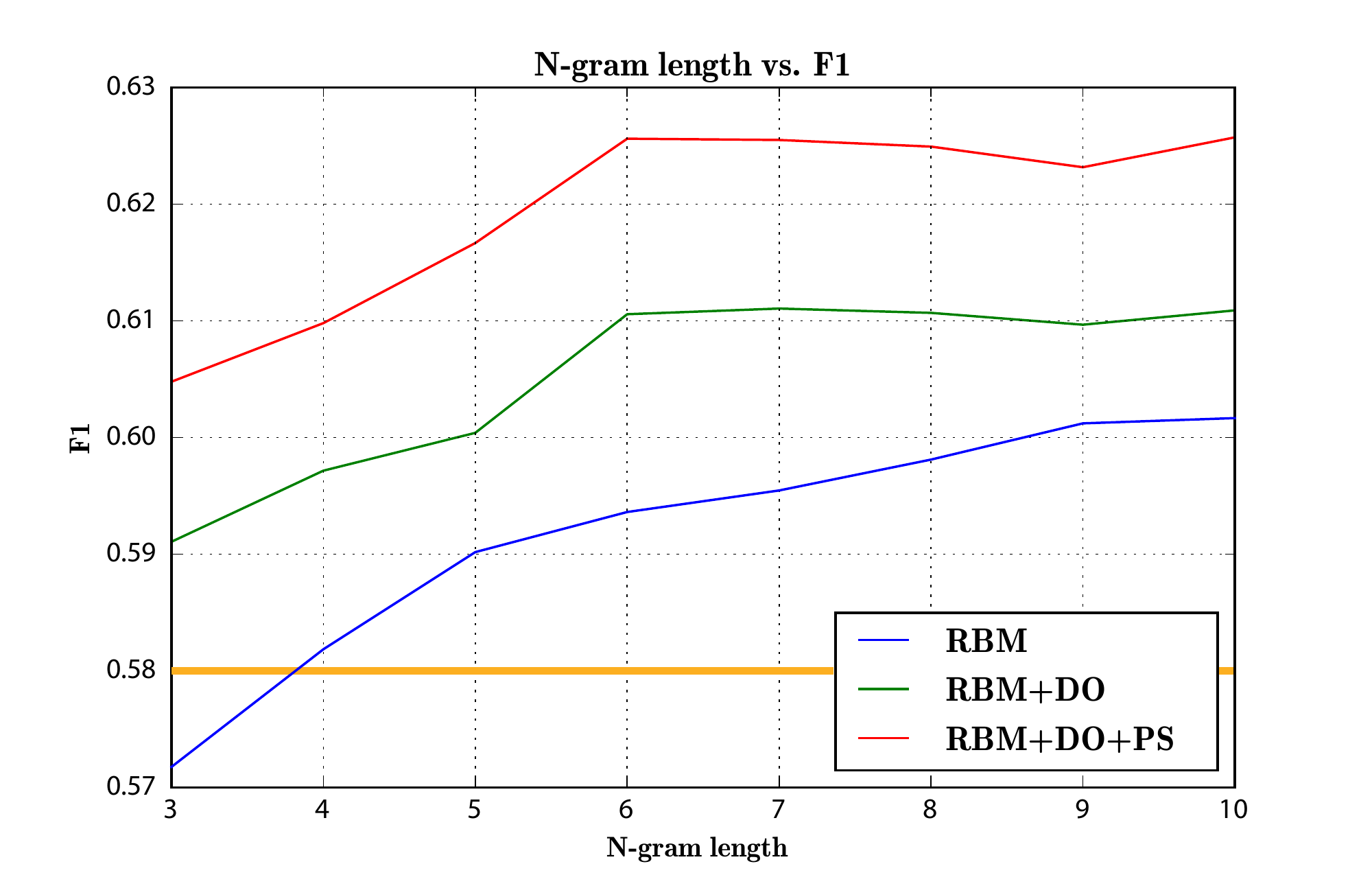}
\caption{F1 scores for different N-gram lengths and methods. The orange horizontal line marks the baseline of the probabilistic IDyOM model~\cite{Pearce:2010ij}.}
\label{fig:f_scores}
\end{figure}

\section{Results and Discussion}\label{sec:results}
We tested three different representations for pitch, yielding the following F1 scores for 10-grams (without dropout and pseudo-supervised training): \emph{absolute pitch} (0.582), \emph{interval} (0.600), and the absolute value of interval (i.e. $|$\emph{interval}$|$) plus \emph{contour} (0.602). The latter representation was chosen for our experiments, as it showed the best performance. Not surprisingly, relative pitch representations lead to better results, as they reduce the number of combination possibilities in the input. Even though the difference in F1 score between \emph{interval} and $|$\emph{interval}$|$ \emph{plus contour} representation is not significant, it still shows that it is valid to decompose viewpoints into their elementary informative parts. Such an approach, next to reducing the input dimensionality, may also support the generalization ability of a model (e.g., $|$interval$|$ representation in music may help to understand the concept of inversion).

Figure \ref{fig:f_scores} shows the F1 scores for different N-gram lengths and methods. By using dropout, the F1 score increases considerably, as dropout improves the generalization abilities of the RBM. With the pseudo-supervised approach, again a significant improvement of the classification accuracy can be achieved. This is remarkable, considering that no additional information was given to the FFNN, the improvement was based solely on ``context-sensitive smoothing''.

\begin{figure}
   \centering
   \includegraphics[width=\linewidth]{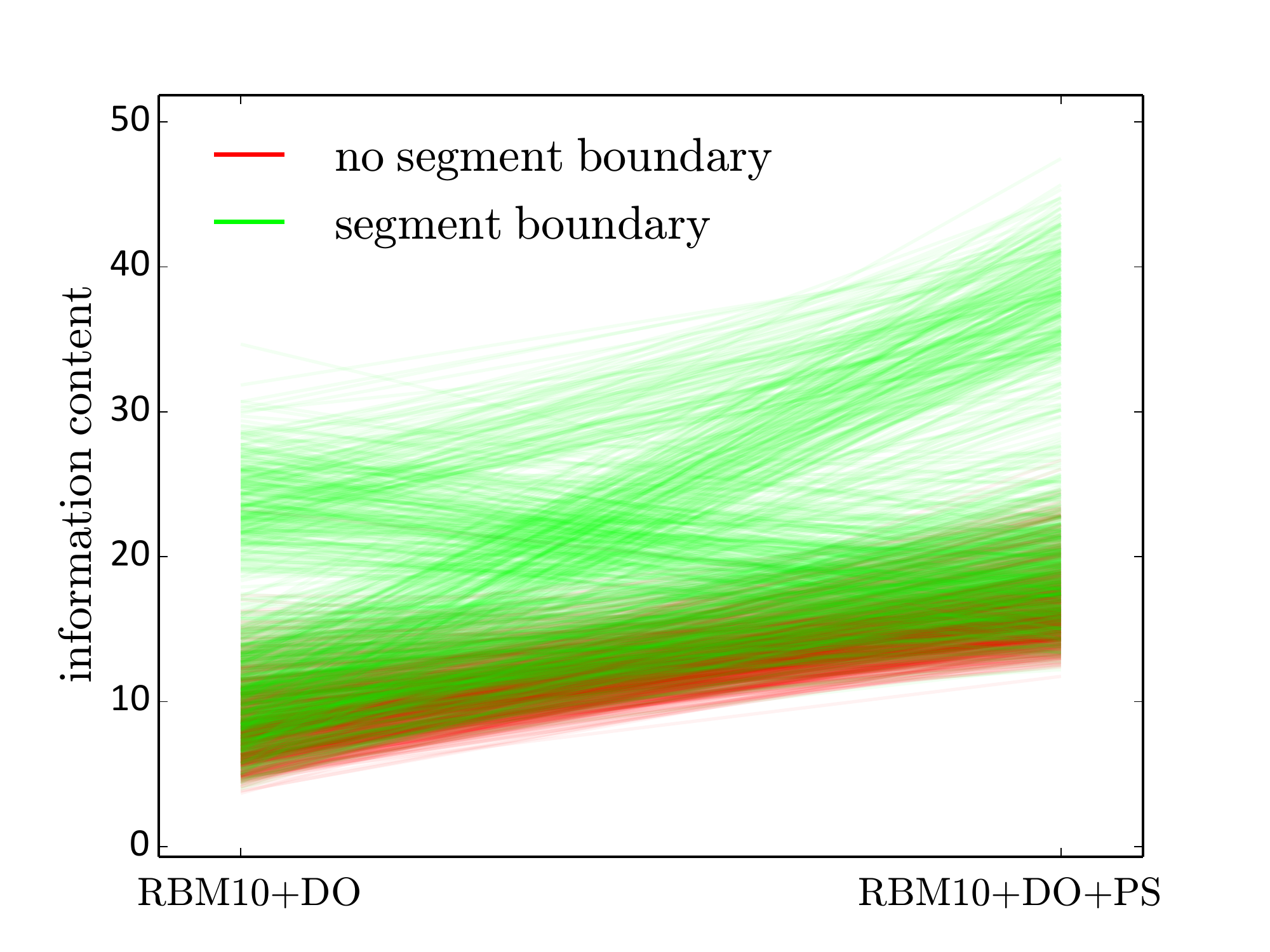}
   \caption{The effect of pseudo-training on estimated IC values; Line segments connect IC values estimated directly from the probabilistic model (RBM10+DO) with the corresponding IC values after pseudo-training (RBM10+DO+PS); Green lines indicate music events that mark a segment boundary, red lines indicate those that do not.}
   \label{fig:trends}
\end{figure}

Figure \ref{fig:trends} shows the adaptation of single IC values through pseudo-supervised optimization. Some previously true positives are erroneously regularized downwards (green lines from upper left to lower right), while some previously false negatives are correctly moved upwards (green lines from lower left to the upper right). Quantitative tests show that our method increases IC values at boundaries more often than it decreases them. In general, if the initial BSP curve is correct in most cases, in pseudo-supervised training, such regularities are detected and utilized.

Table~\ref{results} shows prediction accuracies in terms of precision, recall, and F1 score, both for our method and for the various alternative approaches mentioned in Section~\ref{sec:Related}. The table shows that with the proposed method (RBM10+DO+PS), an information-theoretic approach is now on a par with a Gestalt-based approach (LBDM), while Grouper still provides the best estimates of melodic segment boundaries. However, Grouper exploits additional domain knowledge like musical parallelism, whereas the LBDM model, as well as (RBM10+DO+PS), are pure representatives of the Gestalt-based paradigm and the information-theoretic paradigm, respectively.

The \emph{GPR 2a} method is a simple rule that predicts a boundary whenever a rest occurs between two successive notes. Note how \emph{GPR 2a} accounts for a large portion of the segment boundaries (approx. 45\%). This implies that the challenge is mainly in recognizing boundaries that do not co-occur with a rest. For boundaries without rests, the pseudo-supervised approach yields an improvement of $3.7\%$ in the F-score, while boundaries indicated by a rest did not improve any more (as for those boundaries the initial approach already yields an F-score of $0.99$).

\begin{table}[t]
\centering
\begin{tabular}{@{}llllc@{}}
\toprule
Model & Precision & Recall & F1 \\ \midrule
Grouper & 0.71 & 0.62 & 0.66  \\
LBDM & 0.70 & 0.60 & 0.63 \\
\textbf{RBM10+DO+PS} & \textbf{0.80} & \textbf{0.55} & \textbf{0.63} \\
RBM10+DO & 0.78 & 0.53 & 0.61 \\
RBM10 & 0.83 & 0.50 & 0.60 \\
IDyOM & 0.76 & 0.50 & 0.58 \\
GPR 2a & 0.99 & 0.45 & 0.58 \\ \midrule
GPR 2b & 0.47 & 0.42 & 0.39 \\
GPR 3a & 0.29 & 0.46 & 0.35 \\
GPR 3d & 0.66 & 0.22 & 0.31 \\
PMI & 0.16 & 0.32 & 0.21 \\
TP & 0.17 & 0.19 & 0.17 \\ \midrule
Always & 0.13 & 1.00 & 0.22 \\
Never & 0.00 & 0.00 & 0.00 \\
\bottomrule
\end{tabular}
\caption{Results of the model comparison, ordered by F1 score.}
\label{results}
\end{table}

\section{Conclusion}\label{sec:concl-future-work-seg-mix}
In this chapter, an RBM-based unsupervised probabilistic method for segmentation of melodic sequences was presented.
In contrast to other statistical methods, our method does not rely on frequency counting and thereby circumvents problems related to data sparsity.
Furthermore, we showed how a technique we call \emph{pseudo-supervised} training improves the prediction accuracy of the method.
We used the information content (IC) of musical events (estimated from the probabilistic model) as a proxy for the actual target to be predicted (segment boundaries). 
With these \emph{pseudo targets}, we trained a feed-forward neural network. 
We showed that segment boundaries estimated from the output of this network are more accurate than boundaries estimated from the \emph{pseudo targets} themselves.

In this study, we used the IC as estimated from an RBM, but the pseudo-supervised approach may benefit from including IC estimates from other models, such as IDyOM~\cite{pearce05:_const_evaluat_statis_model_melod}, which, in addition, uses a short-term memory. Besides, there are other probabilistic architectures, such as conditional RBMs~\cite{Taylor:2006vl}, that seem appropriate for estimating IC values from data. Furthermore, although the focus of this chapter has been on IC, it is intuitively clear that IC is not the only factor that determines the perception of segment boundaries in melodies. Future experimentation is necessary to determine whether (combinations of) other information-theoretic quantities are also helpful in detecting melodic segment boundaries. Finally, we wish to investigate whether there are further problems where our method could be beneficial. In general, pseudo-supervised optimization could improve features which are noisy either because of the way they are calculated or because of noise in the data upon which the features are based.


\clearpage

\chapter{Top-Down Imposition of Higher-Level Structure}\label{sec:constraints}

\begin{abstract}
In this chapter, we introduce a method for imposing higher-level structure on generated, polyphonic music.
A Convolutional Restricted Boltzmann Machine (C-RBM) as a generative model is combined with gradient descent constraint optimization to provide further control over the generation process.
Among other things, this allows for the use of a ``template'' piece, from which some structural properties can be extracted, and transferred as constraints to newly generated material.
The sampling process is guided with Simulated Annealing in order to avoid local optima, and find solutions that both satisfy the constraints and are relatively stable concerning the C-RBM.
Results show that with this approach it is possible to control the higher-level self-similarity structure, the meter, as well as tonal properties of the resulting musical piece while preserving its local musical coherence.
\end{abstract}

\section{Introduction}\label{sec:introduction}
For centuries, mathematical formalisms have been used to generate musical material \cite{Kirchmeyer:1968vk}.
Since computers can automate such processes, automatic music generation has become a small, but a steadily emerging field in Artificial Intelligence and Machine Learning.
Nevertheless, automatic music generation as a problem is far from solved: musical outputs created by artificial systems are regarded as a curiosity by human listeners at best, but all too often they are taken as a direct offense to our sense of musical aesthetics.
This sensitivity to violations of even the most subtle musical norms illustrates how complex the problem of (mainly polyphonic) music generation is.
Besides, there are only a few objective evaluation criteria to rigorously test and compare music generation systems, all of which involve human judgment \cite{jordanous2012standardised,pearce2001towards}.

This is lamentable, not least since successful methods for automatic music generation would be of considerable commercial interest to the music, gaming and film industries.
Moreover, potential applications that have remained unexplored as of yet, including adaptive music in cars or fitness applications, could personalize music and thus provide a completely new listening experience.

In line with a global surge in deep learning and neural network modeling over the past decade, several studies address the task of music modeling as a form of sequence learning, in which musical pieces are formulated as a time series of musical events, using state-of-the-art sequence models such as Recurrent Neural Networks (RNN) and Long Short-Term Memory (LSTM).
For restricted genres or representations such as monophonic folk melodies~\cite{sturm16:_music_trans_model_compos_using_deep_learn}, symbolic chord sequences, or drum tracks~\cite{choi16:_text_lstm_autom_music_compos} and even in polyphonic music with clearly defined melodic voices, such as Bach chorales~\cite{boulanger2012modeling,hadjeres2017deepbach,liang2017automatic,huang2017counterpoint},
 sequence modeling approaches yield impressive results that are sometimes hard to distinguish from human-composed material.

However, in more complex musical material, such as piano music from the classical (e.g., Mozart) or romantic period (e.g., Chopin, Liszt), not to mention orchestral works, important musical characteristics may defy straight-forward time series modeling approaches.
\emph{Tonality} for example, is the characteristic that music is perceived to be in a particular (possibly time-variant) \emph{musical key}, implying that some pitches are regarded as more stable than others.
Although the perception of musical key is a complex topic in itself, there is evidence that an essential determining factor is the frequency of occurrence of pitches in the piece~\cite{Smith:2004in}.

In addition to tonality, \emph{meter} is a vital aspect of music.
Perception of meter is the sensation that musical time can be divided into equal intervals at different levels and that positions that coincide with the start of higher-level intervals have more importance than those coinciding with lower levels.
Analogous to the perception of musical key, the perception of meter is in part related to the distribution of musical events over time~\cite{palmer90:_mental_repres_music_meter}. 

Lastly, music often transmits a sense of coherence over the course of the piece, in that it has a \emph{structural organization} in which motifs (small musical patterns), but also larger units such as phrases, melodies or complete sections of the music are repeated throughout the piece, either literally or in an altered form.
This characteristic is reflected in the \emph{self-similarity matrix} of the music, where entry $(i, j)$ expresses the similarity between the music at positions $i$ and $j$.
This coherence by way of repeating and developing musical material throughout the piece is arguably one of the aspects of music that make listening and re-listening a valuable experience to human listeners.

Important musical characteristics such as these are not straight-forward to capture using a simple sequence modeling approach unless the musical material is restricted or simplified as mentioned above.
For example, it is a challenge for current models to generate music that is diverse and interesting, and at the same time induces a stable musical key over an extended period.
It is even more challenging to generate music that exhibits the hierarchical organizational structure common in human-composed music.
For instance, the rather common pattern of a melodic line from the opening of a piece being repeated as the conclusion of that piece is difficult to capture, even if state-of-the-art sequence models like LSTMs are capable of learning long-term dependencies in the data.
Models that fail to capture these higher-level musical characteristics may still produce music that on a short timescale sounds very convincing, but on longer stretches of time tends to sound like it wanders, and misses a sense of musical direction.

In this work, we do not address the problem of learning the discussed properties from musical data.
Instead, our contribution is a method to enforce such properties as constraints in a sampling process.
We start from the observation stated above, that neural network models in the various forms that have recently been proposed are adequate for learning the \emph{local} structure and coherence of the musical surface, that is, the \emph{musical texture}.
The strategy we propose here uses such a neural network (more specifically, a Convolutional Restricted Boltzmann Machine, see Section~\ref{sec:rbm_const}) as one of several components that jointly drive an iterative sampling process of music generation.
This model is trained on musical data and is used to ensure that the musical texture is similar to that of the training data.

The other components involved in the sampling process are cost functions that express how well higher-level constraints like tonal, metrical and self-similarity structure are satisfied in the musical material at each stage in the process.
By performing \emph{gradient descent} on these cost functions, the sampling process is driven to produce musical material that better satisfies the constraints.
The desired shape of these higher-level structural constraints on the piece is not hard-coded in the cost functions but is instantiated from an existing piece.
As such, the existing piece serves as a \emph{structure template}.
The generation process then results in a re-instantiation of that template with new material.
Through recombination of structural characteristics from a musical piece that is not part of the neural network's training data, the model is forced to produce novel solutions.

We refer to the above process as \emph{constrained sampling}.
Informally, it can be imagined as a musical drawing board that is initially filled with random pitches at random times, and where the neural network model, as well as each of the constraints, take turns to slightly tweak the current content of the drawing board to their liking.
This process continues until the musical content can no longer be tweaked to better satisfy the model and constraints jointly.

We believe this approach provides a novel and useful contribution to the problem of polyphonic music generation.
Firstly, it takes advantage of the strengths of state-of-the-art deep learning methods for data modeling.
The combination with multi-objective constraint optimization compensates for the weaknesses of these methods for music generation, mentioned above.
Moreover, it provides high-level user control\footnote{The user has control over the generation process by choice of the template piece, but also more directly by manipulation of the structure templates extracted from the template piece. This aspect is beyond the scope of the current study.} over the generation process, and allows for relating the generated material to existing pieces, both of which are interesting from a musical point of view.

In addition to the description of the constrained sampling approach to music generation, the goal of the present study is to validate the approach in several ways.
First, we present a qualitative discussion of generated musical samples, illustrating the effect of the constraints on the musical result.
We show that although the constraints and the neural network embody different objectives, the evolving musical material produced by the constrained sampling process tends to approximate these different objectives simultaneously.
Furthermore, we adopt \emph{Information Rate} as an independent measure of musical structure \cite{wang2015pattern}, in order to assess the effect of the repetition structure constraint, and compare our approach to two variants of the state-of-the-art RNN-RBM model for polyphonic music generation \cite{boulanger2012modeling}.
This comparison shows that the constrained sampling approach substantially increases the Information Rate of the produced musical material over both unconstrained approaches (including the RNN-RBM variants), implying a higher degree of structure.

The chapter is structured as follows:
Section~\ref{sec:related_work_const} gives an overview of related models and computational approaches to music generation.
Section~\ref{sec:Method_const} describes the components involved in the constrained sampling approach, which is subsequently introduced in Section~\ref{sec:cs}.
Section~\ref{sec:experiment} describes the experimental validation of the constrained sampling approach in the context of Mozart piano sonatas.
We discuss the empirical findings in Section~\ref{sec:results-discussion_const} and give conclusions and future perspectives in Section~\ref{sec:future}.

\section{Related work}\label{sec:related_work_const}

Early attempts using neural networks for music generation were reported in \cite{Todd:1989bb},
where monophonic melodies were encoded in pitch and duration and an RNN was trained to predict upcoming events.
In \cite{mozer1994neural}, an RNN system called CONCERT was proposed, and first systematic tests on how well local and global musical structure (e.g.
AABA) of simple melodies could be learned, were made.
Also, chords were used to test if this facilitates the learning of higher-level structure, but the results were not convincing.
This was one of the first experiments which showed the difficulties of learning structure in music.

More recently, \citet{Eck:2002:FLM:870511} trained a Long Short-Term Memory (LSTM) network (a state-of-the-art RNN variant), jointly on a single chord sequence along with several different melodies.
This is an example of a harmonic template which guides a melodic improvisation.
Chords and melody notes were separated in the input and output connections so that the model could not mix up harmony and melody notes.
That way, the LSTM could overfit on the single chord sequence and generalize on the monophonic melodies.
In a polyphonic setting, common RNNs are not suitable for generation in a random walk fashion as the distribution at time $t$ is conditioned only on the past, but it would be necessary to consider the full joint distribution also for all possible settings in $t$.

This limitation was overcome by the RNN-RBM model for polyphonic music generation introduced in \cite{boulanger2012modeling}, and the similar LSTM Recurrent Temporal RBM (LSTM-RTRBM) model proposed in \cite{lyu2015modelling}.
In those architectures, the recurrent components ensure temporal consistency, while the RBM component is used for sampling a plausible configuration in $t$.
Our contribution lies between the before-mentioned LSTM approach where a higher-level structure is imposed by using a template, and the RNN-RBM approach, where the ability of an RBM to model low-level structure is utilized.
Further methods to constrain generated material by pre-defining voices to guide the sampling process are introduced in \cite{hadjeres2017deepbach} (based on LSTMs), and \cite{huang2017counterpoint} (based on Convolutional Neural Networks), both of which generate Bach chorales.

Another approach that uses a probabilistic model and constraints is called ``Markov constraints'' \cite{pachet2011}, which allows for sampling from a Markov chain while satisfying pre-defined hard constraints.
This is conceptually similar to our method, but we use a different probabilistic model and soft constraints.
Our method is more flexible in defining new constraints, and it is of linear runtime, while Markov constraints are more costly, but also more exact.
\citet{herremans2016morpheus} use a constrained variable neighborhood search to generate polyphonic music obeying a tension profile and the repetition structure from a template piece.
Furthermore, \citet{barbieri2011regularized} uses soft constraints to incorporate a-priori-information in a Gibbs sampling process for a User Rating Profile model.

\citet{cope1996experiments} explicitly imposes higher-level structure in a generation process.
So-called SPEAC identifiers are used to generate music in a given tension-relaxation scheme.
Another example of generating structured material is that in \cite{eigenfeldt2013evolving}, where Markov chains and evolutionary algorithms are used to generate repetition structure for Electronic Dance Music.
\citet{collins2016developing} use Markov chains together with structure schemes and specific methods for handling transitions between repeating segments in order to generate structured music.
Similarly, \citet{whorleytransformational} use a transformational approach to generate Bach chorales, and \citet{conklin_semiotic} generates chords using Markov chains and pre-defined repetition structures.
A Hierarchical Variational Autoencoder for music generation, able to learn hierarchical tonal structure, is proposed in \cite{roberts2017hierarchical}.

A method similar to our approach is that of \citet{gatys2016image} for image style transfer.
They also use gradient descent on the input for satisfying multiple objectives (approximating a gram-matrix defining the style, as well as the initial picture defining the structure).
In contrast to our method, there is no probabilistic model involved.
Different solutions for the same objectives are merely due to the initialization of the input with random noise.
Application of the method to music would, however, not allow for the control needed to comply with some music-specific properties like self-similarity or the somewhat strict rules of musical tonality.

Examples of connectionist generation approaches with constraints in other domains are that in \cite{Graves:2013ua} where biasing and priming is used in LSTMs to control the generation of sequences of handwritten text, and in \cite{Taylor:2006vl} where a conditional RBM is used to generate different human walking styles.
In such problems, the number of variables is fixed and lower than in music generation, and structural properties like repetition are either not a property of the data (handwritten text) or periodic (walking), whereas polyphonic music exhibits complex structure in multiple hierarchical levels.

\section{Method}\label{sec:Method_const}
\begin{sidewaysfigure*}
\centering\footnotesize
\includegraphics[width=\textwidth]{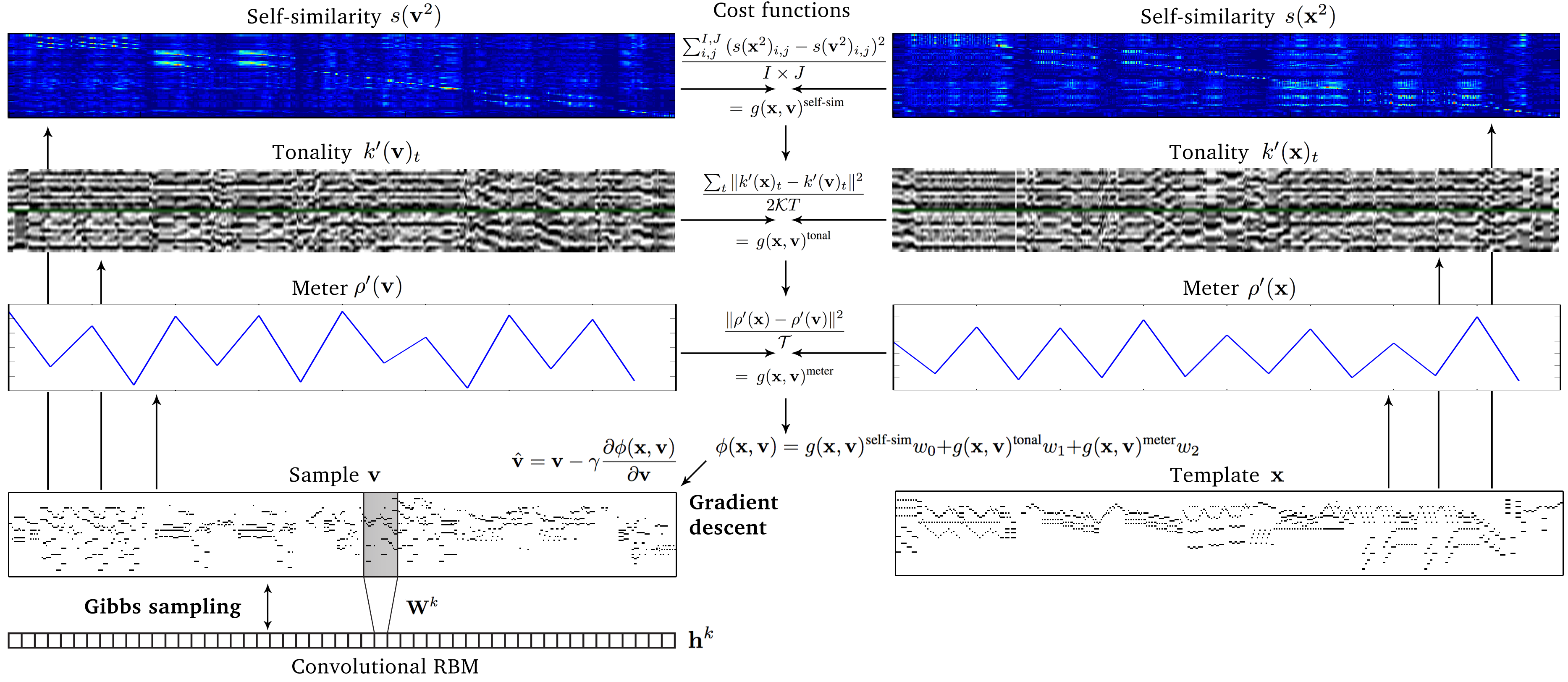}
\caption{Constrained sampling using an existing piece $\mathbf{x}$ as a structure template.
A randomly initialized sample $\mathbf{v}$ is alternately updated with Gibbs sampling (GS) and gradient descent (GD).
In GD, the error $\phi(\mathbf{x}, \mathbf{y})$ between structural features of $\mathbf{x}$ and $\mathbf{v}$ is lowered, in GS the training data distribution is approximated.
The Convolutional RBM consists of visible layer $\mathbf{v}$ and hidden layer $\mathbf{h}$.
The filter $\mathbf{W}^k$ is shared among all units in feature map $\mathbf{h}^k$.
Depicted equations are also given in Section~\ref{sec:gd}.}
\label{fig:cs}
\end{sidewaysfigure*}

In this Section, we describe the methods used to create musical output.
We start by describing the C-RBM used for sampling new content (Section~\ref{sec:rbm_const}).
The gradient descent (GD) method used to impose constraints on the sampling process is introduced in Section~\ref{sec:gd}.
The complete process, referred to as Constrained Sampling (CS) and depicted in Figure~\ref{fig:cs}, is introduced in Section~\ref{sec:cs}.

\subsection{Convolutional Restricted Boltzmann Machine (C-RBM)} \label{sec:rbm_const}
\begin{figure*}
\centering
\includegraphics[width=1.\textwidth]{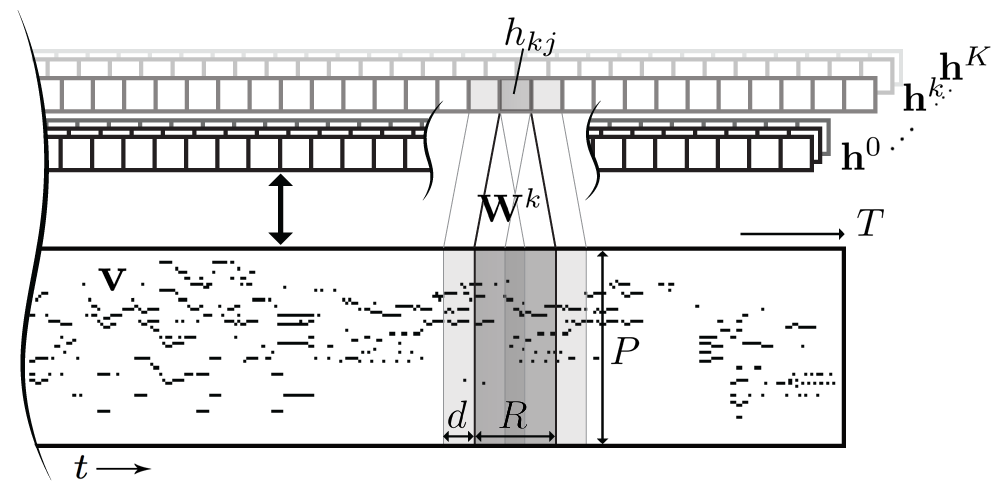}
\caption{Illustration of a C-RBM with strided convolution (using stride $d$) in the time dimension $t$ of a music piece  $\mathbf{v} \in \mathbb{R}^{T \times P}$ in two-dimensional piano roll representation using $K$ one-dimensional feature maps $\mathbf{h}^k$ where all units in a map share their weights $\mathbf{W}^k \in \mathbb{R}^{R \times P}$ and their bias $b_k$ (bias not depicted in the illustration).}
\label{fig:crbm_detail}
\end{figure*}

A C-RBM \cite{lee2009convolutional}
is a two-layered stochastic version of a convolutional neural network with binary units, as known from \citet{lecun1989backpropagation}.
In our setting, the visible layer with units $\mathbf{v} \in \mathbb{R}^{T \times P}$, where $0 \leq v_{tp} \leq 1$, constitutes a piano roll representation (see Section~\ref{sec:repr}) with time $1 \leq t \leq T$ and midi pitch number $1 \leq p \leq P$.
All units in the hidden layer belonging to the $k$th feature map share their weights (i.e.
their filter) $\mathbf{W}^k \in \mathbb{R}^{R \times P}$ and their bias $b_k \in \mathbb{R}$, where $R$ denotes the \emph{filter width} (i.e.
the temporal expansion of the receptive field), and each filter covers the whole midi pitch range $[1, P]$.
We convolve only in the time dimension, which is padded with $R/2$ zeros on either side (the reason for this design decision is given at the end of this section).
We use a stride of $d$, meaning the filters are shifted over the input with step size $d$.
This results in a hidden layer $\mathbf{h} \in \mathbb{R}^{K \times (T/d)}$, where $0 \leq h_{kj} \leq 1$ and $j \in 0 \dots T/d$.
See Figure~\ref{fig:crbm_detail} for an illustration of the C-RBM used in our experiments.

We train the C-RBM with Persistent Contrastive Divergence \cite{tielemanPcd2008} aiming to minimise the free energy function
\begin{equation}\label{equ:energy_const}
\mathcal{F}(\mathbf{v}) = - \sum_{t}\mathbf{a} \, \mathbf{v}^t - \sum_{k,j} \log \left( 1 + e^{\left(b_k + (\mathbf{W}^k * \mathbf{v})_{j\times d} \right)}\right)
\end{equation}

\noindent for training instances $\mathbf{v}$, where $\mathbf{a} \in \mathbb{R}^{P}$ and $\mathbf{b} \in \mathbb{R}^K$ are bias vectors, and $*$ is the convolution operator.
Note that in two-dimensional convolution, each feature map usually has a scalar as bias (e.g.
$b_k$) because all positions in a feature map are assumed to be equivalent.
However, since we convolve only in the time dimension, and since there is a non-uniform distribution over the pitch dimension, we define the bias for the input feature map $\mathbf{v}$ as a vector $\mathbf{a}$ of length $P$.

The probability of a unit being active depends on the full configuration of the opposing layer.
When updating hidden units $\mathbf{h}$ and visible units $\mathbf{v}$, each unit is randomly chosen to be active (i.e.
$1$) or inactive (i.e.
$0$) with probabilities
\begin{equation}\label{eq:ph}
P(h_{kj}=1 \mid \mathbf{v}) = \sigma \big( \big( \sum_{r,p}^{R,P}{W_{r,p}^k \times v_{j\times d+r-\frac{R}{2},p} \big) +b_k \big) }
\end{equation}
and

\begin{equation}\label{eq:pv}
P(v_{tp}=1 \mid \mathbf{h}) = \sigma\big( \big( \sum_{r,k}^{R/d,K}{\tilde{W}_{r\times d,p}^k \times h_{t+r-\frac{R}{2d}}^k\big) +a_p \big),}
\end{equation}
where $\tilde{\mathbf{W}}^k$ denotes the horizontally flipped weight matrix.
Note that it is also valid to propagate such probability values through the network (i.e.
calculate the activation probabilities of one layer based on the \emph{probabilities} of the opposing layer).

A sample can be drawn from the model by randomly initializing $\mathbf{v}$ (following the \emph{standard uniform distribution}), and running block Gibbs sampling (GS) until convergence.
To this end, hidden units and visible units are alternately updated given the other.
In doing so, it is common to sample the states of the hidden units for the top-down pass, but use the probabilities of the visible units for the bottom-up pass.
After an infinite number of such Gibbs sampling iterations, $\mathbf{v}$ is an accurate sample under the model.
In practice, convergence is reached when $\mathcal{F}(\mathbf{v})$ stabilises.

The reason for convolving only in the time dimension is that there are correlations between notes over the whole pitch range.
In a one-layered setting with 2D convolution, the filter height (i.e.
the expansion of filters in the pitch dimension) is typically limited, for example, to one octave.
In that case, correlations would only be learned between notes within one octave.
Learning correlations over a wider range would usually be the role of higher layers in a neural network stack.
However, in order to show the principle of constrained sampling, it is sufficient to use only one layer with 1D convolution, which is also advantageous for limiting the overall complexity of the architecture.

\subsection{Imposing Constraints with Gradient Descent}\label{sec:gd}
When sampling from a C-RBM, the solution is randomly initialized and converges to an accurate sample of the data distribution after many steps (see Section~\ref{sec:rbm_const}).
During this process, we repeatedly adjust the current solution $\mathbf{v}$ towards satisfying a desired higher-level structure regarding some musical properties.
To this end, we subject $\mathbf{v}$ (i.e.
the input, not the model parameters) to a Gradient Descent (GD) optimization process aiming to minimize a differentiable cost function $\phi(\cdot)$ using learning rate $\gamma$ as
\begin{equation}\label{equ:gd}
\hat{\mathbf{v}} = \mathbf{v} - \gamma \frac{\partial \phi(\mathbf{x},\mathbf{v})}{\partial \mathbf{v}},
\end{equation}
where $\mathbf{x} \in \mathbb{R}^{T\times P}$, $0 \leq x_{tp} \leq 1$, is a template piece from which we want to transfer some structural properties to our sample $\mathbf{v}$.
After every GD update, we set each entry
$\hat{v}_{tp} = \min(1, \max(0,\hat{v}_{t,p}))$, to ensure $\hat{\mathbf{v}} \in [0,1]^{T\times P}$.
The cost function may consist of several terms $g_d(\mathbf{x},\mathbf{v})$ (weighted with factors $w_d$), each defining a soft constraint which is to be imposed on the sample:

\begin{equation}\label{equ:cost}
\phi(\mathbf{x},\mathbf{v}) = g_0(\mathbf{x},\mathbf{v}) w_0 + \dots + g_{D-1}(\mathbf{x},\mathbf{v}) w_{D-1}.
\end{equation}

Note that $\mathbf{x}$ and $\mathbf{v}$, as representations of a musical score, could be assumed to be binary, but we define them as \emph{continuous} variables.
This is because we want to store continuous results of the GD optimization in $\mathbf{v}$, as well as intermediate probabilities during Gibbs sampling.
Defining $\mathbf{x}$ as a continuous variable is a generalization towards encoding note intensities or note probabilities, making it possible to express relative importance between notes.

In the following, we will introduce three constraints we tested in our experiments.
Note that the method is not limited to those constraints, and can be extended with additional terms which are differentiable with respect to $\mathbf{v}$.

\subsubsection{Self-Similarity Constraint}\label{sec:selfsim}
The purpose of the self-similarity constraint is to specify the repetition structure (e.g.
AABA) in the generated music piece, using a template \emph{self-similarity matrix} as a target.
Such a self-similarity representation is particularly useful because it also provides \emph{distances} between any two parts of a piece.
Thus, the degree of similarity, including substantial dissimilarity, may be encoded, too.
Such a representation abstracts from the actual musical texture and is therefore to a large extent content-invariant.
This allows for transferring the similarity structure in different hierarchical levels between pieces of different style, tonality, or rhythm.

\begin{figure*}
\centering
\includegraphics[width=\textwidth]{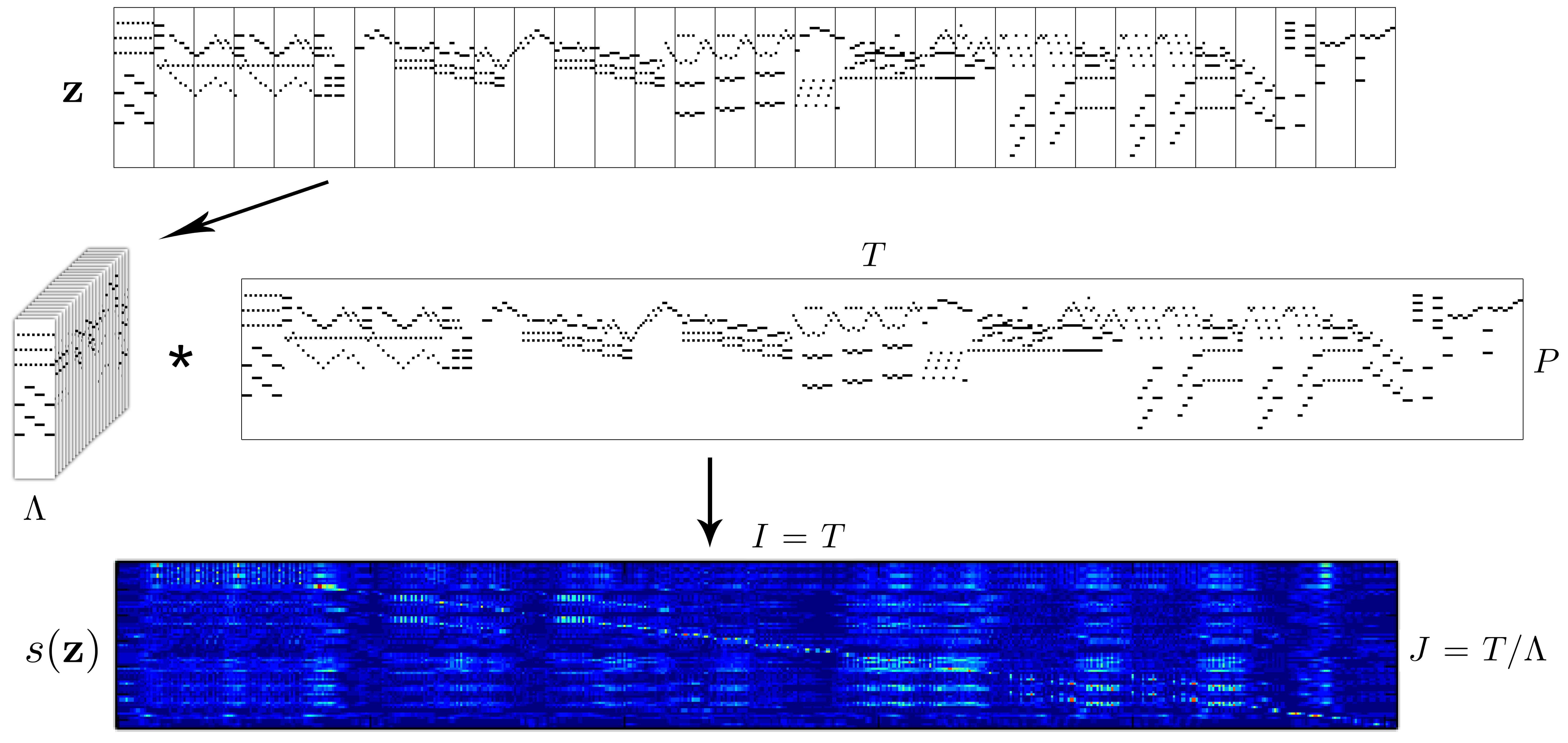}
\caption{Depiction of calculating the self-similarity matrix $s(\mathbf{z}) \in \mathbb{R}^{I \times J}$ using convolution.
A music piece in piano roll representation $\mathbf{z} \in [0,1]^{T \times P}$ is horizontally tiled, and those tiles are used as filters for a convolution with $\mathbf{z}$.
The response for a single filter constitutes a single line in the resulting self-similarity matrix.
Low to high response is depicted in a range from darker blue to brighter red colors.}
\label{fig:selfsim1}
\end{figure*}

A self-similarity matrix $s(\mathbf{z}) \in \mathbb{R}^{I \times J}$ for an arbitrary music piece $\mathbf{z} \in [0,1]^{T \times P}$ in piano roll representation is calculated by tiling $\mathbf{z}$ horizontally in tiles of width $\Lambda$ and by using them as 2-D filters for a convolution over the time dimension of $\mathbf{z}$ (see Figure~\ref{fig:selfsim1}).
Therefore $I = T$ and $J = T / \Lambda$, and we calculate a single entry at position $i,j$ of the self-similarity matrix as

\begin{equation}
s(\mathbf{z})_{i,j} = \sum_{\lambda,p}^{\Lambda, P}{z_{j \times \Lambda + \lambda, p}z_{i+\lambda,p}}.
\end{equation}

To impose the self-similarity constraint, we minimize the mean squared error (MSE) between a target self-similarity matrix of the squared template piece $s(\mathbf{x}^2)$ and the self-similarity matrix of the squared intermediate solution $s(\mathbf{v}^2)$ as

\begin{equation}\label{equ:sqerrselfsim}
g(\mathbf{x},\mathbf{v})^{\text{self-sim}} = \frac{\sum_{i,j}^{I,J}{(s(\mathbf{x}^2)_{i,j} - s(\mathbf{v}^2)_{i,j})^2}}{I \times J}.
\end{equation}

The reason for squaring $\mathbf{x}$ and $\mathbf{v}$ is that it leads to higher stability in the optimization because it reduces low-intensity noise and it adds contrast to the resulting self-similarity matrix.
We also tried to represent transposed repetition as a constraint using two-dimensional convolution.
However, we found that this leads to a perfect reconstruction of the template piece, as such a self-similarity representation fully specifies the musical texture.

\subsubsection{Tonality Constraint}\label{sec:tonality}

\begin{figure}[t]
\begin{center}
\footnotesize
\includegraphics[width=1.\linewidth]{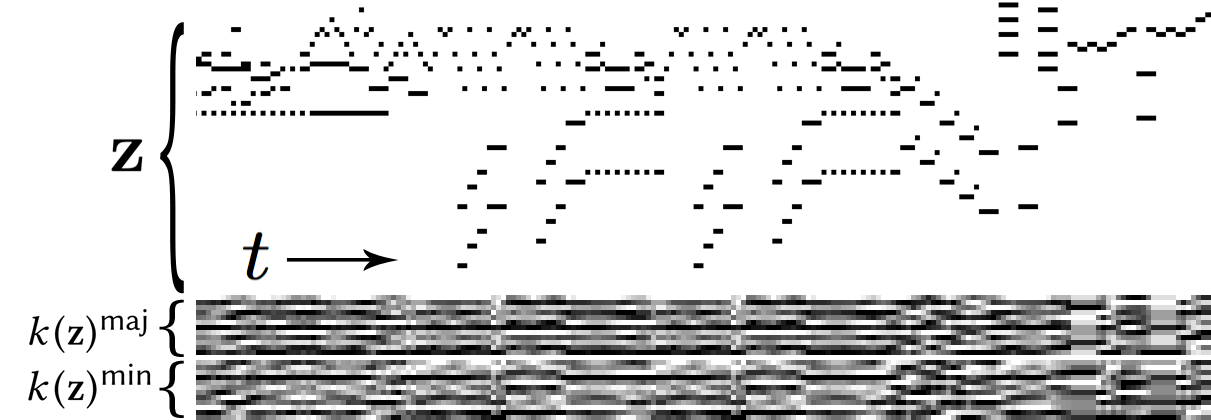}
\caption{Example of key estimation vectors over time.
$k(\mathbf{z})^\text{maj}$ represent estimations for $12$ possible major keys and $k(\mathbf{z})^\text{min}$ represent estimations for $12$ possible minor keys, where the pitch classes constituting the tonic are ordered from the top to the bottom.
Bright pixels represent high strength, and dark pixels represent low strength of the corresponding key.
For example, the upper most line in $k(\mathbf{z})^\text{maj}$ represents the estimation strength of the C major key over time, the third line represents the strength of the D major key, etc.}
\label{fig:tonality}
\end{center}
\end{figure}

Tonality is another fundamental higher-order property in music.
It describes perceived tonal relations between notes and chords.
This information can be used to, for example, determine the \emph{key} of a piece or a musical section.
A key is characterized by a tonal center (the pitch class that is considered to be central, e.g., C, or A\raisebox{.45mm}{$\sharp$}), and a mode (the subset of pitch classes that form part of the key, e.g., \emph{major} or \emph{minor}).
The distribution of pitch classes in the musical texture within a (temporal) window of interest is an essential factor in the perceived key of that window.
Different window lengths $M$ may lead to different key estimates, constituting a hierarchical tonal structure.
A popular method to estimate the key in a given window is to compare the distribution of pitch classes in the window with so-called key profiles $u^{\text{mode}}$ (i.e., paradigmatic relative pitch-class strengths for specific modes; the profiles are invariant to changes of tonal center).
In \cite{Temperley:2001wm}, key profiles for \emph{major mode} $u^{\text{maj}}$ and \emph{minor mode} $u^{\text{min}}$ are defined as

\begin{equation*}
u^{\text{maj}} = (5,\ 2,\ 3.5,\ 2,\ 4.5,\ 4,\ 2,\ 4.5,\ 2,\ 3.5,\ 1.5,\ 4)^\top,
\end{equation*}
\begin{equation*}
u^{\text{min}} = (5,\ 2,\ 3.5,\ 4.5,\ 2,\ 4,\ 2,\ 4.5,\ 3.5,\ 2,\ 1.5,\ 4)^\top,
\end{equation*}

\noindent where the numerical values express the strengths of the different pitch classes that make up the key.
We use these two key profiles as filters for a music piece $\mathbf{z} \in [0,1]^{T \times P}$.
By repeating them $M$ times in the \emph{time dimension}, we obtain a filter for a window of size $M$.
By repeating them $O = P/12$ times in the \emph{pitch dimension}, we extend the filters over all octaves represented in $\mathbf{z}$.
When shifted in the pitch dimension with shifts $\kappa \in 0 \dots \mathcal{K}-1$ we obtain a filter for each of the $\mathcal{K} = 12$ possible keys.
If we choose the profile for a specific mode $u^{\text{mode}}$, an estimation window size $M$, and the number of octaves $O$ represented by $\mathbf{z}$, we obtain a key estimation vector $k(\mathbf{z})_t^\text{mode} \in \mathbb{R}^\mathcal{K}$ at time $t$ for all shifts $\kappa$ as

\begin{equation}
k(\mathbf{z})_t^\text{mode} = \sum_{m,o,i}^{M,O,I}{u^{\text{mode}}((i+\kappa)\bmod I) \cdot z_{t+m,i+o*12}},
\end{equation}
for $I = 12$ entries in key profile $u^{\eta}$, and $\cdot$ denotes the common multiplication of scalars.
Subsequently, we concatenate the key estimation vectors of both modes, $k(\mathbf{z})_t^\text{maj}$ and $k(\mathbf{z})_t^\text{min}$, to obtain a combined estimation vector $k(\mathbf{z})_{t} \in \mathbb{R}^{2\mathcal{K}}$ in $t$, which is finally normalized as
\begin{equation}
k'(\mathbf{z})_{t} = \frac{k(\mathbf{z})_{t} - \min(k(\mathbf{z})_{t}) \mathbf{I}}{\max(k(\mathbf{z})_{t}) - \min(k(\mathbf{z})_{t})},
\end{equation}
where $\mathbf{I}$ is a vector of ones of length $2\mathcal{K}$.\footnote{Even though the derivatives of $\min(\cdot)$ and $\max(\cdot)$ are not guaranteed to be always defined, in practice these cases are hardly ever a problem in gradient descent, and are typically dealt with in software frameworks for symbolic differentiation such as Theano~\cite{2016arXiv160502688short}.}
Figure~\ref{fig:tonality} depicts the resulting concatenated key estimation vectors.
Using these vectors, we may impose a specific tonal progression on our solution by minimizing the MSE between the target estimate $k'(\mathbf{x})_t$ and the estimate of our current solution $k'(\mathbf{v})_t$ such that:

\begin{equation}
g(\mathbf{x},\mathbf{v})^{\text{tonal}} = \frac{\sum_t\|k'(\mathbf{x})_{t} - k'(\mathbf{v})_{t}\|^2}{2\mathcal{K}T}.
\end{equation}

\subsubsection{Meter Constraint}
The meter (e.g.
$3/4$, $4/4$, $7/8$) defines the \emph{duration} and the perceived \emph{accent patterns} in regularly occurring bars of a music piece.
For example, in a $4/4$ meter, one can expect relatively strong accents on the first and the third beat of a bar.
We impose the meter extracted from a template piece on our sample, to obtain a degree of global rhythmic coherence.

Perceived accent patterns depend on the relative occurrence of note onsets in a bar, on the intensity of played notes, or the length of notes starting at the respective positions of a bar.
However, note intensities are not encoded in our data, and it is not obvious how to incorporate note durations in our differentiable cost function.
Therefore, we use note onsets only.
To this end, we constrain the relative occurrence of note onsets within a bar to follow that of a template piece.
Abiding by such a distribution helps the generated material to keep implying a regular meter.

The onset function $\omega(\cdot)$ results from a discrete differentiation over the time dimension of an arbitrary music piece in piano roll representation $\mathbf{z} \in [0,1]^{T \times P}$.
We rectify that result (as we are not interested in note offsets), and sum over the pitch dimension:
\begin{equation}
\omega(\mathbf{z},t) = \sum_p^P\max(0, {z_{t,p} - z_{t-1,p}}).
\end{equation}

In order to calculate the relative occurrences of onsets within a bar, the length $\mathcal{T}$ of a bar has to be pre-defined.
We count the number of onsets occurring on the respective positions of all bars in the music piece.
This is, we sum up all values of distance $\mathcal{T}$ in the onset function $\omega(\cdot)$ as
\begin{equation}
\rho(\mathbf{z})_{\tau} = \sum_{\mu}^{T/\mathcal{T}}{\omega(\mathbf{z}, \tau + \mu*\mathcal{T})},
\end{equation}
where $\tau \in 0 \dots \mathcal{T}-1$ is the position in a bar.
In our experiments, we use a resolution of $16th$ notes in the representation and the template is in $4/4$ meter, therefore $\mathcal{T} = 16$.

To keep the function independent of the absolute number of onsets involved, $\rho(\mathbf{z})$ is standardized by subtracting its mean $\overline{\rho(\mathbf{z})}$ and dividing through its standard deviation $\sigma(\rho(\mathbf{z}))$, resulting in zero mean and unit variance:
\begin{equation}\label{equ:onsetdist}
\rho'(\mathbf{z}) = \frac{\rho(\mathbf{z}) - \overline{\rho(\mathbf{z})}}{\sigma(\rho(\mathbf{z}))}.
\end{equation}

\begin{figure}
\begin{center}
\footnotesize
\includegraphics[width=.8\linewidth]{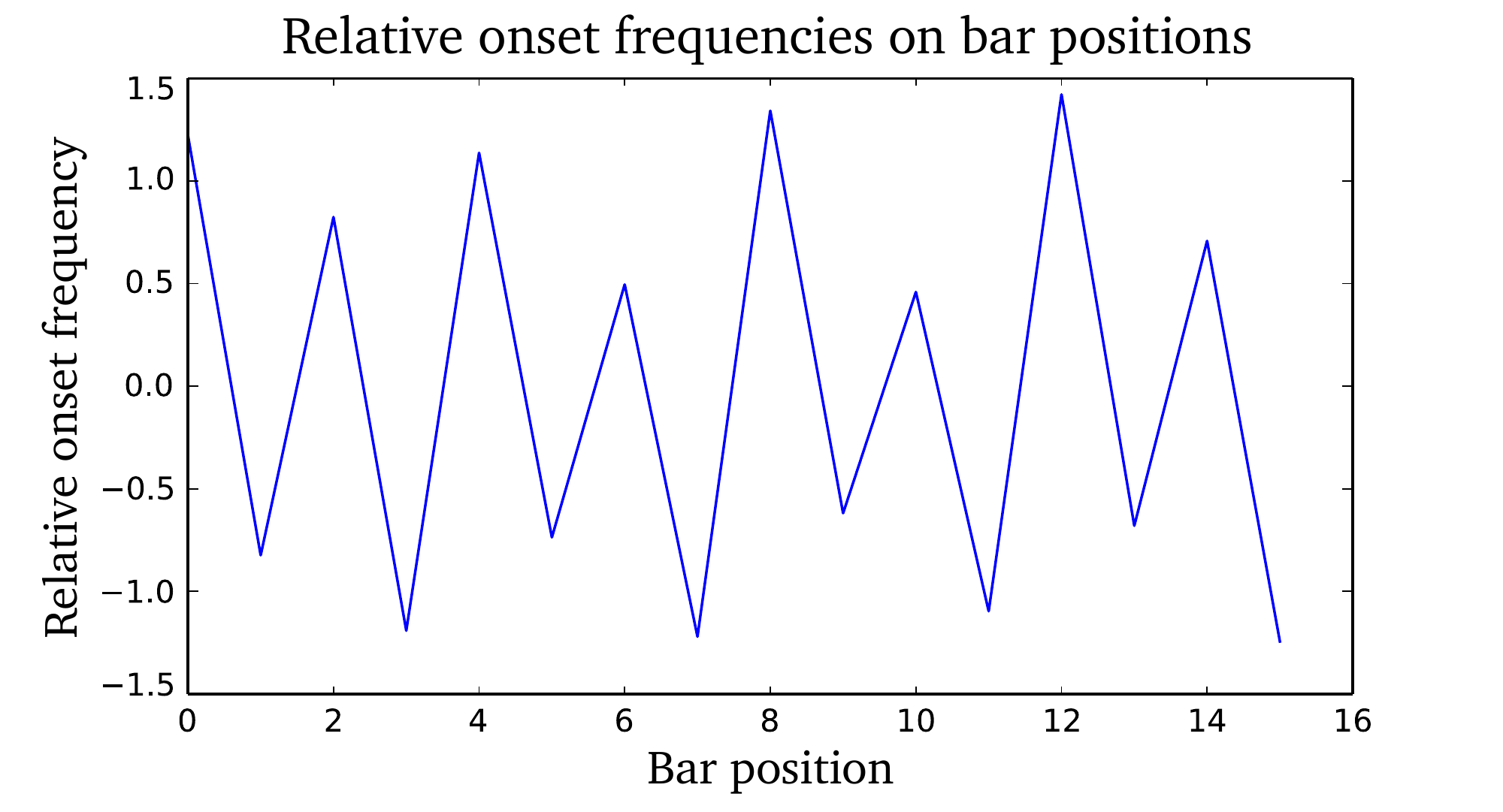}
\caption{Relative (standardized) onset frequencies $\rho'(\mathbf{z})$ on bar positions of a music piece as obtained from Equation~\ref{equ:onsetdist}.}
\label{fig:onsets}
\end{center}
\end{figure}

A standardized onset distribution is plotted in Figure~\ref{fig:onsets}.
Finally, we minimize the MSE between a standardized onset distribution $\rho'(\mathbf{x})$ and that of our intermediate solution $\rho'(\mathbf{v})$ as

\begin{equation}
g(\mathbf{x},\mathbf{v})^{\text{meter}} = \frac{\|\rho'(\mathbf{x}) - \rho'(\mathbf{v})\|^2}{\mathcal{T}}.
\end{equation}

\section{Constrained Sampling}\label{sec:cs}

In this Section, we describe how the C-RBM is used as a generative model to produce musical textures that resemble that of human-composed music, and combined with the soft constraints described above, to enforce additional tonal, meter and self-similarity structure on those textures.

\begin{sidewaysfigure*}
\centering
\includegraphics[width=1\textwidth]{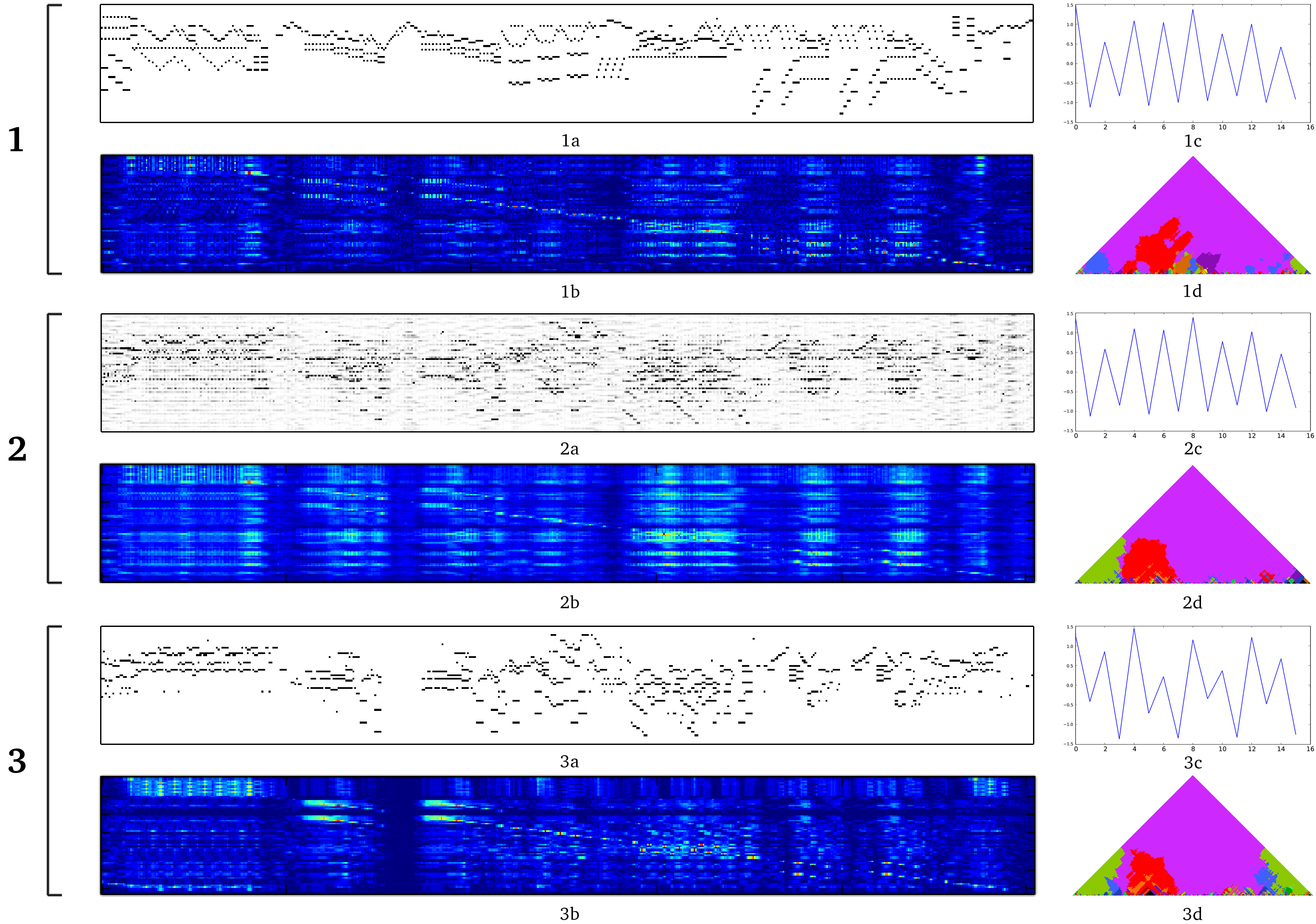}
\caption{Illustration of constrained sampling.
(1) Template piece, (2) Intermediate sample after the GD phase, (3) Sample after the GS phase.
Figures in each group: (a) Piano roll representation, (b) Self-similarity matrix, (c) Onset distribution in $4/4$ meter, (d) Keyscape (see Section \ref{sec:keyscape} for an explanation).
After the GD phase (2), the target higher-level properties imposed as constraints are relatively well approximated.
Due to limited training data and the stochastic nature of Gibbs sampling, after the GS phase the higher-level properties are more dissimilar again.}
\label{fig:iteration}
\end{sidewaysfigure*}

The method proposed here has several practical merits.
First, a C-RBM can take any input as a starting point for (further) sampling.
This allows for local ``mutations'' of intermediate solution candidates in a heuristic process like Simulated Annealing (see Section~\ref{sec:simulatedannealing}) and facilitates the controlled exploration of the search space.
Second, in a C-RBM continuous values in the input are interpreted as probabilities.
This facilitates external guidance through a gradual adaption of note probabilities in a directed gradient descent (GD) optimization process.
For illustration, Figure~\ref{fig:iteration}[2a] shows an example of a piano roll after a GD phase.
The grey tones (non-zero probabilities) in the background of the piano roll will influence the subsequent sampling step from the C-RBM.  
Third, the solution is sampled as a single instance (i.e.
all notes in a music piece are updated simultaneously) and temporal dependencies are modeled in a bi-directional manner.
That way, global constraints can be imposed by iterative adaption of local structures.

\subsection{Example Scheme and Details}

In Figure~\ref{fig:cs}, an overview of Constrained Sampling (CS) is shown.
During the constrained sampling process we alternate between a GS phase with the one-layered C-RBM (Section~\ref{sec:rbm_const}), and a GD optimization phase on the cost functions (Section~\ref{sec:gd}).
In each phase, typically multiple iterative updates take place, and the sampling results are sensitive to the balance struck between the GS and GD phases, in terms of the number of updates performed in each phase.

The numbers proposed in the following CS sampling scheme have been found to work well in our experiments.
The scheme may have to be adapted to work well with other training settings (e.g., different C-RBM architectures, or different constraints), and is mainly for illustrative purposes.
In Algorithm~\ref{alg:cs}, the whole process including Simulated Annealing (see Section~\ref{sec:simulatedannealing}) is shown.

\begin{algorithm}[t]
\KwData{

$\mathbf{x} \in [0,1]^{T \times P}$ -- Template Piece

$\mathbf{v} \in [0,1]^{T \times P}$ -- Random (standard uniform dist.)
$\mathbf{\hat{v}} = \mathbf{v}, N = 250, M = 15$
}

\For{$i \in 1 \dots N$}{
$\mathbf{v}' \leftarrow \mathbf{v}$

$\mathbf{v} \leftarrow$ 20 GD steps using Eq.~\ref{equ:gd} with $\gamma = 10$

\For{$j \in 1 \dots M$}{
$\mathbf{v} \leftarrow$ 100 GS steps using $\mathbf{v}$

$\mathbf{v} \leftarrow$ 1 GD step using Eq.~\ref{equ:gd} with $\gamma = 5$
}
\tcc{Simulated Annealing}
$T_i = 1 - i/N$

$r_e, r_c \leftarrow$ random values $\in [0,1]$

\If {$r_e < \exp\left(-\frac{\mathcal{F}'(\mathbf{v}) - \mathcal{F}'(\mathbf{v}')}{T_i}\right)$ \textbf{or} $r_c < \exp\left(-\frac{\phi'(\mathbf{x},\mathbf{v})- \phi'(\mathbf{x},\mathbf{v}')}{T_i}\right)$}{
$\mathbf{v} \leftarrow \mathbf{v}'$
}
\tcc{Store best solution so far}
\If {$\frac{\mathcal{F}'(\mathbf{v}) + \phi'(\mathbf{x},\mathbf{v})}{2} < \frac{\mathcal{F}'(\mathbf{\hat{v}}) + \phi'(\mathbf{x},\mathbf{\hat{v}})}{2}$}
{$\mathbf{\hat{v}} \leftarrow \mathbf{v}$}
}
\Return
$\mathbf{\hat{v}}$
\caption{Constrained Sampling.
Number of iterations represent an example scheme, as used in the experiments.} \label{alg:cs}
\end{algorithm}

Starting from a random uniform noise in $\mathbf{v}$, we alternate $20$ GD steps using learning rate $\gamma = 10$ (i.e.
GD phase, see Figure~\ref{fig:iteration}[2a] for a result of this phase), and $1500$ GS steps (i.e.
GS phase, see Figure~\ref{fig:iteration}[3a] for a result of this phase).
We consider this \emph{one} constrained sampling iteration.
We found that results improve when, during the GS phase, after every $100$ GS steps we execute $1$ GD step with learning rate $\gamma = 5$.
After 250 CS iterations, the sample with the minimum average value of the standardized GD cost and the standardized free energy over the whole CS process is chosen (see Section~\ref{sec:simulatedannealing} on standardizing the cost and free energy functions).

\begin{figure}[t]
\centering\footnotesize
\includegraphics[width=.8\linewidth]{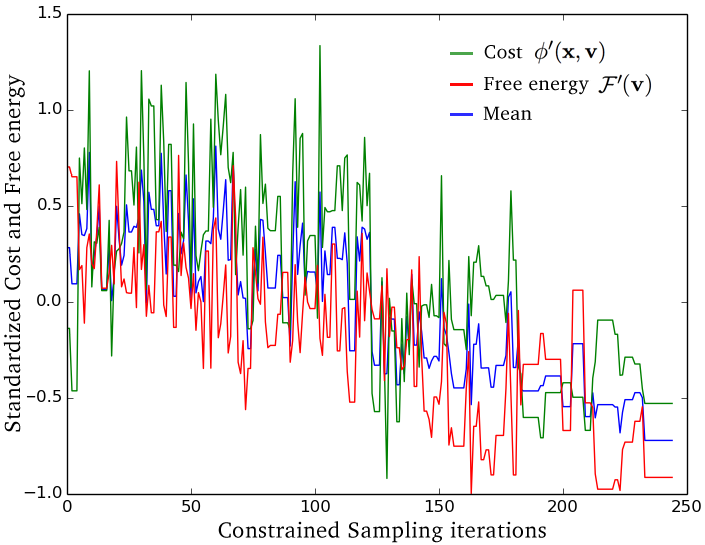}
\caption{Standardised cost, free energy and their mean in a constrained sampling process over 250 iterations.
Periods of constant cost (horizontal line segments) in later iterations are a result of Simulated Annealing, where some unfavorable solution candidates are rejected.}
\label{fig:cost_curve}
\end{figure}

\begin{figure}[t]
\centering\footnotesize

\includegraphics[width=.7\linewidth]{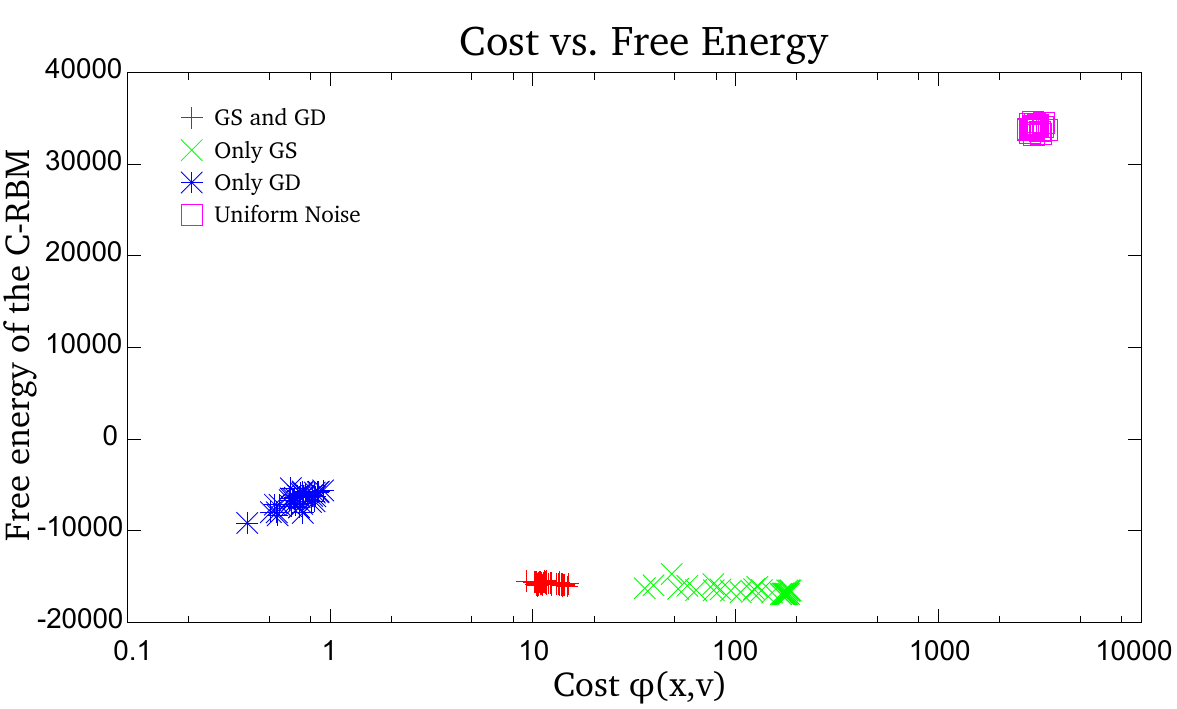}

\caption{Influence of Gibbs sampling (GS) and gradient descent (GD) on free energy $\mathcal{F}(\mathbf{v})$ (see Equation~\ref{equ:energy_const}) and cost $\phi(\mathbf{x},\mathbf{v})$ (see Equation~\ref{equ:cost}).
Using only GS results in low free energy but relatively high cost.
Using only GD, the cost is very low, but the free energy is high.
When using GS and GD, both methods compete, resulting in a trade-off between low cost and low free energy although we choose enough GS steps in the GS phase to always return to a ``meaningful'', low free energy state.
For reference we test against random uniform noise, resulting in very high free energy and cost.
For each cluster, 50 data points were generated with the trained C-RBM model (for GS) and the cost function (for GD) used in our experiment (see Section~\ref{sec:experiment})}.
\label{fig:costvsfe}
\end{figure}

During CS, in the C-RBM the free energy is to be reduced (i.e.
a high probability solution is to be found), while in GD optimization the objective function is to be minimized (see Figure~\ref{fig:cost_curve} for a plot of the curves).
As the two models used compete in approximating their objectives (see Figure~\ref{fig:costvsfe}), their mutual influence has to be balanced.
In addition to using Simulated Annealing to prevent strong deteriorations of the solution concerning the objectives (see Section~\ref{sec:simulatedannealing}), some parameters need to be carefully adjusted.

The main parameters for balancing the models are the number of GD and GS steps used in a CS iteration, as well as the learning rate and the relative weighting of the cost terms in the GD optimization (see Tab.~\ref{tab:weights} for weightings used in our experiments).
In general, the optimal number of steps in each model is inversely proportional to the size of the training corpus.
The more training data, the more possible solutions can be sampled by the probabilistic model making it easier to satisfy constraints imposed by the GD optimization.
Conversely, with a model trained on little data, more GS steps are necessary in order to find another low free energy solution after being distracted by the GD phase.

Although we do not provide a formal convergence proof, all experiments show a joint decrease of the various quantities to be minimized (the C-RBM free energy, and the cost-functions of each of the constraints). Convergence is reached when both the gradient descent cost and the free energy of the C-RBM reach a minimum. When reaching equilibrium in an RBM, the visible unit configuration (the sample) keeps changing during further sampling while the free energy remains at the minimum. Therefore, with our method convergence is reached concerning the overall (average) cost, but not concerning a final solution in v.

\subsection{Simulated Annealing}\label{sec:simulatedannealing}
Due to the interdependency between the sampling process of the probabilistic model and the GD optimizer, it can easily happen that good intermediate solutions deteriorate again by further sampling.
Simulated Annealing (SA) helps to find good minima by preventing sampling steps which would lower the solution quality too much (see Algorithm~\ref{alg:cs} for the integration of SA in CS).
After each constrained sampling (CS) iteration, we evaluate the SA equation to obtain the probability
\begin{equation}
p_k(\mathbf{v},\mathbf{v}',i) = \exp\left(-\frac{f(\mathbf{v}) - f(\mathbf{v}')}{T_i}\right)
\end{equation}
of keeping solution candidate $\mathbf{v}$, where $\mathbf{v}'$ is the previous solution.
We evaluate this equation twice after each CS iteration.
The first time, $f(\boldsymbol{\cdot})$ is the \emph{standardised} RBM free energy function $\mathcal{F}'(\boldsymbol{\cdot})$ (see Equation~\ref{equ:energy_const}) and the second time, $f(\boldsymbol{\cdot})$ is the \emph{standardised} GD cost function $\phi'(\boldsymbol{\cdot})$ (see Equation~\ref{equ:cost}).
For each of the two resulting probabilities, we generate a random number between $0$ and $1$, and evaluate if it is smaller than the respective probability.
If this is the case for both random numbers, we go on with solution candidate $\mathbf{v}$, otherwise, we return to the former solution $\mathbf{v}'$.
The most crucial factor for the sensitivity of SA is the variance of $f(\boldsymbol{\cdot})$ over all solutions, where a higher variance leads to smaller probabilities for acceptance of a solution.
Therefore, we standardise $\mathcal{F}(\boldsymbol{\cdot})$ and $\phi(\boldsymbol{\cdot})$, resulting in $\mathcal{F}'(\boldsymbol{\cdot})$ and $\phi'(\boldsymbol{\cdot})$, to obtain comparable probabilities in SA.
This is done by scaling both functions to approximately zero mean and unit variance, based on the observed values during the experiments.
As annealing scheme we use $T_i = 1 - i/N$.
In Figure~\ref{fig:cost_curve} the standardized curves over a CS process are depicted.
In later iterations, Simulated Annealing causes periods of constant cost, as some solution candidates are rejected.

\begin{table}
\begin{center}
\begin{tabular}{@{}ll@{}}
\toprule
Constraint & $w_d$ \\ \midrule
Self-similarity & 1.5 \\
Tonality & 5.0 \\
Meter & 0.5 \\
\bottomrule
\end{tabular}
\vspace{4mm}
\caption{Relative weightings $w_d$ of the terms used in the GD objective function $\phi(\mathbf{x},\mathbf{v})$ (see Section~\ref{sec:gd}).}
\label{tab:weights}
\end{center}
\end{table}

\section{Experiment}\label{sec:experiment}
This section describes an experimental validation of the method described in Sections~\ref{sec:Method_const} and \ref{sec:cs}.
In Section~\ref{sec:corpus} and Section~\ref{sec:repr}, we introduce the training data and the data representation scheme, respectively.
Section~\ref{sec:training_const} describes the training of the C-RBM.
In Section~\ref{sec:qu_evaluation} we use the Information Rate (IR) adopted from \cite{wang2015pattern} to quantify the structural organization of music.
With the IR metric, we compare our constrained sampling approach with other state-of-the-art polyphonic music generation methods and with original pieces of Mozart.
Lastly, Section~\ref{sec:procedure} briefly describes the procedure followed to produce musical material for qualitative evaluation.

\subsection{Training Data}\label{sec:corpus}
We use MIDI files encoding the scores of the second movement of three Mozart piano sonatas, as encoded in the Mozart/Batik data set~\cite{widmer2003discovering}: Sonata No. 1 in C major, Sonata No. 2 in F major and Sonata No. 3 in B flat major.
When applying a (major) tonality constraint, we want to make sure that there is enough training data for the probabilistic model in any possible (major) key.
Otherwise, in the GS phase, an intermediate solution might always be changed back from a key imposed by the GD optimization to the closest key available in the training data.
Therefore, we transpose each piece into all possible keys, which also helps to reduce sparsity in the training data.
This results in a training corpus size of $15144$ time steps (of sixteenth note resolution).

\subsubsection{A Note on Training Data Set Size}\label{sec:overfitting}
A widely shared insight in machine learning is that more data is better when training neural network models.
In general, it is easy to see why this is the case since larger amounts of data provide a richer coverage of the relations to be learned in the data.
However, depending on the intended purpose of the model, there may be exceptions to this rule.
In the present study, where the primary purpose of the model is to generate plausible musical textures in the style of Mozart piano sonatas, we have found that the set of all available training data (34 pieces) is likely too small for the C-RBM model to approximate the data distribution well enough to produce samples of high musical quality.
A pragmatic trade-off we have chosen in this case is to reduce the size of the training data to a few pieces, and let the model slightly overfit those pieces.
This will improve the musical quality of the samples, at the cost of increased local resemblances of generated samples to fragments of the training data.

\subsection{Data Representation}\label{sec:repr}
We transform MIDI data in a binary piano roll representation of $T=512$ time steps over a range of $P=64$ pitches (MIDI pitch number $28$-$92$), using a temporal resolution of sixteenth notes (see Figure~\ref{fig:iteration}[1a]).
Notes are represented by active units (black pixels), and note durations are encoded by activating units up to the note offset.
If two notes directly follow each other at the same pitch, they cannot be distinguished anymore.
Thus, the first note is shortened by a sixteenth note if possible (i.e.
if it is longer than a sixteenth note); otherwise the merger has to be accepted.
Note that a temporal subdivision of sixteenth notes cannot represent all rhythmic patterns in the data without distortion.
For example, the durations \{1/12, 1/12, 1/12\} of 1/8 note triplets (as contained in the Sonata No. 1) change to \{1/16, 1/8, 1/16\} using this representation.
We accept this bias, as it does not hinder our efforts to test the influence of constraints on a generated texture.

\subsection{Training}\label{sec:training_const}
We train a single C-RBM using Persistent Contrastive Divergence (PCD)~\cite{tielemanPcd2008} with $10$ fantasy particles, using learning rate $15 \times 10^{-4}$.
Compared to standard Contrastive Divergence~\cite{hinton06}, the PCD variant is known to draw better samples.
One training instance has a length of $T = 512$, and we use a batch size of $1$.
The \emph{filter width} R (see Section~\ref{sec:rbm_const}) is set to 17, and we convolve only in the time dimension with stride 4, using 2048 hidden units.

We apply the well-known L1 and L2 weight regularization with strengths $8 \times 10^{-4}$ and $1 \times 10^{-2}$, respectively, to prevent overfitting and exploding weights.
Also, we use the max-norm regularization \cite{srebro2005rank}, which is additional protection against exploding weights when using high learning rates.
We also use sparsity regularization as introduced in \cite{Lee:2007uz}, to increase sparsity and selectivity in the hidden unit activations, leading to a better generalization of the data.
When training with PCD, single neurons may always be active, independent of the presented input.
Therefore we reset (i.e.
randomize) the weights of any neuron which exceeds the threshold of $0.85$ average activation over the data.

\subsection{Quantitative Evaluation}\label{sec:qu_evaluation}
Based on the observation that probabilistic models can generate meaningful low-level structure but struggle in obeying some higher-level structure, the focus of this study is to increase the structural organization of the generated material. In the important case of self-similarity structure, a critical property in music is the balance of repetition and variation. This ratio is expressed by an information theoretic measure called Information Rate (IR). It is the mutual information between the present and the past observations and is maximal when repetition and variation are in balance. Thus, the IR is minimal for random sequences, as well as for very repetitive sequences. It has been shown that it provides a meaningful estimator on musical structure, for instance in parameter selection for musical pattern discovery \cite{wang2015pattern}.

For a given sequence $v_0^N = \{v_0,v_1,v_2,\dots,v_N\}$, the average IR is defined by
\begin{equation}
IR(v_0^N) = \frac{1}{N}\sum_n^N{H(v_n) - H(v_n \mid v_0^{n-1})},
\end{equation}
where $H(v)$ is the entropy of $v$, which is estimated based on the statistics of the sequence \emph{up to event} $v_n$.
We approximate $H(v_n \mid v_0^{n-1})$ using a first-order Markov Chain, and $H(v_n)$ by counting identical time slices.
It may seem counterintuitive at first sight to utilize a first-order model for measuring the higher-level structure of a piece.
Arguably, a low-order estimation yields too optimistic IR values, as the conditional entropy tends to be underestimated.
However, the initial idea of contrasting the prior entropy of events with their conditional entropy is still applicable using a first-order entropy estimation.
That is, a high IR is achieved when specific events occur rarely but are very likely given their direct predecessors---a situation which occurs mainly in sequences with higher-level repetition structure.
Note that the IR does not provide a measure for the overall musical quality of the evaluated sequences, but only for the aspect of self-similarity structure from an information theoretic point of view.

\subsubsection{Model Comparison}
In addition to using the C-RBM without constraints, we use the RNN-RBM \cite{boulanger2012modeling}, a state-of-the-art polyphonic music generation model, as a baseline for the quantitative evaluation. Furthermore, we replace the RNN portion of the RNN-RBM with Gated Recurrent Units (GRUs, Cho \emph{et al.}, 2014) resulting in a GRU-RBM. Both recurrent models are trained on the same data as the C-RBM, as described in Section \ref{sec:corpus}\footnote{Samples from the RNN-RBM and the GRU-RBM can be listened to on Soundcloud under \url{http://www.soundcloud.com/pmgrbm}}.

\begin{figure*}
\centering\footnotesize
\includegraphics[width=\linewidth]{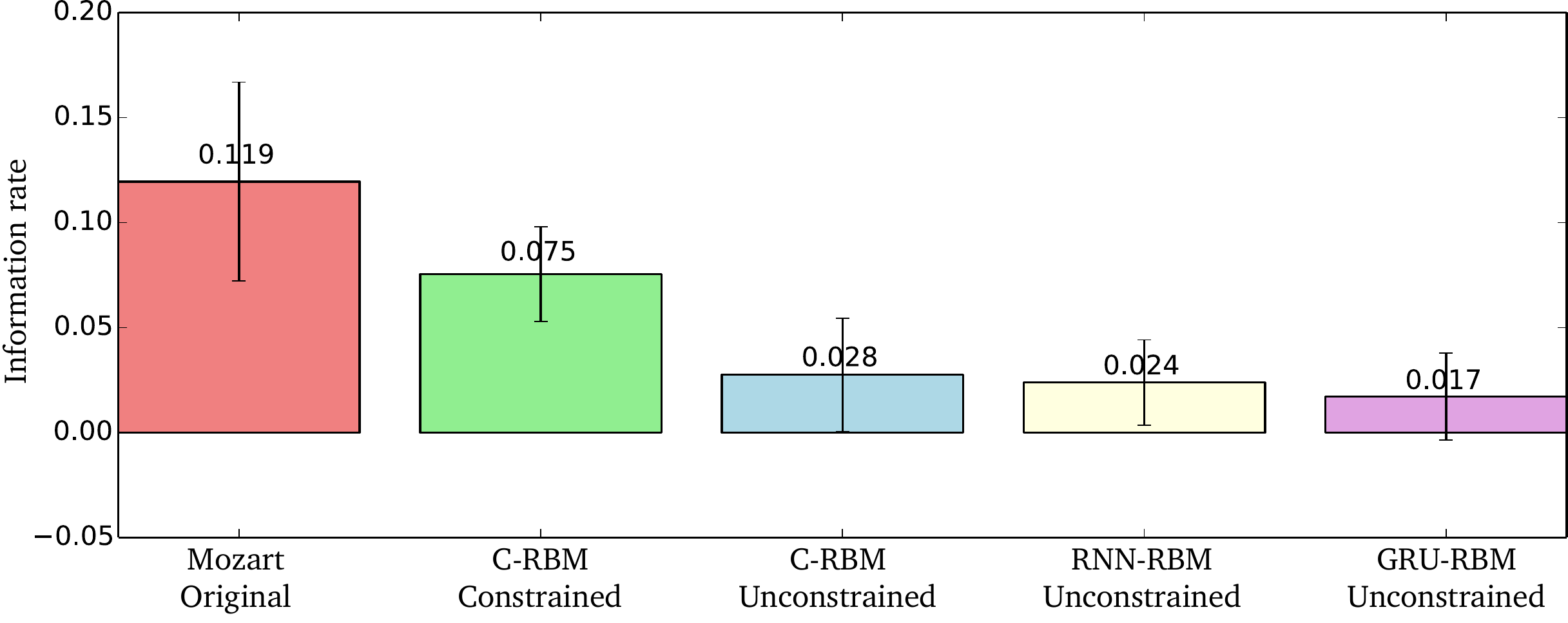}
\vspace{-0.2cm}
\caption{Box plot showing Average Information Rates for $34$ original Mozart piano sonatas, $102$ C-RBM samples with structure constraints, $102$ C-RBM samples without constraints, $102$ samples from an RNN-RBM and $102$ samples from a GRU-RBM.
Whiskers show standard deviations.}
\label{fig:info_rate}
\end{figure*}

We compare the average Information Rates between original Mozart piano sonatas (all $34$ pieces of the Mozart/Batik data set, Widmer, 2003), C-RBM constrained samples using the original pieces as structure templates (3 samples per original piece resulting in $102$ samples with different lengths), $102$ C-RBM unconstrained samples, $102$ RNN-RBM unconstrained samples and $102$ GRU-RBM unconstrained samples. The $102$ unconstrained samples per model are created by generating three samples for each original piece in the length of the original piece. For results on this comparison see Figure~\ref{fig:info_rate}, for a discussion see Section \ref{sec:results-discussion_const}.

\subsubsection{Further Measures for Evaluating Musical Structure}
Information theory could provide additional quantitative measures for the evaluation of structure in music.
An essential basis for that is Information Content (IC), a measure of the predictability of an event in a specific context.
Prior research has shown that IC can act as a kernel for determining segment boundaries (see \cite{Pearce:2010ig} and Chapter \ref{sec:segment}).
Evaluating the plausibility of IC over time in generated sequences could, therefore, constitute an adequate measure for the evaluation of musical structure.

Moreover, the principle of uniform information density (UID) is a theory originating in linguistics.
It is based on the proposal of  \citet{shannon2001mathematical} that for optimal data flow through a noisy channel, the transferred information density (i.e., the IC per time step) should be as uniform as possible.
It was shown that speakers intuitively follow these rules to keep the processing effort of the receiver at a moderate level \cite{jaeger2007speakers,aylett2006language}.

Recent research provides evidence that UID could also account for structural decisions in music composition, where it implies that the average IC in any window of fixed duration over a musical piece should be constant.
For example, \citet{temperley2014information} states ``There is a tendency that when an intervallic pattern is repeated with alterations, the alterations tend to lower the probability of the pattern rather than raising it''.
When a musical passage is repeated, its Information Content (i.e., the listener's surprise) declines.
The finding mentioned above provides some evidence that in such cases, surprising alterations should be inserted to keep the UID constant.

Since the IR is sufficient for evaluating our results, we do not use IC and UID.
Nevertheless, they seem promising as further evaluation measures to quantify structure in generated music.

\subsection{Qualitative Evaluation}\label{sec:procedure}
The C-RBM is trained as described in Section~\ref{sec:training_const} on the Mozart Sonatas (see Section~\ref{sec:corpus}).
After that, we pick a template piece (the first movement of the piano sonata No.
6 in D major) and generate constrained samples, as introduced in Section~\ref{sec:cs}.
For the weights used to balance the different terms in the GD cost function, please see Tab.~\ref{tab:weights}.
In the self-similarity constraint (see Section~\ref{sec:selfsim}), we use a window size $\Lambda$ of $8$ (i.e.
half a bar), and for the tonality constraint we use an estimation window width $M$ of $4$ (see Section~\ref{sec:tonality}).
Figure~\ref{fig:selfsim} shows some resulting samples, which are discussed in detail in Section \ref{sec:results-discussion_const}.

\subsubsection{Keyscape}\label{sec:keyscape}
\begin{figure*}
\vspace{-0.5cm}
\centering\footnotesize
\includegraphics[width=.95\linewidth]{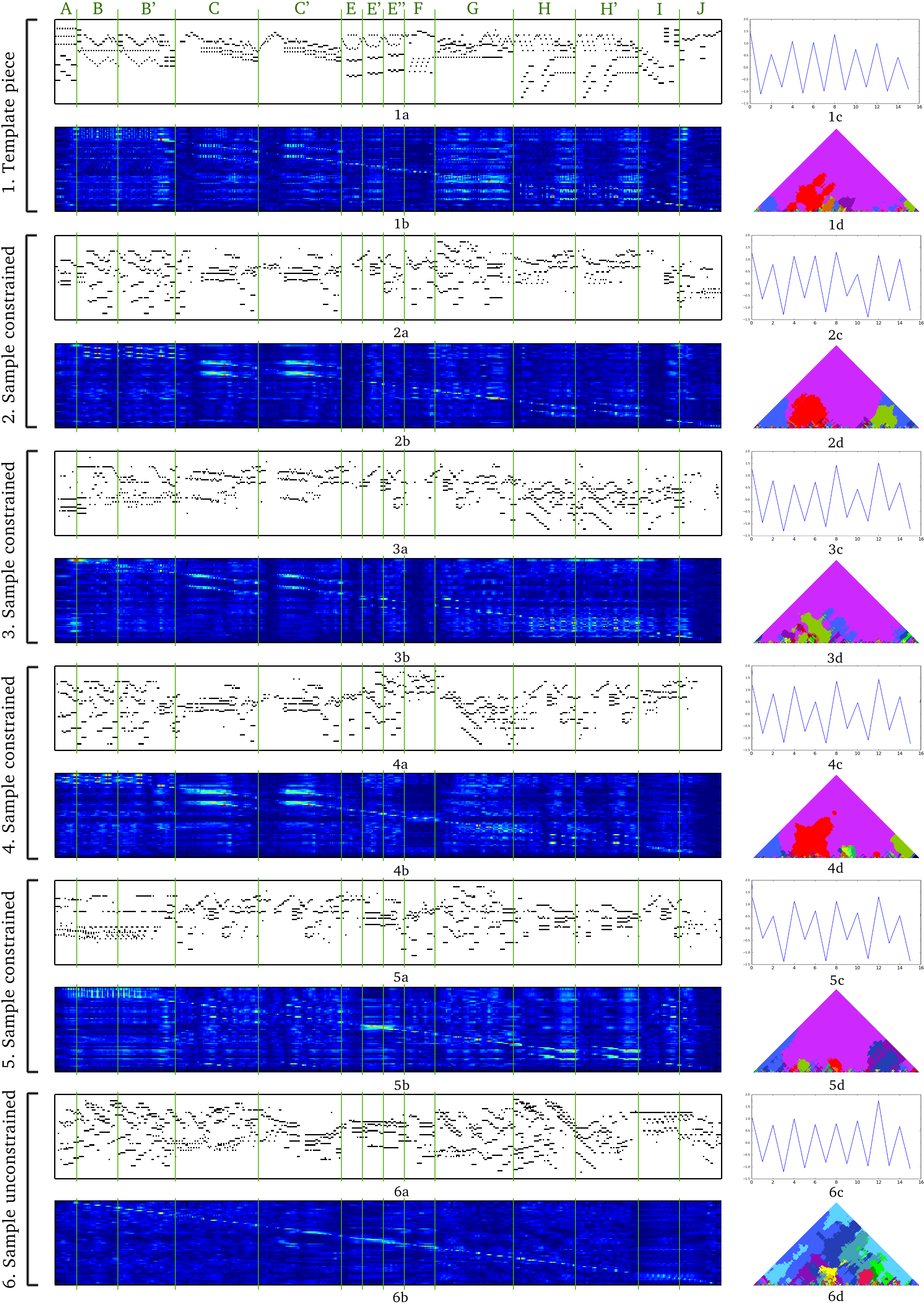}
\vspace{-0.2cm}
\caption{Template piece (1), Constrained samples (2 to 5) and an unconstrained sample as baseline (6).
Figures in each group: (a) Piano roll representation, (b) Self-similarity matrix, (c) Onset distribution in $4/4$ meter, (d) Keyscape.
By constrained sampling, the template piece's self-similarity and tonal structure, as well as the onset distributions, are transferred to the generated solutions 2 to 5.
The unconstrained sample (6) at the bottom was sampled without constraints, and thus does not reflect the structure of the template piece.}
\label{fig:selfsim}
\end{figure*}

We use keyscapes to illustrate the tonality of the pieces in Figure~\ref{fig:iteration} and Figure~\ref{fig:selfsim}.
A keyscape illustrates the tonal context over a musical piece, where each key receives a distinct color.
We use the humdrum mkeyscape tool by David Huron, which analyses the musical piece with the Krumhansl-Schmuckler key-finding algorithm \cite{krumhansl1990} in different levels of detail.
The top of the pyramid depicts the key estimation for the entire piece, while towards the base the analysis is based on ever smaller window sizes.
Each scale has a distinct color assigned to it, and the keyscape is colored according to the most predominant scale estimation.

\section{Results and Discussion}\label{sec:results-discussion_const}
\subsection{Quantitative Evaluation}
Figure~\ref{fig:info_rate} shows average Information Rates (IRs) for original Mozart piano sonatas and for samples from different models (see Section~\ref{sec:qu_evaluation}), where higher IRs indicate more distinct self-similarity structures.
It should be pointed out in advance that sampling from a probabilistic model introduces some sampling noise which increases the predictive entropy and therefore lowers the IR.
It is difficult to judge to what extent sampling noise on the one hand, and difficulties of the model to adapt to given constraints, on the other hand, lead to the significantly lower IR of the constrained samples compared to the IR of the original Mozart transcriptions.
Nevertheless, it is clear from the results firstly that the IR of the original music is higher than that of the generated music, and secondly that the models \emph{without constraints} produce music with lower IR than the constrained C-RBM does.
Note that the latter point is a non-trivial result, since the self-similarity constraint does not explicitly optimize the Information Rate (neither do the tonal or meter constraints, obviously), but encourages similarity or dissimilarity between the music at specific positions.
This result is in accordance with the initial observation that the baseline models fail to generate higher-level self-similarity structure.
Due to their gating mechanism, GRUs are usually better at learning long-term dependencies than regular recurrent units.
The fact that the GRU-RBM does not perform better than the RNN-RBM shows that GRUs also have problems in modeling the content-invariant self-similarity property.

\subsection{Qualitative Evaluation}
Figure~\ref{fig:selfsim} shows piano roll representations for the template piece (Figure~\ref{fig:selfsim}[1a]), four generated samples that were constrained with properties from the template piece (Figure~\ref{fig:selfsim}[2a] to Figure~\ref{fig:selfsim}[5a]) and a baseline sample generated without constraints from the template piece (Figure~\ref{fig:selfsim}[6a]).
The corresponding constraints for each musical piece are depicted in the respective figures b-d.
The repetition structure is labelled on top of the figure with the typical uppercase letters used to describe musical form, and the section boundaries are indicated with vertical, green lines.\footnote{All samples illustrated in Figure~\ref{fig:selfsim} can be listened to on Soundcloud under \url{http://www.soundcloud.com/pmgrbm}}

We chose the constrained samples by creating 20 solutions and picking the best four concerning the minimum average value of the standardized GD cost and the standardized free energy over a constrained sampling process (see Section~\ref{sec:simulatedannealing} on standardizing the cost and free energy functions).
Thus, results are selected which satisfy the given constraints better, rather than according to their musical quality.
Empirically we found that the musical quality in our setting increases when loosening the influence of the constraints, as this allows the probabilistic model to create more plausible samples (e.g.
the examples in Figure~\ref{fig:selfsim} sometimes lack appropriate transitions between different sections which are an effect of both constraint satisfaction and limited training data).

By approximating the self-similarity matrix of the template piece, some aspects of the repetition structure were convincingly transferred to the constrained samples (see Figure~\ref{fig:selfsim}[1b] to Figure~\ref{fig:selfsim}[5b]).
For example, the exact repetitions C / C' and H / H' occur in every sample.
It is interesting to see how the extension of B to B' is solved.
Especially in the samples depicted in Figure~\ref{fig:selfsim}[2] and Figure~\ref{fig:selfsim}[3], the extension of B is realized by musical textures consistent with the immediate past.
In the sample in Figure~\ref{fig:selfsim}[5], the model did not produce satisfactory results for phrase B and B' in a musical sense, although it found a solution which is self-similar over that time period and therefore satisfies that self-similarity constraint to a certain degree.

Parts E / E' / E'' are special cases, because even though they are very similar at first sight, they are transposed repetitions which cannot be captured by the self-similarity matrix as it is currently defined.
In the self-similarity matrix of the template, we see that each of those ``E'' sections is more or less similar or dissimilar to different regions in the piece.
Also, we note that each repetition has the length of one bar.
When comparing these ``E'' sections with those in the samples, we recognize the limits of the method concerning temporal resolution.
The C-RBM has a filter length of one bar, which is too wide for sampling three bars with different requirements concerning similarity while keeping a plausible low-level structure.
Therefore, in some samples, the generated patterns span the whole, or at least two of the ``E'' sections.

Part G in the repetition structure is similar to most parts of the piece, as can be seen from the bright areas over the full height of the respective self-similarity matrices.
In the samples, this is realized by choosing textures which are also similar to most parts.
Part J, in contrast, is very dissimilar to most areas of the template piece.
Probably due to limited training data, this sometimes results in empty areas in the samples.
Except for an apparent similarity in B and B', which is not reflected in the self-similarity matrix, the unconstrained baseline sample does not follow the repetition structure of the template piece.

The onset distributions (see Subplots \emph{c} in Figure~\ref{fig:selfsim}), which are plots resulting from Equation~\ref{equ:onsetdist}, are sometimes rather dissimilar to the onset distribution of the template.
This shows that it is not easy to approximate this global property.
One reason for this may be that it is a property which summarizes the complete music piece in only a few values, which makes it easy in the GD optimization to approximate by distributing small changes over the whole sample.
Those are, however, locally not strong enough to be kept during GS.
Incorporating note durations for emphasizing onsets of longer notes would lead to more characteristic onset distributions, which could further lead to more substantial local changes in the probability of notes in the piano roll.
However, in the onset distributions, there is a tendency of the peaks at position $0$ and $8$ to be higher than the others, which corresponds to the tendency in the onset distribution of the template piece.
Note that the reason for every second value in the distributions being low is not the meter constraint but the stride of one beat in the convolution.
This provides the model with a regular grid allowing it to learn that the probability of an onset is lower at every second time step.
Therefore, those values are also low for the unconstrained baseline piece.

The keyscape (see Section~\ref{sec:keyscape}) for each sample is depicted in the respective subplots d.
We can see that the main key (i.e.
A major) of the template piece got well transferred to the constrained samples, as the colors of the upper areas of the keyscapes (purple) match exactly.
Towards the lower areas of the keyscapes, the colors of some samples do not correlate with those of the template piece's keyscape.
However, especially the modulation to E major in the second quarter of the piece, depicted in red, and the blue area at the beginning of the piece (D major) are to some degree approximated in Figure~\ref{fig:selfsim}[2d] and Figure~\ref{fig:selfsim}[4d].
In sample Figure~\ref{fig:selfsim}[3], the green area indicates a F\raisebox{.45mm}{$\sharp$} minor scale, which is similar to the E major scale (red) of the template piece (i.e.
there is a difference in one note, namely D/D\raisebox{.45mm}{$\sharp$}).
In general, the tonal structure of constrained samples is more stable than that of the unconstrained baseline sample, where the keyscape indicates tonal incoherence.

As mentioned above, the illustrated samples are the best four of $20$ concerning the overall cost.
The most obvious shortcoming of samples not selected because of the higher cost is that they do not satisfy some of the given structural constraints on a local level.
This includes the failure to reproduce a repetition at specific positions, or erroneously modulating into a key which does not occur in the template piece.
However, a closer inspection of such cases shows that incorrect keys are often closely related to the desired keys, for instance, the parallel minor/major key.
Parallel minor/major keys have most of their pitches in common, but they are expected to follow a different distribution.
Another common problem in the non-optimal samples is that they show areas without any notes, which is probably a symptom of contrasting objectives of GD and GS.
We found that the C-RBM is very sensitive to changes in the higher-level parameters of the system.
Other models amenable to GS could lead to a more stable functioning, like LSTMs used with GS in \cite{hadjeres2017deepbach}.

\section{Conclusion and Future Work}\label{sec:future}
Music is typically highly structured at both lower and higher levels.
State-of-the-art sequence models such as RNNs and LSTMs have been successfully used to generate music in restricted settings, but in more complex musical material, such as piano music from the classical or romantic period, not to mention orchestral works, important musical characteristics such as tonal, metrical and self-similarity structure tend to defy straight-forward time series modeling approaches.

The method for music generation presented here addresses this problem by combining a stochastic neural network for sampling plausible musical textures at a local level with soft constraints that impose higher-level structure regarding meter, tonality and self-similarity structure, obtained from a template piece.

The experimental validation of the proposed method reveals that the generated music possesses a stronger degree of structural organization (as measured by Information Rate (IR)) than unconstrained models, including a state-of-the-art RNN-RBM model for polyphonic music generation.
A qualitative analysis of some generated music supports this finding and clearly reveals repeated (but not identical) musical patterns, as well as global similarities in tonal structure to the template piece.

The empirical results also reveal some shortcomings.
Firstly, while imposing constraints with the proposed method helps to generate high-level structure, a meaningful low-level structure can currently only be generated when the model is trained on relatively small amounts of data.
Overcoming this drawback may require more powerful generative models---amenable to some form of Gibbs sampling---as an alternative to the C-RBM.
We hypothesize that generative models can only generalize well on the low-level structure if they can explicitly represent (transposed) repetition.
A promising approach to this is proposed in the next chapter, where it is shown that relations between musical sections can be learned and represented as so-called ``mapping codes''.

Secondly, when listening to the generated musical samples, it is clear that the tonal, meter and self-similarity constraints presented here are by no means fully elaborated nor exhaustive.
For example, more specialized constraints---like a differentiable formalization of the IR measure---could optimize sequences to obey desired structural properties directly.
Perhaps most importantly, what is currently missing is a constraint that enforces musical closure at boundaries of structural units.
Without such a constraint, the music contains repeated musical structures, but these structures are hard to identify perceptually because their boundaries are not marked by salient musical cues (such as harmonic resolution).
That said, the proposed constrained sampling approach is general enough to accommodate this and possibly other constraints.

\newcommand{\subsubsubsection}{\vspace{.5cm}\noindent\textbf}

\chapter{Learning Musical Structure by Learning Transformations}\label{sec:structure}
In Section \ref{sec:constraints}, we proposed to generate music with higher-level structure using external constraints in the sampling process.
A more elegant solution would combine both the generative model and structural constraints into a generative model with implicit awareness of structural properties.
In this chapter, we start from the observation that structured music can be described by sets of transformations between musical sections and show that a Gated Autoencoder (GAE) can learn such musical transformations.
We then adapt the GAE to musical sequence learning, first by employing predictive training, and then by combining the GAE with an RNN.
The competitive results in prediction tasks suggest that learning musical transformations is a powerful concept in music modeling and that in architectures like the GAE both generation and structure learning can be combined in a single, integrated solution.

\section{Introduction}\label{sec:introduction_full}

An important notion in western music is that of structure: the phenomenon that a musical piece is not an indiscriminate stream of events, but can be decomposed into temporal segments, often in a hierarchical manner.
An important factor determining what we regard as structural units is the relation of these units to each other.
The most apparent relation is the literal repetition of a segment of music on multiple occasions in the piece.
However, more complex music tends to convey structure not only through repetition, but also through other types of relations between structural units, such as chromatic or diatonic transpositions, or rhythmical changes.
We refer to these relations in general as \emph{transformations}.

\begin{figure}
\includegraphics[trim=1cm 21.5cm 1cm 1cm, clip, width=1.\linewidth]{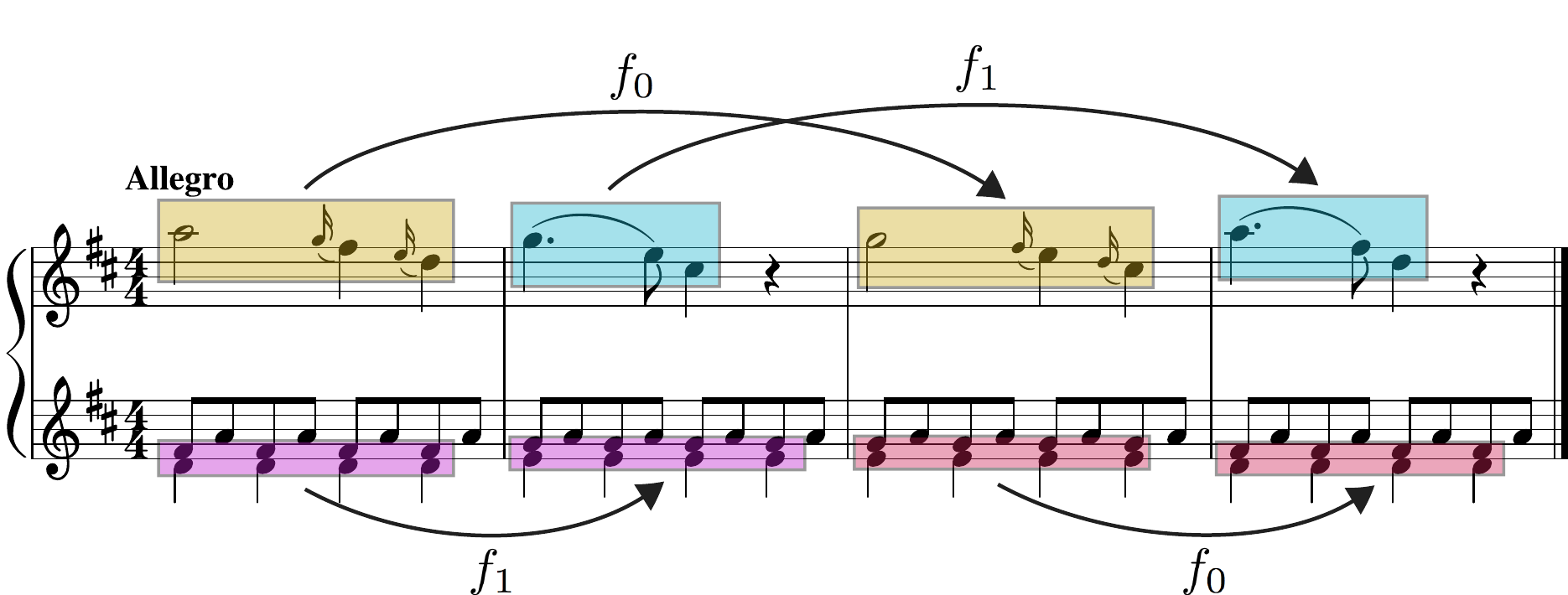} \\
\vspace{-3mm} \\
\includegraphics[trim=.1cm 0cm .6cm 0cm, clip, width=1.\linewidth]{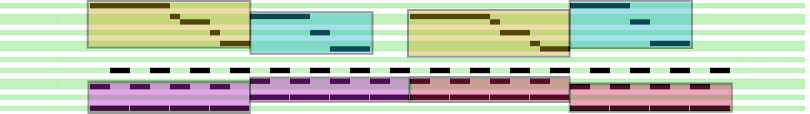}
\caption{Beginning of the ``Rondo in D major'' (K. 485) by W.
A.
Mozart in Western music notation and in a piano roll representation, where transformation functions $f_n$ effect diatonic transpositions (best viewed in color).
In the piano roll (bottom), green horizontal lines mark the diatonic pitches in the scale of D major.}
\label{fig:rondo}
\end{figure}

Figure~\ref{fig:rondo} shows an example of mid-level structure induced by transformations in an excerpt of a rondo by W.~A.~Mozart where related phrases are marked by boxes in the same color (note that this interpretation is not unique).
The yellow boxes mark the input and output of a function $f_0$ performing a diatonic transposition by -1 scale step.
The same function is applied in the bass section, marked with red boxes.
Likewise, $f_1$ performs a diatonic transposition by +1 scale step in the melody section (blue boxes) and the bass section (purple boxes).
The transformations defined in $f_0$ and $f_1$ constitute structural relationships which may be applied to any musical material and in that sense are \emph{content-invariant}.
Note how this view on music as a collection of basic musical material that is transformed in a variety of ways can provide very concise and schematically simple representations of a musical piece.
Finding such a representation may be a goal in itself (music analysis), or it may serve as the basis for other tasks, such as music summarization, music classification, and similarity estimation, music generation or computer-assisted composition.

There are some fundamental challenges to automated detection of structure in music, however.
Firstly, although there are a few common types of transformations such as chromatic and diatonic transposition, it is not easy to give an exhaustive list of transformation types that should be considered when comparing potential structural units of a musical piece.
Ideally, rather than relying on human intuition, we would like to infer transformations from actual music that can account for a large portion of structure in that music.
A second challenge is that composers may use the concept of transformation as a compositional tool, but in their artistic freedom, they may perform further ad-hoc alterations at will. 
The resulting musical material may thus exhibit \emph{approximate} transformations, deviating from exact transformations in subtle but unpredictable ways.

State of the art methods for the motivic analysis of music, such as the compression-based methods discussed in~\cite{louboutin16:_using}, do not directly address these issues.
We believe that a more data-driven approach is needed, where transformations are learned from data in an unsupervised manner.
This naturally leads us to consider connectionist models for unsupervised learning, such as Restricted Boltzmann Machines (RBMs)~\cite{hinton06}, and autoencoders~\cite{bengio-2009}.
A particularly promising approach towards unsupervised learning of content-invariant musical transformations are bilinear models like Gated Autoencoders (GAEs) \cite{memisevic2011gradient} or Factored Boltzmann Machines \cite{memisevic2010learning}.
By employing \emph{multiplicative interactions}, these models can learn co-variances between pairs of data instances.
It was shown that GAEs are effective at encoding transformations between image pairs, such as spatial translation and rotation of objects \cite{memisevic2013learning}.

In this chapter, we incrementally extend the original GAE architecture in order to better adapt it to music processing and structure learning.
First, we investigate its ability to learn transformations from given pairs of musical fragments (i.e., a symmetrical setting), more specifically n-grams of vertical slices from a piano roll representation.
A comparison of the GAE with an RBM architecture demonstrates the GAE's particular suitability for learning transformations in music (see Section \ref{sec:gae}).

Second, we modify the GAE architecture and training procedure to obtain a predictive setting.
Thereby, transformations between musical notes---intervals---can be effectively learned.
Interval representations are transposition-invariant and are therefore useful, for example, to detect repeated themes and sections in music (see Section \ref{sec:interval}).

Third, we combine the GAE with an RNN in order to model the temporal succession of intervals.
We show that the thereby obtained Recurrent Gated Autoencoder (RGAE) is more general than common RNNs because it yields improved prediction performance and can better learn structural aspects of music (i.e., repetition and transposition, see Section \ref{sec:rgae}).

\section{Related Work}\label{sec:related_work_gae}

In the following, we shall cover the related work of this chapter, ordered according to the sections of the chapter.
First, related work and the origin of the underlying model---the GAE---is presented.
Second, works intending to perform analogy making by learning relations between data instances are discussed.
Finally, related work to invariance learning and (musical) sequence prediction with GAEs is covered.

\textbf{GAEs} utilize \emph{multiplicative interactions} to learn correlations between or within data instances and are also referred to as bi-linear models.
Bi-linear models are two-factor models whose outputs are linear in either factor when the other is held constant, a property which also applies to the GAE.
The method was inspired by the Correlation Theory of the Brain \cite{von1994correlation}, where it was pointed out that some cognitive phenomena cannot be explained with the conventional brain theory and an extension was proposed which involves the correlation of neural patterns.

This principle was further used in the work of \citet{adelson1985spatiotemporal}, where motion patterns in the three-dimensional $x$-$y$-$t$ space are modeled by filter pairs receptive to distinct orientations in that space.
This work shows that mapping units in a GAE function similar to \emph{complex cells} in the visual cortex, which gives rise to critical perceptual processes like detecting fluent motion from a series of static images.

In machine learning, this principle was deployed in \emph{bi-linear} models, for example, to separate person and pose of face images \cite{tenenbaum2000separating}.
\cite{olshausen2007bilinear} proposed another variant of a bi-linear model in order to learn objects and their optical flow.
Due to its similar architecture, the Gated Boltzmann Machine (GBM) \cite{memisevic2007unsupervised,memisevic2010learning} can be seen as a direct predecessor of the GAE.
The GAE was introduced by \cite{memisevic2011gradient} as a derivative of the GBM, as standard learning criteria (e.g., minimizing cross-entropy or mean-squared error) became applicable through the development of Denoising Autoencoders \cite{vincent2010stacked}.

GAEs were further used  to learn transformation-invariant representations for classification tasks \cite{ICML2012Memisevic_105}, for parent-offspring resemblance \cite{dehghan2014look}, for learning to negate adjectives in linguistics \cite{rimell2017learning}, for activity recognition with the Kinekt sensor \cite{mocanu2015factored}, in robotics to learn to write numbers \cite{droniou2014learning}, for learning multi-modal mappings between action, sound, and visual stimuli \cite{droniou2015deep}, and for modeling facial expression \cite{susskind2011modeling}.

\textbf{In Section \ref{sec:gae}}, we test the GAE in learning transformations between musical fragments.
The problem of detecting musical relations falls in the class of \emph{analogy-making}  \cite{hofstadter95}, an essential capability of the human brain in which the objective is to produce a data instance X given the three instances A, B, C, and the query “X is to A as B is to C.”
\citet{nichols2012musicat} shows how analogy-making also plays a vital role in music cognition, and ``Musicat'', a musical analogy-making model is presented.
Identifying analogies between musical sections exhibiting the same transformation is a first step towards identifying similarities between transformed musical objects (i.e.
by utilizing \emph{transformation-invariant} representations \cite{memisevic2013aperture}).

The problem of relating data instances is also tackled by some deep-learning methods, like in \cite{reed2015deepanalogy}, where visual analogy-making is performed by a deep convolutional neural network using a supervised strategy.
Siamese architectures are used by \citet{bromley1993signature} for signature verification, and by \citet{chopra2005learning} to identify identical persons from face images in different poses.
With a GAE, a mapping between two inputs is learned explicitly.
In contrast, with conventional deep-learning methods, relations are often indirectly qualified by comparing representations of related instances.
For example, it is a common approach to construct a space in which operations like addition and subtraction of data vectors imply some semantic meaning \cite{mikolov2013distributed,pennington2014glove}.

An analogy-making model similar to the GAE is the Transforming Autoencoder introduced by \citet{Hinton:2011wm}.
In contrast to the GAE, this model is supervised, as during training the respective transformations have to be specified in a parameterized way (e.g., rotation angle or distance of shift in image transformation).
Another related model is the Spatial Transformer Network, which transforms an input image (conditioned on itself) before passing it on to classification layers \cite{jaderberg2015spatial}.

Very few works exist on bi-linear models applied to music and audio.
A form of a bi-linear model was applied to learn co-variances within spectrogram data for music similarity estimation \cite{schluter2011music}.

\textbf{In Section \ref{sec:interval}}, we show how the GAE can be used for learning transposition-invariant interval representations in music.
Transposition-invariance is also achieved in \cite{meredith2002algorithms}, by transforming symbolic music into point-sets, in which translatable patterns are identified.
In \cite{marolt2008mid}, an approach for calculating transposition-invariant mid-level representations from audio, based on the 2-D power spectrum of melodic fragments, is introduced.
Similarly, a method to calculating non-invertible but interpretable interval representations from audio is given in \cite{walters2012intervalgram}, where chromagrams which are close in time are cross-correlated to obtain local pitch-invariance.
In contrast to these methods, our approach is invertible and learns filters to correlate spectrogram bins of variable time lags without the need for preprocessing.

\textbf{In Section \ref{sec:rgae}}, we propose the Recurrent Gated Autoencoder (RGAE) and show its potential for sequence prediction, and the learning of possibly transposed repetitions of short musical fragments.
In sequence modeling, the GAE was utilized to learn co-variances between subsequent frames in movies of rotated 3D objects \cite{memisevic2013aperture} and to predict accelerated motion by stacking more layers in order to learn higher-order derivatives \cite{michalski2014modeling}.
The latter method is very similar to the one proposed in Section \ref{sec:rgae}, as it also learns transformations from sequential data.
It assumes constant transformation between all time steps in a sequence, and the input and output of a model instance have the same size.
In contrast, we use different dimensionalities between input and output, and we do not assume constant transformation but rather learn \emph{sequences of transformations} using an RNN.

Probabilistic n-gram models specialized on learning to predict monophonic pitch sequences include IDyOM \cite{pearce05:_const_evaluat_statis_model_melod}, and \cite{DBLP:conf/ismir/LanghabelLTR17}. Both employ multiple features of the musical surface.
In this chapter, we do not compare the RGAE with these models, but rather with other recurrent neural network models, which use the same feature representation as the RGAE.
In particular, we compare the RGAE to the currently best performing recurrent connectionist sequence model, the Recurrent Temporal Discriminative Restricted Boltzmann Machine (RTDRBM) \cite{cherla2016neural}.
Its architecture is similar to the well-known RTRBM proposed in \cite{DBLP:conf/nips/SutskeverHT08}, but it employs a different cost function.

For structured sequence generation, Markov chains together with pre-defined repetition structure schemes were employed in \cite{collins2016developing}, where specific methods for handling transitions between repeating segments were proposed; in \cite{pachet2017sampling}, where an approach to a controlled creation of variations was introduced; in \cite{conklin_semiotic}, where chords were generated, obeying a pre-defined repetition structure.
A constrained variable neighborhood search to generate polyphonic music obeying a tension profile and the repetition structure from a template piece was proposed in \cite{herremans2016morpheus}.
In \cite{eigenfeldt2013evolving}, Markov chains and evolutionary algorithms were used to generate repetition structure for Electronic Dance Music.
Recently, connectionist architectures have been proposed able to reproduce higher-level structural characteristics of the training data \cite{DBLP:conf/iclr/Huang19}.

\section{Learning Transformations of Musical Material}\label{sec:gae}

%
%
%

\begin{abstract}
As stated in the introduction of this chapter (see Section \ref{sec:introduction_full}), structural relationships are often based on \emph{transformations} of musical material, like chromatic or diatonic transposition, inversion, retrograde, or rhythm change.
In this section, we study the potential of two unsupervised learning techniques---Restricted Boltzmann Machines (RBMs) and Gated Autoencoders (GAEs)---to capture pre-defined transformations from constructed data pairs.
More specifically, we define common musical transformations (chromatic transposition, diatonic transposition, tempo change and retrograde), and construct n-gram pairs accordingly (see Section~\ref{sec:ExperimentTransform}).
Such a controlled setting provides ground-truth data and allows for performing a proof-of-concept on the general suitability of GAEs in music, and should be regarded as a first step towards identifying and learning transformations implicit in a music corpus.

We evaluate the models by using the learned representations as inputs in a discriminative task where for a given type of transformation (e.g., diatonic transposition), the specific relation between two musical patterns must be recognized (e.g., an upward transposition of $3$ diatonic steps, see Section~\ref{sec:discuss_discriminative}).
Furthermore, we measure the error of the models when reconstructing transformed musical patterns (see Section~\ref{sec:discuss_recon}).
We also test the performance of the GAE in applying learned relations to data instances not seen during training (i.e., analogy-making, see Section~\ref{sec:discuss_gen}).

In Section~\ref{sec:Method_gae} we describe the GAE and the RBM models used in our experiments.
The experiment setup including data preparation, used model architectures, and training is introduced in Section~\ref{sec:ExperimentTransform}.
Results are presented and discussed in Section~\ref{sec:results-discussion_gae} and Section~\ref{sec:concl-future-work0} gives an outlook on future work and challenges in musical transformation learning.

We find that it is challenging to learn musical transformations with the RBM and that the GAE is much more adequate for this task since it can learn representations of specific transformations that are largely content-invariant.
We believe these results show that transformation-learning models such as GAEs may provide the basis for more encompassing music analysis systems, by endowing them with a better understanding of the structures underlying music.
\end{abstract}

\subsection{Method}\label{sec:Method_gae}

\subsubsection{Gated Autoencoder} \label{sec:autoencoder}
\begin{figure}[t]
\begin{center}
\includegraphics[width=.5\linewidth]{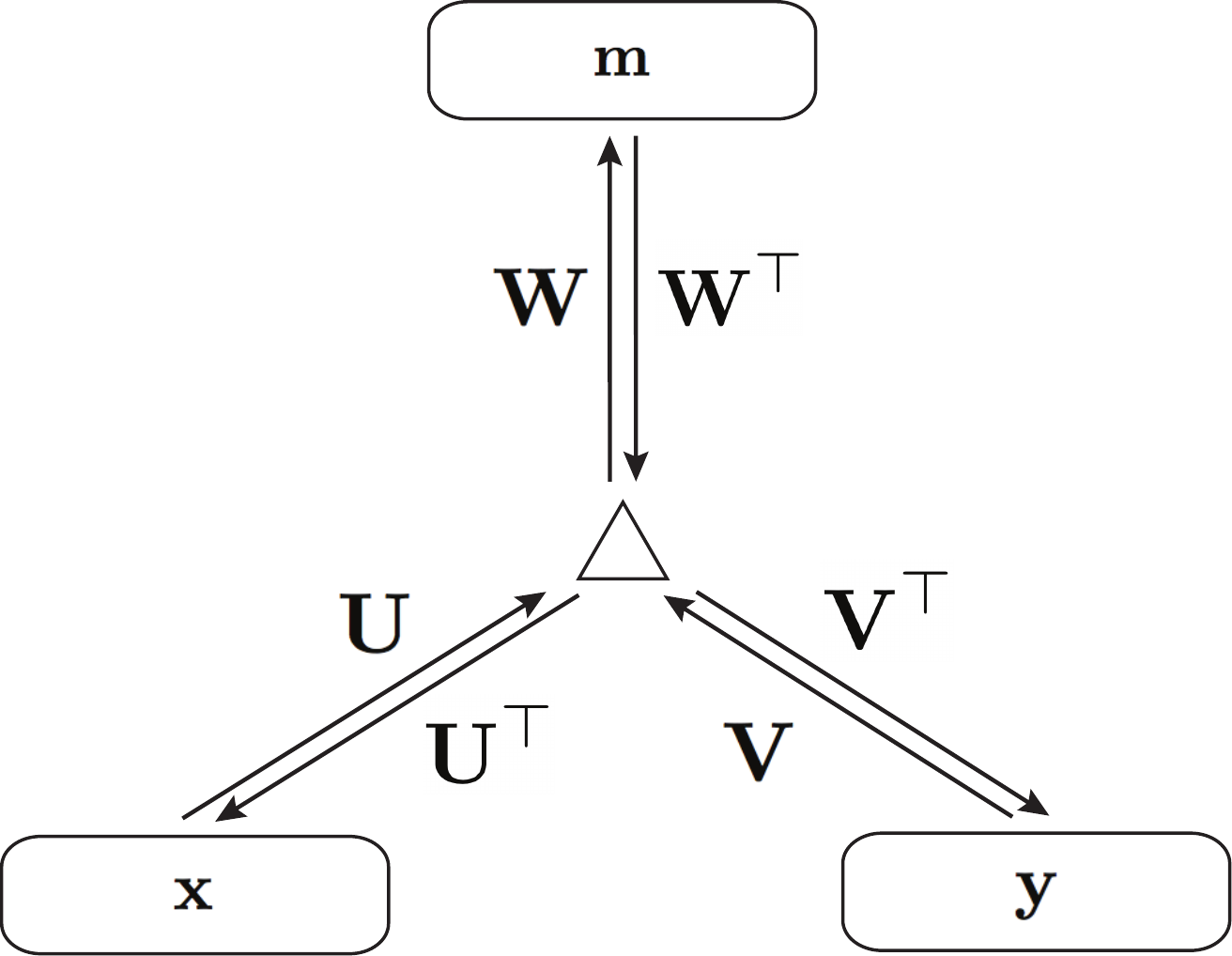}
\caption{Schematic illustration of the Gated Autoencoder.}
\label{ga}
\end{center}
\end{figure}

\begin{figure}
\begin{center}
\includegraphics[width=\linewidth]{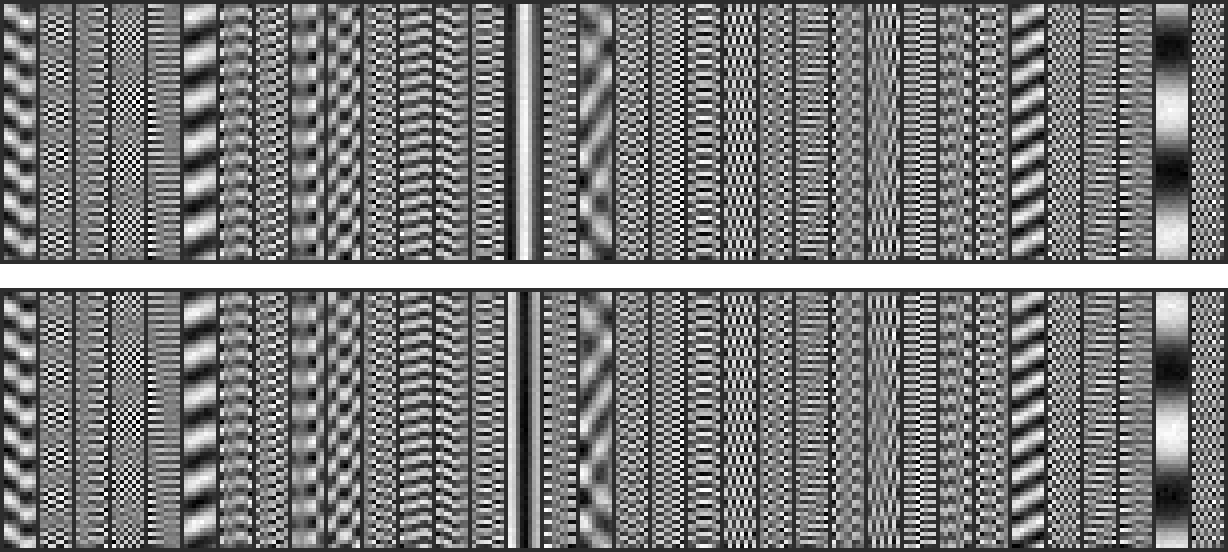}
\end{center}
\caption{Some complementary filters manually selected from $\mathbf{U}$ (top) and $\mathbf{V}$ (bottom) learned from \emph{transposed} pairs of musical n\=/grams.}
\label{fig:filters_gae}
\end{figure}

A GAE learns mappings between data pairs $\mathbf{x},\mathbf{y}\in\mathbb{R}^{P}$ using latent mapping units $\mathbf{m}\in\mathbb{R}^{L}$ as
\begin{equation}\label{eq:gamap_gae}
\mathbf{m} = \sigma (\mathbf{W}(\mathbf{Ux} \cdot \mathbf{Vy})),
\end{equation}
where $\mathbf{U}, \mathbf{V} \in \mathbb{R}^{P \times O}$ and $\mathbf{W} \in \mathbb{R}^{O \times L}$ are weight matrices, and $\cdot$ denotes the element-wise (Hadamard) product (see Figure~\ref{ga} for an illustration).
In our experiments, $\sigma$ is the sigmoid function, but other non-linearities, like the square-root or softplus are also reported in the literature \cite{adelson1985spatiotemporal,alain2013gated,ICML2012Memisevic_105}.
The element-wise product between filter responses $\mathbf{Ux}$ and $\mathbf{Vy}$, leads to filter pairs in $\mathbf{U}$ and $\mathbf{V}$ encoding \emph{correspondences} between the inputs $\mathbf{x}$ and $\mathbf{y}$.
GAEs are very expressive in representing transformations, with the potential to represent any orthogonal transformation (including any permutation in the input space---''shuffling pixels'' \cite{memisevic2013learning}).
The resulting filter pairs show transformation-specific features, like phase-shifted Fourier components when learning transposition between n\=/grams of polyphonic music in our experiments (see Figure~\ref{fig:filters_gae}).
Such phase-shifted filter pairs are then sensitive to corresponding shifts between pairs of data instances.

An interesting aspect of a GAE is its symmetry when solving Equation \ref{eq:gamap_gae} for its inputs, allowing for a reconstruction of an input given the other input and a mapping code.
The reconstruction for binary data is given by

\begin{equation}\label{recon1}
\mathbf{\tilde{x}} = \sigma (\mathbf{U^\top} (\mathbf{W}^\top \mathbf{m} \cdot \mathbf{Vy})),
\end{equation}
and likewise
\begin{equation}\label{recon2}
\mathbf{\tilde{y}} = \sigma (\mathbf{V}^\top (\mathbf{W}^\top \mathbf{m} \cdot \mathbf{Ux})).
\end{equation}
A commonly used training loss \cite{memisevic2013learning,michalski2014modeling} is the symmetric reconstruction error
\begin{equation}\label{eq:recon_symm}
\mathcal{L} = ||\mathbf{x} - \mathbf{\tilde{x}}||^2 + ||\mathbf{y} - \mathbf{\tilde{y}}||^2.
\end{equation}

\begin{figure}
\begin{tabular}{cc}
\includegraphics[trim=31mm 25mm 25mm 25mm, clip, width=.45\linewidth]{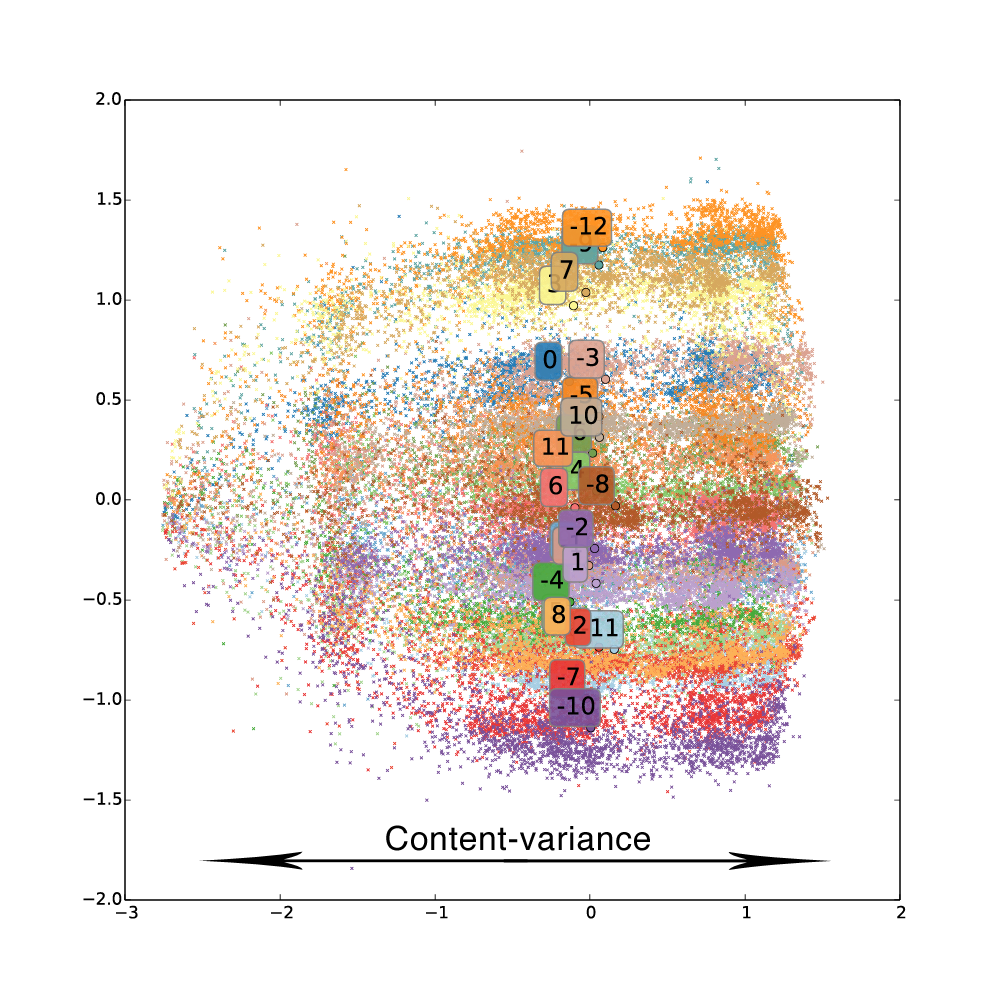} &
\hspace{.45cm}
\includegraphics[trim=31mm 25mm 25mm 25mm, clip, width=.45\linewidth]{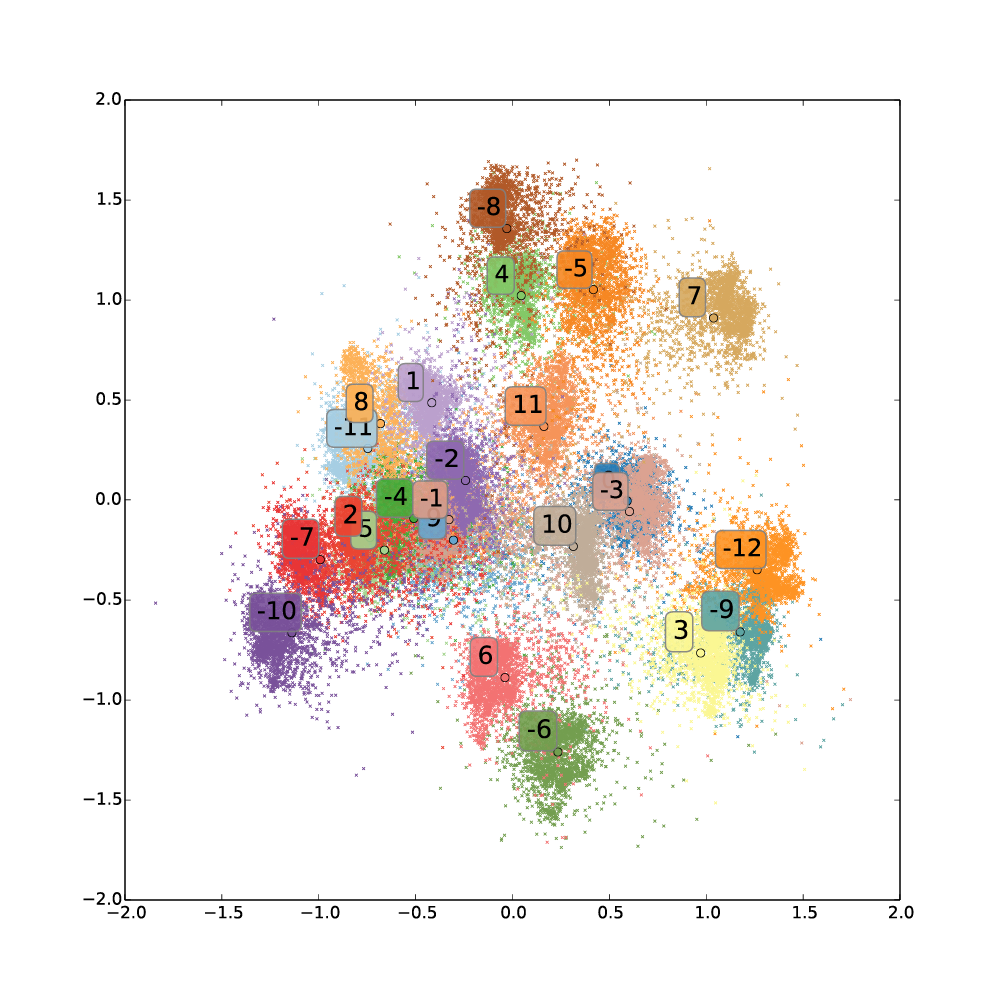} \\
a & b \\
\end{tabular}
\caption{First two principal components (a) and second and third principal component (b) of mappings resulting from unsupervised learning of pairs exhibiting the \emph{chromatic transposition} (TransC) relation.
Points are colored, and cluster centers are named according to the different classes within the TransC relation type (i.e.
distances of note shifts).}
\label{pca}
\end{figure}

When using additive interactions in neural networks, units resemble logical OR-gates which accumulate evidence.
In contrast, multiplicative interactions resemble AND-gates that can detect co-occurrences.
Multiplicative interactions enable the model to ignore input that does not exemplify a known transformation, making it less content-dependent \cite{memisevic2013learning}.
Ideally, learned representations encode fully content-invariant transformations between data pairs.
In practice, there is always some content-dependence in the resulting codes (see Figure~\ref{pca}, where the variance along the first principal component is content-dependent, as it is not correlated with transformation classes). A solution for improving content-invariance in GAEs is given in \cite{lattner2017improving}.

\subsubsection{Restricted Boltzmann Machine} \label{sec:rbm_gae}
An RBM is an energy-based stochastic neural network with two layers, a visible layer with units $\mathbf{v} \in \{0,1\}^r$ and a hidden layer with units $\mathbf{h} \in \{0,1\}^q$ \cite{Hinton:2002ic}.
The units of both layers are fully interconnected with weights $\mathbf{W} \in \mathbb{R}^{r \times q}$, while there are no connections between the units within a layer.
Given a visible vector $\mathbf{v}$, the free energy of the model is calculated as

\begin{equation}\label{equ:energy}
\mathcal{F}(\mathbf{v}) = - \mathbf{a}^\intercal\mathbf{v} - \sum_i \log \left( 1 + e^{\left(b_i + \mathbf{W}_i \mathbf{v} \right)}\right),
\end{equation}

\noindent where $\mathbf{a} \in \mathbb{R}^r$ and $\mathbf{b} \in \mathbb{R}^q$ are bias vectors, and $\mathbf{W}_i$ is the $i$-th row of the weight matrix.
The model is trained by minimizing the free energy given the training data, and by maximizing the free energy for samples from the model---a method called Contrastive Divergence \cite{Hinton:2002ic}.

Reconstruction is performed by projecting a data instance into the hidden space, followed by a projection of the resulting hidden unit activations back into the input space. The reconstruction cross-entropies---as presented in Section \ref{sec:discuss_recon}---arise from the error between the input before and after the projection.

\subsection{Experiment}\label{sec:ExperimentTransform}
The current experiment aims to test the performance of a GAE in learning musical relations between pairs of n\=/grams.
As a baseline, we train an RBM with approximately the same number of parameters on \emph{concatenated} pairs of n\=/grams, like in \cite{susskind2011modeling}.
In order to have a controlled setting and labels for the transformation classes, we construct pairs where one item is an original n\=/gram selected at random from a set of polyphonic Mozart sonatas (see Figure \ref{results_generate}(A) for n\=/gram examples), and the other item is constructed by applying a randomly selected transformation to this n\=/gram.
The experiment is unsupervised in that there are no targets or labels when training the GAE, but we only present one type of transformation per training session.
The implicit goal of the GAE is to cluster the classes \emph{within} each type of transformation.
For example, within the transformation type \emph{chromatic transposition} (TransC), the classes are the different semitone intervals by which the items in a pair are shifted relative to each other.

We use the resulting representations as an input to a classification feed-forward Neural Network (FFNN), to assess how discriminative they are concerning the given classes.
Furthermore, we measure the reconstruction cross-entropies for both models, and we apply transformations extracted from single examples to new data.

\subsubsection{Data}\label{sec:TrainingDataInterval}
We use MIDI files encoding the scores of Mozart piano sonatas from the Mozart/Batik data set~\cite{widmer2003discovering}.
From those pieces, we create binary piano-roll representations, using the MIDI pitch range $[36,100]$ and a temporal resolution of 1/16th note.
From those representations we create $8$-grams (i.e., any consecutive eight 1/16th slices from the piano-roll representation) and for each $8$-gram we construct a counterpart which exhibits the respective transformation.
In order to remove potential peculiarities of the corpus concerning absolute shifts and keys, we start by shifting the notes of the original n\=/grams by $k$ semitones taken randomly from the set $[-12,11]$.
In the following, we describe how the respective counterparts are constructed.

\subsubsubsection{Chromatic Transposition (TransC)}
Chromatic transposition is performed by shifting every pitch in a set of notes by a fixed number of semitones (see
Figure \ref{results_generate}(1)).
For each n\=/gram, we construct a counterpart by shifting all notes by $k$ semitones taken from the set $[-12,11]$.
Note that this also includes the transposition by $0$ semitones, which is the particular case ``exact repetition''.
The resulting $24$ classes for testing the discriminative performance of our models are consequently the elements of the set $[-12,11]$.

\subsubsubsection{Diatonic Transposition (TransD)}
Diatonic transposition is the transposition within a key.
That is, any pitch in a set of notes is shifted by a fixed number of \emph{scale steps} (i.e.
transposition using only allowed notes, e.g.
depicted as green lines in Figure \ref{fig:rondo}).
This operation may change the intervals between notes (see Figure~\ref{results_generate}(2) for examples of diatonic transposition).
At first, we estimate the key for each n\=/gram by choosing the one with the fewest accidentals from the set of keys for which all pitches of the n\=/gram are ``allowed'', omitting n\=/grams which do not fit in a key.
Using this method, we obtain n\=/grams which are assigned to a unique key and can be unambiguously transformed into another key.
That way, a unique transformation can be learned for any constructed instance pair.
We create counterparts for each n\=/gram by shifting each pitch by $k$ \emph{scale steps} of the estimated scale, taken from the set $[-7,6]$.
Thus, the resulting $14$ classes the models are trained on are the elements of the set $[-7,6]$.

\subsubsubsection{Half time / double time (Tempo)}
``Half time'' means doubling the duration of all notes, and ``double time'' is the inverse operation.
We realize the half time relation by scaling each n\=/gram to double its width and then taking only the first half of the result (see
Figure \ref{results_generate}(3)), and for the inverse operation, we swap the thus obtained n\=/gram pairs.
Consequently, we assess the performance of the models in discriminating between the relations \{double time, half time, identity\}.

\subsubsubsection{Retrograde (Retro)}
In music, retrograde means to reverse the temporal order of given notes.
We can simply create pairs exhibiting the retrograde relation by flipping n\=/grams horizontally.
This results in two classes: \{retrograde, identity ($\neg$retrograde)\}


\subsubsection{Model Architectures}
\subsubsubsection{Architecture of the GAE and the RBM}
In order to obtain comparable results for the GAE and the RBM, for each test we choose the same number of parameters and layers for each model.
The GAE may be seen as a two-layered model, where the factors constitute the hidden units of the first layer, and the mapping units constitute the hidden units of the second layer.
Consequently, we also use a two-layered RBM architecture where the number of hidden units in the first layer is equal to the number of factors in the GAE, and the number of hidden units in the top layer is equal to the number of mapping units in the GAE.
We test three different model sizes for every type of transformation.
The smallest model (128/64) has 128 units in the first layer and 64 units in the second layer.
The next bigger model has 256/128 units, and the largest model has 512/256 units in the respective layers.

\subsubsubsection{Architecture of the Classification FFNN}
For all classification tasks, the same architecture is used: A three-layered feed-forward Neural Network (FFNN) with $512$ Rectified Linear units (ReLUs) in the first layer, $256$ ReLUs in the second layer, and Softmax units in the output layer.
The size of the output layer is equal to the number of classes.

\begin{landscape}
\begin{table*}
\centering
\setlength{\tabcolsep}{.5em}
\renewcommand{\arraystretch}{1.3}
\begin{tabular}{rrrrrrrrrrrrrrrrr}
\toprule
& & \multicolumn{14}{c}{predicted interval} & $\sum$\\
& & -7 & -6 & -5 & -4 & -3 & -2 & -1 & 0 & 1 & 2 & 3 & 4 & 5 & 6 &$\text{diag}$ \\
\midrule
\multirow{15}{-2em}{\rotatebox{90}{target interval}} & -7 & 7187 & 1 & \tikzmark{e1}1 & 1 & 3 & \tikzmark{c1}3 & 0 &  \tikzmark{a}10 & 1 & \tikzmark{c2}2 & 1 & 0 & 4 & 0 &  \\
& -6 & 6 & 6943 & 2 & 14 & 2 & 7 & 9 & 1 & 30 & 1 & 30 & 0 & 7 & 6 & \textbf{0} \\
& -5 & \tikzmark{e2}15 & 21 & 7043 & 14 & 26 & 13 & 6 & 3 & 3 & 29 & 0 & 6 & 10 & 3 & \textbf{10} \\
& -4 & 1 & 10 & 3 & 7055 & 7 & 5 & 0 & 0 & 6 & 1 & 17 & 0 & 5 & 0 & \textbf{10} \\
& -3 & 2 & 1 & 6 & 55 & 7090 & 6 & 11 & 0 & 2 & 3 & 8 & 34 & 1 & 9 & \textbf{11} \\
& -2 & \tikzmark{c4}5 & 4 & 3 & 16 & 11 & 6960 & 10 & 2 & 4 & 0 & 12 & 2 & 20 & 0 & \tikzmark{d2}\textbf{52} \\
& -1 & 1 & 8 & 1 & 1 & 6 & 3 & 7050 & 5 & 6 & 3 & 6 & 7 & 13 & 24 & \textbf{3} \\
& 0 & \tikzmark{a1}7 & 0 & 2 & 0 & 0 & 4 & 0 & 7124 & 0 & 6 & 0 & 3 & 3 & 0 & \tikzmark{b}\textbf{164} \\
& 1 & 0 & 46 & 3 & 14 & 7 & 5 & 18 & 6 & 7028 & 0 & 23 & 1 & 4 & 13 & \textbf{28} \\
& 2 & \tikzmark{c3}13 & 1 & 25 & 3 & 16 & 8 & 3 & 0 & 2 & 7082 & 15 & 16 & 4 & 2 & \tikzmark{d1}\textbf{59} \\
& 3 & 1 & 5 & 0 & 12 & 10 & 9 & 5 & 2 & 5 & 0 & 6972 & 39 & 6 & 1 & \textbf{33} \\
& 4 & 1 & 2 & 6 & 0 & 12 & 3 & 6 & 2 & 3 & 7 & 19 & 7027 & 1 & 1 & \textbf{29} \\
& 5 & 9 & 7 & 5 & 17 & 0 & 89 & 11 & 8 & 7 & 4 & 48 & 11 & 6874 & 35 & \tikzmark{f1}\textbf{117} \\
& 6 & 2 & 21 & 5 & 4 & 17 & 6 & 41 & 1 & 36 & 6 & 7 & 23 & 4 & 7053 & \textbf{135} \\
\multicolumn{3}{l}{\rule{0pt}{2.5ex}$\sum\text{diag}$} &\textbf{2} & \textbf{30} & \textbf{13} & \textbf{12} & \tikzmark{d3}\textbf{58} & \textbf{7} & \tikzmark{b1}\textbf{232} & \textbf{32} & \tikzmark{d4}\textbf{104} & \textbf{42} & \textbf{30} & \tikzmark{f2}\textbf{158} & \textbf{141} & \textbf{98488} \\
\bottomrule
\end{tabular}
\begin{tikzpicture}[overlay, remember picture, red, yshift=.25\baselineskip, shorten >=.5pt, shorten <=.5pt]
\draw [->] ($(pic cs:a)+(0pt,3pt)$) -- ($(pic cs:b)+(0pt,5pt)$);
\draw [->] ($(pic cs:a1)+(0pt,4pt)$) -- ($(pic cs:b1)+(0pt,8pt)$);
\end{tikzpicture}
\begin{tikzpicture}[overlay, remember picture, blue, yshift=.25\baselineskip, shorten >=.5pt, shorten <=.5pt]
\draw [->] ($(pic cs:c1)+(0pt,3pt)$) -- ($(pic cs:d1)+(0pt,3pt)$);
\draw [->] ($(pic cs:c4)+(0pt,3pt)$) -- ($(pic cs:d4)+(0pt,8pt)$);
\end{tikzpicture}
\begin{tikzpicture}[overlay, remember picture, orange, yshift=.25\baselineskip, shorten >=.5pt, shorten <=.5pt]
\draw [->] ($(pic cs:c2)+(0pt,3pt)$) -- ($(pic cs:d2)+(0pt,4pt)$);
\draw [->] ($(pic cs:c3)+(0pt,5pt)$) -- ($(pic cs:d3)+(0pt,6pt)$);
\end{tikzpicture}
\begin{tikzpicture}[overlay, remember picture, green, yshift=.25\baselineskip, shorten >=.5pt, shorten <=.5pt]
\draw [->] ($(pic cs:e1)+(0pt,3pt)$) -- ($(pic cs:f1)+(0pt,4pt)$);
\draw [->] ($(pic cs:e2)+(0pt,5pt)$) -- ($(pic cs:f2)+(0pt,8pt)$);
\end{tikzpicture}
\vspace{0.3cm}
\footnotesize
\caption{Confusion Matrix for classifier FFNN trained on representations of GAE TransD 128/64.}
\vskip-2ex
\label{results_confusion}
\end{table*}
\end{landscape}

\subsubsection{Training}
In the following, we report how the GAE, the RBM, and the FFNN are trained in our experiments.
For each transformation type, we train on $490\, 000$ pairs using a validation set of $10\, 000$ pairs.
The final testing is done on a test set of size $100\, 000$ per transformation type.

\subsubsubsection{Training the Gated Autoencoder}
The model parameters of the GAE are trained using stochastic gradient descent to minimize the symmetric reconstruction error (see Equation \ref{eq:recon_symm}).
We train the model on the data pairs for $1000$ epochs, using a mini-batch size of $500$, a learning rate of $3\times10^{-5}$ and a momentum of $0.93$.
Learning improves when the input (i.e. $\mathbf{y}$ and $\mathbf{x}$ in Equations \ref{recon1} and \ref{recon2}, respectively) is corrupted during training, as it is done in de-noising autoencoders \cite{vincent2010stacked}. We achieve this by randomly setting $35\%$ of the input bits to zero and training the GAE to reconstruct an uncorrupted, transformed version of it (i.e. $\mathbf{\tilde{x}}$ and $\mathbf{\tilde{y}}$ in Equations \ref{recon1} and \ref{recon2}, respectively). For the first $100$ epochs, the input weights are re-scaled after each parameter update to their average norm, as described by \citet{susskind2011modeling,memisevic2011gradient,ICML2012Memisevic_105}. We use L1 and L2 weight regularization on all weights, and Lee's sparsity regularization \cite{Lee:2007uz} on the mapping units and on the factors.

\subsubsubsection{Training the Restricted Boltzmann Machine}
We train two RBM layers on \emph{concatenated} data pairs with greedy layer-wise training using \emph{persistent contrastive divergence} (PCD)~\cite{tielemanPcd2008}, a variation of the standard \emph{contrastive divergence} (CD) algorithm~\cite{Hinton:2002ic}, which is known to result in a better approximation of the likelihood gradient.
We use a learning rate of $3 \times 10^{-3}$ which we reduce to zero during $300$ training epochs.
We use a batch size of $100$, and we reset (i.e.,
randomize) the weights of neurons, whose average activity over all training examples exceeds $85\%$. We use L1 and L2 weight regularization and the sparsity regularization proposed by~\citet{Goh:2010wi} setting $\mu = 0.08$ and $\phi = 0.75$.

\subsubsubsection{Training the Classification FFNN}
The FFNN is trained in a supervised manner on the latent representations of both the GAE (i.e., its mapping unit configurations) and the RBM (i.e., its hidden unit configurations), resulting from unsupervised training on related data pairs using the categorical cross-entropy loss.
The network is trained for $300$ epochs with a learning rate of $0.005$, a batch size of $100$ and a momentum of $0.93$.
We apply L2 weight regularization on all weights, sparsity regularization as in \cite{Lee:2007uz}, $50\%$ dropout \cite{srivastava2014dropout} on all except the top-most layer, as well as batch normalization \cite{ioffe2015batch}.

\subsection{Results and Discussion}\label{sec:results-discussion_gae}


\subsubsection{Discriminative Performance}\label{sec:discuss_discriminative}
Table~\ref{results_class} shows the mis-classification rates of the FFNN, trained on representations of the GAE and the RBM.
The models are trained separately for each transformation type, chromatic transposition (TransC), diatonic transposition (TransD), half-time/double-time (Tempo), and retrograde (Retro) where for each type, there are several classes. For example, for chromatic transposition, the classes are the 24 transposition distances of the set $[-12,11]$ (see Section~\ref{sec:TrainingDataInterval}).

\begin{table}[t]
\centering
\begin{tabu}{rllll}
\toprule
& \multicolumn{4}{c}{Transformation} \\ \cline{2-5}
\rule{0pt}{2.5ex}  Model  \ \ \ \ &  TransC  \ \  \ \ & TransD  \ \ \ \ & Tempo  \ \ \ \ & Retro  \ \ \\ \midrule
\rowfont{\footnotesize}
 \multicolumn{1}{l}{\ \ \# Classes  \ \ \ \ \rule{0pt}{2ex}} & 24 & 14 & 3 & 2 \\
 \multicolumn{1}{l}{\ \ \textbf{Random \rule{0pt}{3ex}}} & 95.83 & 92.86 & 66.67 & 50.00 \\
 \multicolumn{1}{l}{\ \ \textbf{RBM \rule{0pt}{3ex}}\ \ \ \ \ \ } & & & & \\
 \ \ 128/64  \ \  \ \ & 23.10 & 87.55 & 50.09 & 50.06  \\
 \ \ 256/128  \ \  \ \ & 6.12 & 51.05 & 47.52 & 50.11  \\
 \ \ 512/256  \ \  \ \ & 2.18 & 19.47 & 40.33 & 50.19  \\
 \multicolumn{1}{l}{\ \ \textbf{GAE} \rule{0pt}{2.5ex}} & & & & \\
 \ \ 128/64  \ \  \ \ & 1.88 & 1.51 & 2.47 & 3.24 \\
 \ \ 256/128  \ \  \ \ & \textbf{0.02} & 0.26 & 0.89 & 1.10 \\
 \ \ 512/256  \ \  \ \ & 0.03 & \textbf{0.23} & \textbf{0.11} & \textbf{0.28} \\
\bottomrule
\end{tabu}
\caption{Mis-classified rates (in percent) from the classification Feed-Forward Neural Network trained on representations of the Restricted Boltzmann Machine (RBM) and the Gated Autoencoder (GAE) in different architecture sizes and for different transformation types.
\emph{Random} denotes the random guessing baseline dependent on the number of classes.}
\label{results_class}
\end{table}

The GAE is clearly better at learning discriminative representations of musical relations.
The misclassification rates of the largest architecture (512/256) are below $0.3\%$ for all relations.
In contrast, the RBM is less suitable for unsupervised learning of relations---in the case of Retro, it does not even outperform random guessing.
Note furthermore that in contrast to the GAE, increasing the capacity of the RBM does not always improve results.
This suggests that the architecture itself is a limiting factor:
As the activations of units in an RBM only depend on the additive accumulation of evidence, it attempts to learn all combinations of mutually transformed data instances it is presented with during training. For transposition relations, learning such combinations results in representations which are informative enough that the classification FFNN can infer the transformation class from it. Retrograde and Tempo, however, are too complex transformations to grasp for content-dependent representation learner.

In contrast, the GAE separates the problem in representing characteristics of the input n-grams with additive input connections on the one hand, and modeling transformations using multiplicative interactions on the other hand. Also, when training the GAE on reconstructing an input, the other input provides content information allowing the mapping units to represent only the content-invariant transformation. It is still a non-trivial result that the GAE can learn all pre-defined transformation types. In particular diatonic transposition (TransD) is a very relevant transformation in music and it is a valuable finding that the GAE is capable of learning it.

Table~\ref{results_confusion} shows a confusion matrix for the GAE TransD 128/64 relation.
It is instructive for interpretation of the table to realize that two diatonic transpositions of the same n-gram yield another pair of n-grams that are also diatonically transposed versions of each other.
For instance, the diatonic transposition pairs 0 and -7,  1 and -6, 2 and -5, and so on, all yield a resultant transposition of -7 (a downward octave).
Analog equivalences hold for resultant transpositions of an upward octave, the upward/downward fifth, third, and so on.
The equivalent resultant transpositions have been marked by colored arrows in the table.
The arrows clearly highlight that the model is biased toward confusions \emph{by one octave} from the correct shift (red arrows).
For example, when the actual shift between n\=/gram pairs is $+5$, the classifier frequently estimates a shift of $-2$, which is $7$ scale steps or \emph{one octave} away from the correct target.
Similarly, frequently confused targets are a third (green arrows), a fifth (blue arrows), and an octave plus a third (orange arrows) away from the correct target.
Since the data pairs are uniformly distributed over all classes, it follows that these confusions are caused by covariances \emph{within} single n\=/grams, where thirds, fifths, and octaves occur frequently.
Similarly, the misclassifications of second interval shifts (i.e., the entries directly above and below to the diagonal) most likely result from second intervals in melodies. 
Many of the filters learned by the model are also receptive to intervals of notes which occur sequentially in time (see Figure \ref{pca}, where filters are mostly horizontally oriented), therefore also intervals in melodies are represented in the latent space of the model.

\subsubsection{Reconstruction}\label{sec:discuss_recon}
Table~\ref{results_recon} lists the reconstruction cross-entropies for each architecture versus transformation type.
The error of the largest GAE architecture is about an order of magnitude smaller than that of the best RBM architecture.
Furthermore, the cross-entropies do not substantially improve when increasing the size of the RBM, again suggesting an architectural advantage of the GAE over the RBM.
\citet{olshausen2007bilinear} suggest that the advantage consists in the ability to separate content and its manifestation (the ``what'' and the ``where'' ).
Mapping units only have to encode the "where" components leading to more efficient use of the model parameters.

\begin{table}[t]
\centering
\begin{tabular}{rlllll}
\toprule
& \multicolumn{4}{c}{Transformation} \\ \cline{2-5}
\rule{0pt}{2.5ex}  Model  \ \  \ \ & TransC \ \  \ \  & TransD \ \  \ \  & Tempo \ \  \ \  & Retro \ \ \\ \midrule
\multicolumn{1}{l}{ \ \  \textbf{RBM \rule{0pt}{2ex}}\ \ \ \ \ \ \ \ } & & & & \\
128/64  \ \  \ \ & 0.131 & 0.122 & 0.098 & 0.110 \\
256/128  \ \  \ \  & 0.122 & 0.119 & 0.084 & 0.095 \\
512/256 \ \  \ \  & 0.113 & 0.112 & 0.075 & 0.086 \\
\multicolumn{1}{l}{\ \ \textbf{GAE} \rule{0pt}{2.5ex}} & & & & \\
128/64 \ \  \ \  & 0.026 & 0.032 & 0.025 & 0.033 \\
256/128 \ \  \ \  & 0.018 & 0.024 & 0.015 & 0.017 \\
512/256 \ \  \ \  & \textbf{0.012} & \textbf{0.016} & \textbf{0.007} & \textbf{0.009} \\
\bottomrule
\end{tabular}
\caption{Reconstruction cross-entropies for the Restricted Boltzmann Machine (RBM) and the Gated Autoencoder (GAE) in different sizes and for different transformation types.}
\label{results_recon}
\end{table}

\subsubsection{Generation}\label{sec:discuss_gen}
\begin{figure}[!ht]
\newlength{\figheight}
\centering
\setlength{\figheight}{.300\linewidth}
\setlength{\textfloatsep}{-0cm}
\begin{tabular}{cc|c}
\rule{0pt}{3ex} & A & B \\
1 & \raisebox{-.5\figheight}{\rule{0em}{18ex}\includegraphics[height=\figheight]{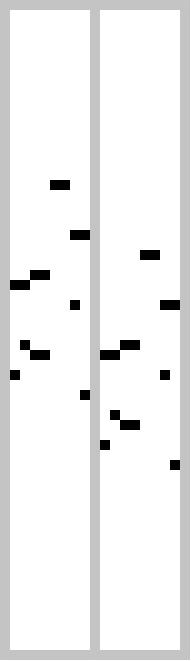}} & \raisebox{-.5\figheight}{\includegraphics[height=\figheight]{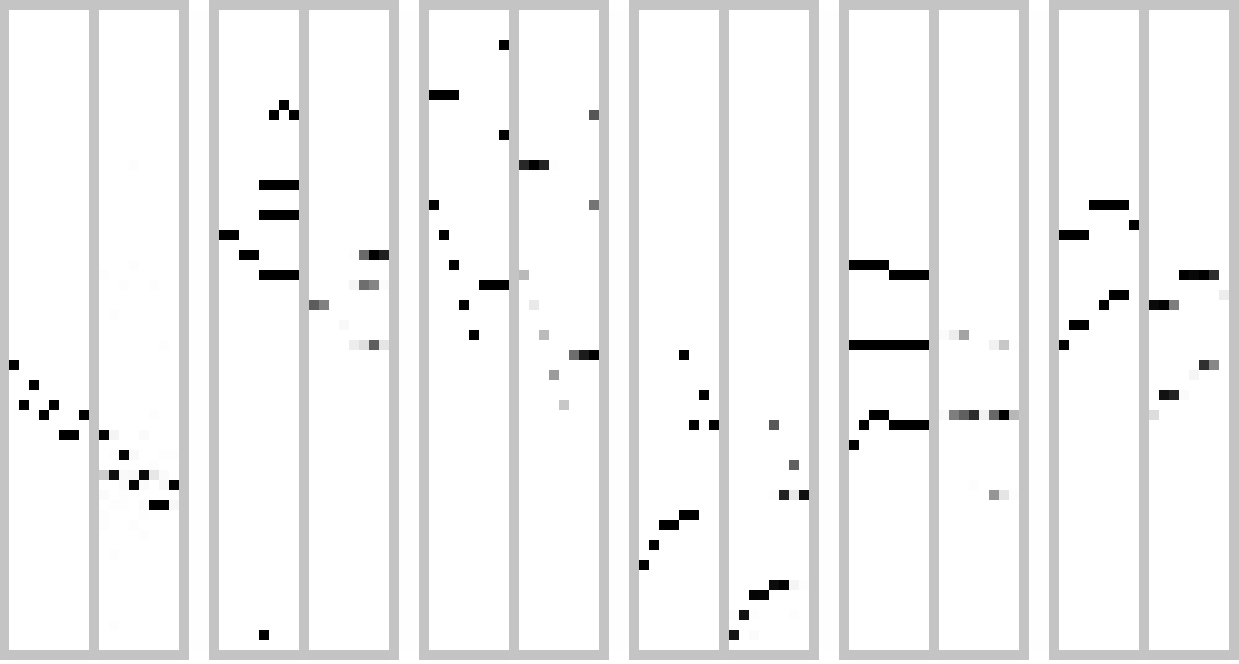}}\\
2 & \raisebox{-.5\figheight}{\rule{0em}{22ex}\includegraphics[height=\figheight]{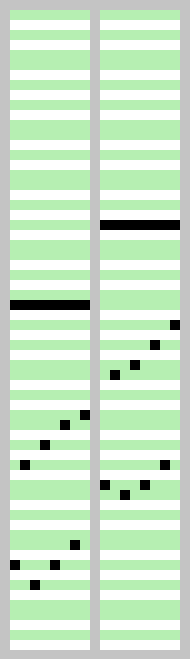}} & \raisebox{-.5\figheight}{\includegraphics[height=\figheight]{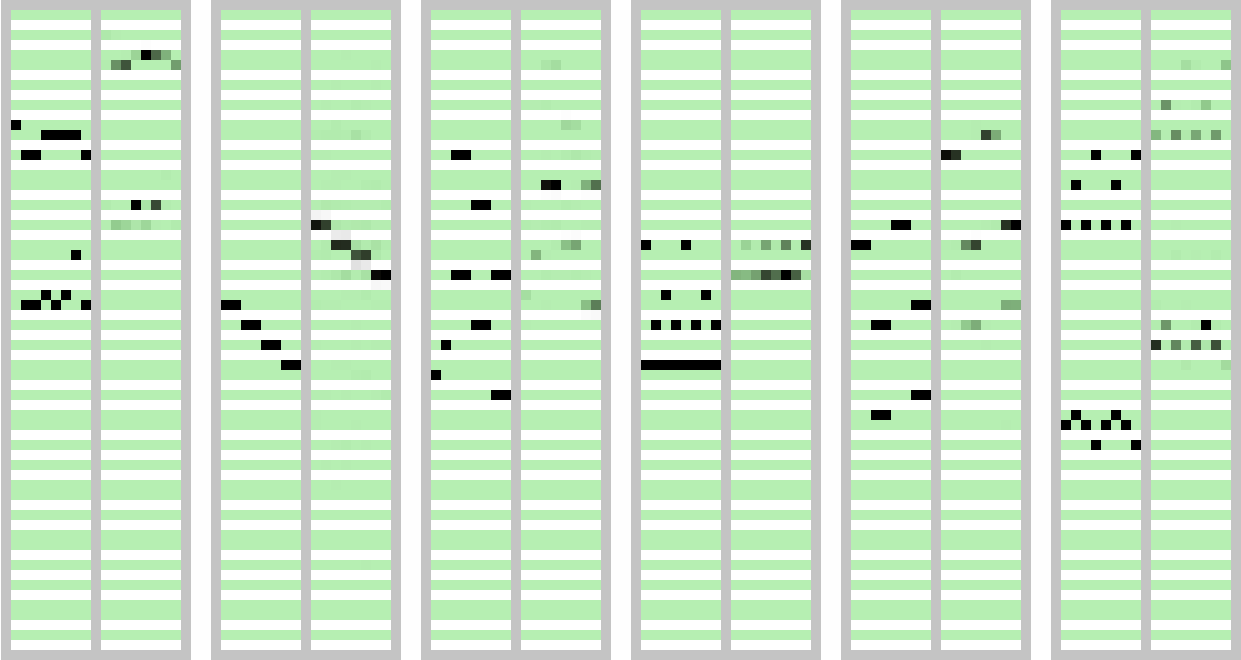}}\\
3 & \raisebox{-.5\figheight}{\rule{0em}{22ex}\includegraphics[height=\figheight]{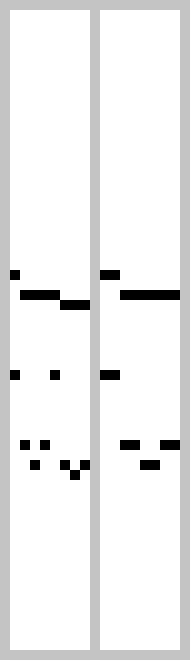}} & \raisebox{-.5\figheight}{\includegraphics[height=\figheight]{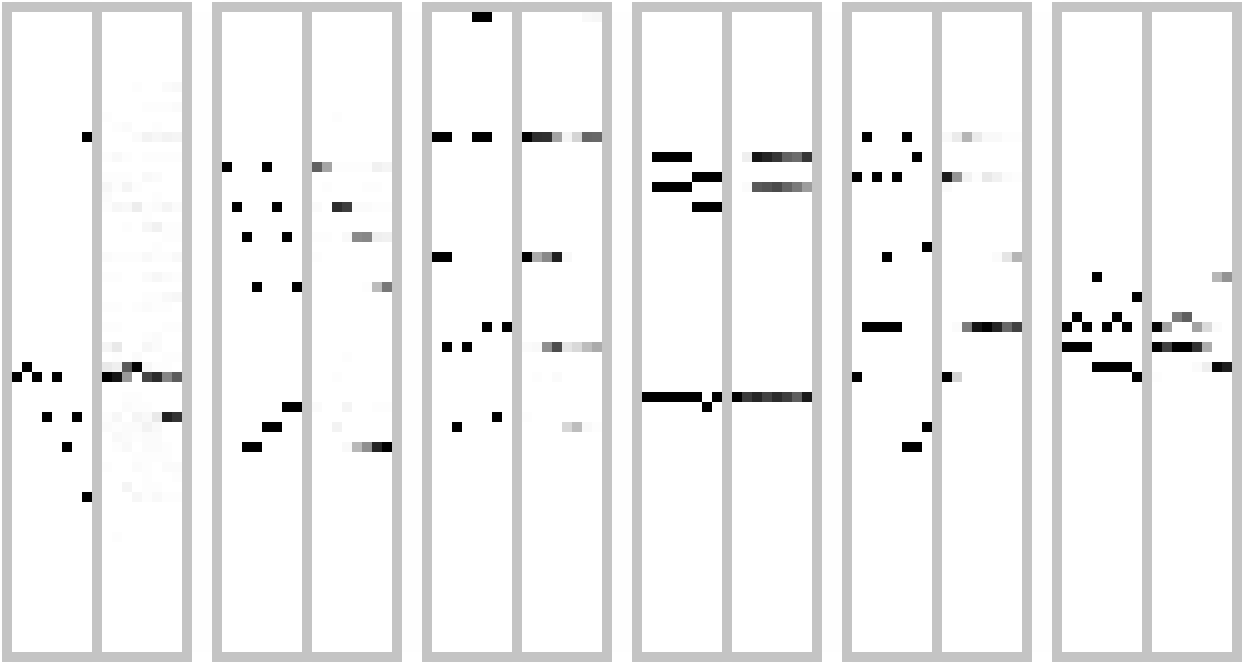}}\\
4 & \raisebox{-.5\figheight}{\rule{0em}{22ex}\includegraphics[height=\figheight]{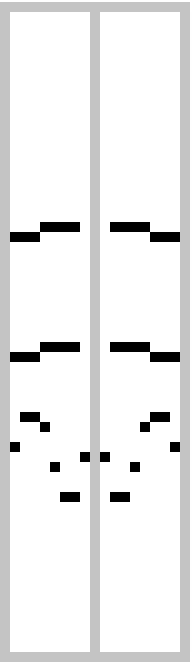}} & \raisebox{-.5\figheight}{\includegraphics[height=\figheight]{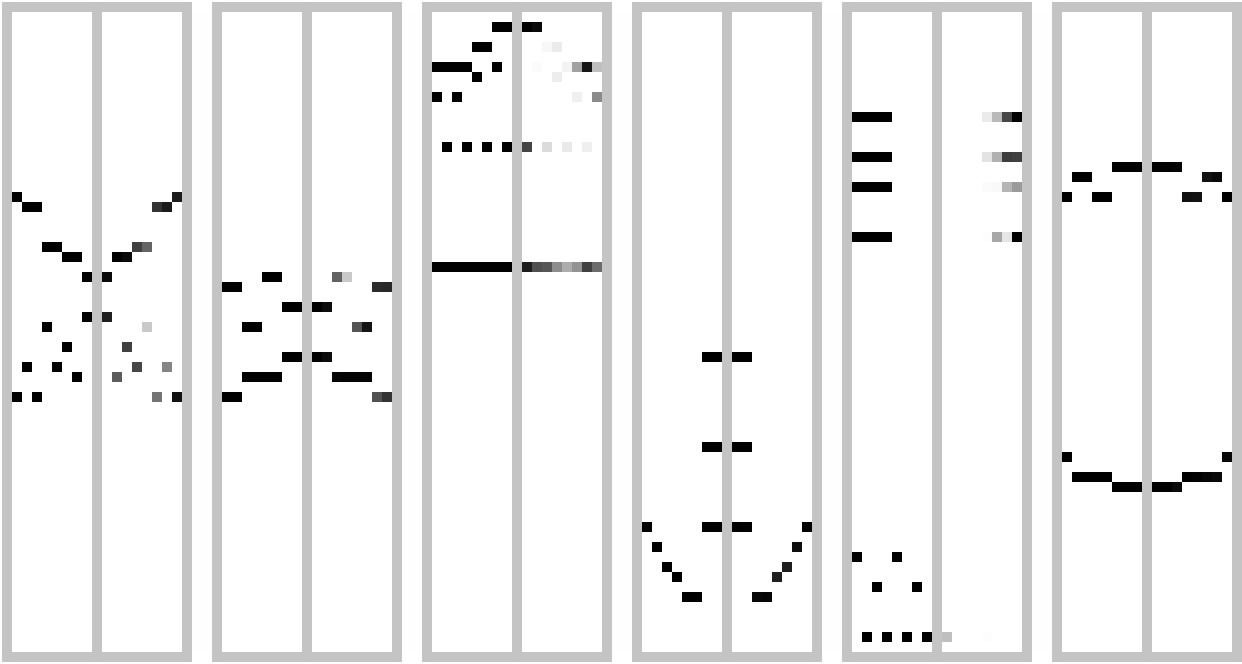}}\\
\end{tabular}
\footnotesize
\caption{Results of the analogy-making task. Transformations are inferred by the GAE from pairs of 8-grams (A) and applied to new n-grams, not seen during training (B, left parts) to generate counterparts with analogous transformations (B, right parts), where the level of blackness indicates the certainty of the model that the respective note is part of the transformed result: 1) Chromatic transposition, 2) diatonic transposition, 3) tempo change (halftime), and 4) retrograde.
Green horizontal lines mark scale notes in diatonic transposition, for visual guidance.
}
\label{results_generate}
\end{figure}

Figure~\ref{results_generate} shows analogy-making examples for the GAE.
Mappings are inferred from template n\=/gram pairs (A) and applied to single instances which were not part of the training corpus.
The resulting pairs (B) should exhibit the same transformation as the template pairs from which the transformation was inferred.
For diatonic transposition (2), the mapping is applied only to instances in the same key as the source instance (i.e.
left instance) of the template pairs, and valid pitches are marked with green lines.

The transformations of the template pairs shown in Figure~\ref{results_generate} are chromatic transposition by $-7$ semitones (1), diatonic transposition by $+5$ scale steps in C major (2), halftime (3), and retrograde (4). Depending on the type of transformation, the results vary in their overall quality.
The examples shown in Figure~\ref{results_generate} have been selected to illustrate both high and low-quality transformations.
We found that diatonic transposition (TransD) transformation was generally of lower quality than the other transformation types.
In low-quality transformations, notes are frequently missing in generated counterparts, which is a result of learned representations being not fully content-invariant.
Figure~\ref{pca} shows that content-dependence in the first principal component, as this component is not correlated with content-invariant classes.
Note that the GAE is ``conservative'' when it comes to generating analogies, in that the reconstruction errors are caused by omitting existing notes, but very rarely by incorrectly introduced notes.
The reason is that factors only take on large values when inputs comply with their transformations (see \cite{ICML2012Memisevic_105} for more details).

\subsection{Conclusion and Future Work}\label{sec:concl-future-work0}
We have evaluated the performance of two connectionist models in learning transformations from pairs of musical n-grams in an unsupervised manner.
We found that the Gated Autoencoder was more effective than a standard RBM, both in an input reconstruction task and in a discriminative task in which the representations learned by the models were used to classify n-gram pairs in terms of the transformation they exhibit.
Ideally, transformations learned by a GAE are fully content-invariant.
We found that this is not the case in practice when training the GAE using the classical objective function.
In other recent work, we show that the content-invariance of learned representations can be improved by a regularization term that explicitly penalizes content-variance~\cite{lattner2017improving}.

The results reported in this section show that when given enough n-gram pairs exhibiting transformations of a specific type, the GAE can learn these transformations.
Future work should assess the suitability of the models to automatically infer transformations from a music corpus.
This poses the further challenge of selecting appropriate n-gram pairs for training the model, as the majority of randomly selected n-gram pairs in a corpus would be unrelated.
A possible way to address this issue is to use the GAE itself to make musical representations locally invariant towards some properties, in order to better detect similar fragments.
For example, local invariance towards transposition can unveil mutually transposed texture-pairs on a global scale (see Section \ref{sec:interval}).
Another way is to use a bootstrapping approach in which the selection of n-gram pairs for training the model is based on some measure defined by the model itself, for example by greedily selecting the pairs for which the score is high (i.e., the energy of the pair is low \cite{im2015scoring}).

\section{Learning Interval Representations from Polyphonic Music Sequences}\label{sec:interval}
%
%

\begin{abstract}
In the previous section, we tested a common Gated Autoencoder (GAE) architecture on n-gram pairs and found that it is well-suited for learning musical transformations---an important step towards modeling musical structure.
In order to achieve a controlled setting, the n-gram pairs were constructed, rather than sampled from real music pieces.
Therefore, it was known which transformations they exhibit and \emph{that} the pairs always exhibit a systematic transformation.
In a real-world setting, there is the chicken-egg situation that for selecting related training pairs in a music piece, an already trained GAE would be necessary to score the relatedness of pairs (see Section \ref{sec:concl-future-work0}).

A different approach is based on the observation that music is a sequential phenomenon, and thus we can rely on some local correlations.
A GAE can learn such regularities and provide us with a transformational view on the musical surface.
In this study, the architecture and loss function of the GAE is modified to obtain a predictive setting, where events are represented as the result of learned transformation functions applied to preceding events.
It turns out that learning such local transformations of musical notes leads to very useful features: \emph{pitch intervals}.

Most concepts in music theory (e.g., scale types and modes, cadences, chord types) are defined in terms of intervals (i.e., relative distances) to reference pitches.
Therefore, when computer models are employed in music tasks, it can be useful to operate on interval representations rather than on the raw musical surface.
Moreover, interval representations are transposition-invariant, valuable for tasks like audio alignment, cover song detection, and music structure analysis.
In this section, we employ a GAE to learn fixed-length, invertible and transposition-invariant interval representations from polyphonic music both in the symbolic domain and in audio.
Applying a k-nearest neighbor (k-nn) classifier in the mapping space, we show that the mappings (i.e., latent representations) learned by the model encode the intervals contained in the polyphonic input data. 
Furthermore, we examine the organization of intervals in the mapping space and show that the model organizes them in a musically meaningful manner.
Finally, based on the learned mappings, transposition-invariant self-similarity matrices are constructed and employed in the MIREX task ``Discovery of Repeated Themes and Sections'', yielding competitive results.

The contributions of this study include
(1) adapting the GAE architecture for sequence learning, in particular using different dimensionality for input and prediction, as well as two mapping layers,  
(2) proposing a novel cost function to support the learning of transposition-invariant interval representations,
(3) showing the usefulness of the learned interval representations for the discovery of musical sections.
Furthermore, to our knowledge, this is the first time that it is shown that a GAE can represent multiple transformations (i.e., multiple intervals) within a data pair at once.

This section is structured as follows.
Section \ref{sec:interval-motiv} provides some additional motivation for relative pitch processing in music.
In Section~\ref{sec:model_interval}, the used architecture is described and in Section~\ref{sec:data-relative}, data is introduced on which the GAE is trained.
The training procedure, including the novel method to support the emergence of transposition-invariance, is proposed in Section~\ref{sec:training_inter}.
The experiments conducted to examine the properties of learned mappings are introduced in Section~\ref{sec:experimentsInverval}, and results and a discussion are given in Section~\ref{sec:res_inter}.

\end{abstract}

\subsection{Motivation for Modeling Relative Pitch Processing}\label{sec:interval-motiv}
The notion of relative pitch is important in music understanding. Many music-theoretical concepts, such as scale types, modes, chord types, and cadences, are defined in terms of relations between pitches or pitch classes.
But relative pitch is not only a music-theoretical construct.
It is common for people to perceive and memorize melodies in terms of pitch intervals (or in terms of \emph{contours}, the upward or downward \emph{direction} of pitch intervals) rather than sequences of absolute pitches.
This characteristic of music perception also has ramifications for the perception of form in musical works. It implies that transposition of some musical fragment along the pitch dimension (such that the relative distances between pitches remain the same) does not alter the perceived identity of the musical material, or at least establishes a sense of similarity between the original and the transposed material.
As such, adequate detection of musical form in terms of (approximately) repeated structures presupposes the ability to account for pitch transposition---one of the most common types of transformations found in music.

Relative pitch perception in humans is currently not well-understood~\cite{mcdermott08d}. For example, there are no established theories on how the human brain derives a relative representation of pitch from the tonotopic representations formed in the cochlea, neither is it clear whether there is a connection between the perception of pitch relations in simultaneous versus consecutive pitches.

Computational approaches to address tasks of music understanding (such as detecting patterns and form in music) often circumvent this issue by representing musical stimuli as sequences of monophonic pitches, after which simply differencing consecutive pitches yields a relative pitch representation.
This approach also works for polyphonic music, to the extent that the music can be meaningfully segregated into monophonic pitch streams.
A drawback of this approach is that it presupposes the ability to segregate musical streams, which is often far from trivial due to the ambiguity of musical contexts.
To take an analogous approach on acoustical representations of musical stimuli is even more challenging since it further depends on the ability to detect pitches and onsets in sound.

In this study, we take a different approach altogether.
We train a Gated Autoencoder (GAE) to learn representations that represent the relation between the music at some time point $t$ and the preceding musical context.
During training, these representations are adapted to minimize the reconstruction error of the music at $t$ given the preceding context and the representation itself.

A crucial aspect of the GAE is its bilinear architecture involving multiplicative connections, which facilitates the formation of relative pitch representations.
We stimulate such representations more explicitly using an altered training procedure in which we transpose the training data using arbitrary transpositions.
The results are two models (for symbolic music and audio) that can map both monophonic and polyphonic music to a sequence of points in a vector space---the \emph{mapping space}---in a way that is invariant to pitch transpositions.
This means that a musical fragment will be projected to the same mapping space trajectory independently of how it is transposed.

We validate our approach experimentally in several ways. First, we show that musical fragments that are nearest neighbors in the mapping space have many pitch intervals in common (as opposed to nearest neighbors in the input space). Then we show that the topology of the learned mapping space reflects musically meaningful relations between intervals (such as the tritone being dissimilar to other intervals). Lastly, we use mapping space representations to detect musical form both for symbolic and audio representations of music, showing that it yields competitive results, and in the case of audio even improves the state-of-the-art.
A re-implementation of the transposition-invariant GAE for audio is publicly available\footnote{see \url{https://github.com/SonyCSLParis/cgae-invar}}.

\subsection{Model}\label{sec:model_interval}
\begin{figure}
\begin{center}
\includegraphics[width=.5\linewidth]{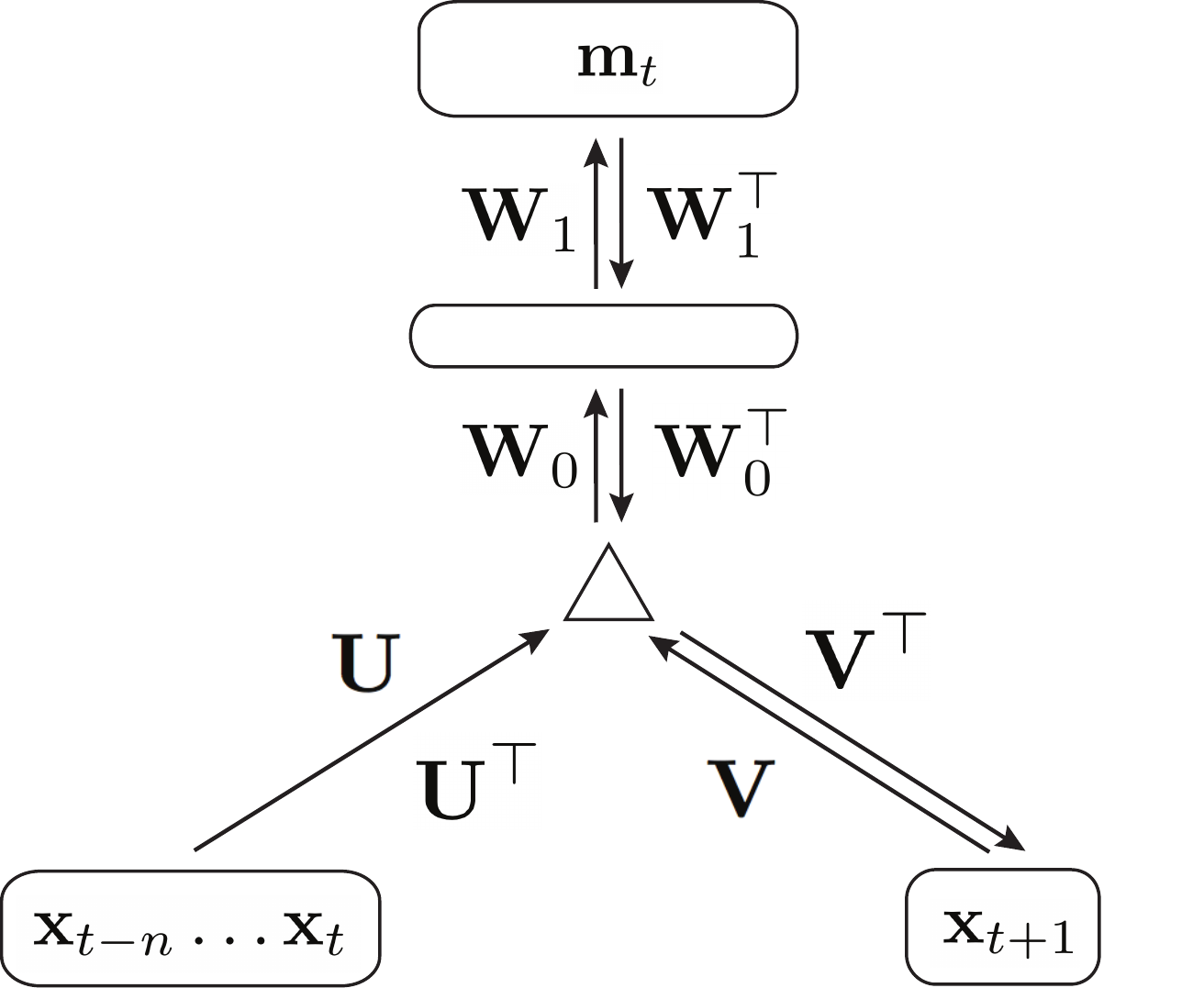}
\caption{Schematic illustration of the Gated Autoencoder (GAE) architecture used in the experiments.}
\label{fig:ga}
\end{center}
\end{figure}

Let $\mathbf{x}_j$ be a vector representing pitches of currently sounding notes (in the symbolic domain) or the energy distributed over frequency bands (in the audio domain), in a fixed-length time interval.
Given a temporal context $\mathbf{x}_{t-n}^{t} = \mathbf{x}_{t-n} \dots \mathbf{x}_{t}$ as the input and the next time step $\mathbf{x}_{t+1}$ as the target, the goal is to learn a mapping $\mathbf{m}_t$ which does not change when shifting $\mathbf{x}_{t-n}^{t+1}$ up- or downwards in the pitch dimension.
A GAE (depicted in Figure~\ref{fig:ga}) is well-suited for this task, modeling the intervals between reference pitches in the input and pitches in the target, encoded in the latent variables of the GAE as mapping codes $\mathbf{m}_j$.
In Section \ref{sec:gae}, we trained the GAE by minimizing the symmetric error when reconstructing the output from the input and vice versa.
In the proposed architecture, we use predictive training and just learn to infer the output from the input.
More precisely, the goal of the training is to find a mapping $\mathbf{m}_t$ for any input/target pair which transforms the input into the given target.
The mapping at time $t$ is calculated as

\begin{equation}\label{eq:gamap}
\mathbf{m}_t = \sigma_h (\mathbf{W}_1 \sigma_h(\mathbf{W}_0(\mathbf{U}\mathbf{x}_{t-n}^{t} \cdot \mathbf{V}\mathbf{x}_{t+1}))),
\end{equation}
where $\mathbf{U}, \mathbf{V}$ and $\mathbf{W}_k$ are weight matrices, $\sigma_h$ is the hyperbolic tangent non-linearity, and we will refer to the learnt mappings $\mathbf{m}_j$ as the \emph{mapping space} of the input/target pairs. 
The operator~$\cdot$ (depicted as a triangle in Figure~\ref{fig:ga}) depicts the Hadamard (or element-wise) product of the filter responses $\mathbf{U}\mathbf{x}_{t-n}^{t}$ and $\mathbf{V}\mathbf{x}_{t+1}$, denoted as \emph{factors}.
This operation allows the model to \emph{relate} its inputs, making it possible to learn interval representations.
We found that using two mapping layers $\mathbf{W}_0$ and $\mathbf{W}_1$ improves the results of our experiments.
In the previous section (see Section \ref{sec:gae}), we use only one layer, as we simply adopt the vanilla GAE architecture from the literature to perform a proof-of-concept for learning musical transformations.
Having said that, we hypothesize that more layers would also improve the results in Section \ref{sec:gae}.
In fact, in our most recent publication \cite{DBLP:conf/waspaa/Lattner19} we use stacks of several (convolutional) layers for the input, target, and mapping pathways (depicted as $\mathbf{U}$, $\mathbf{V}$, and $\mathbf{W}_k$ in Figure~\ref{fig:ga}).

The target of the GAE can be reconstructed as a function of the input $\mathbf{x}_{t-n}^{t}$ and a mapping $\mathbf{m}_t$:
\begin{equation}\label{recony}
\mathbf{\tilde{x}}_{t+1} = \sigma_g(\mathbf{V}^\top (\mathbf{W}_0^\top\mathbf{W}_1^\top \mathbf{m}_t \cdot \mathbf{U}\mathbf{x}_{t-n}^{t})),
\end{equation}
where $\sigma_g$ is the sigmoid non-linearity for binary input and the identity function for real-valued input.

The cost function is defined to penalize the error of reconstructing the target $\mathbf{x}_{t+1}$ given the input $\mathbf{x}_{t-n}^{t}$ and the mapping $\mathbf{m}_t$.
For real-valued sequences, the mean-square error
\begin{equation}\label{eq:cost-mse}
\mathcal{L}_{\text{MSE}}(\mathbf{x},\mathbf{\tilde{x}}) = \frac1N\sum_{n=1}^N\ (x_n - \tilde x_n )^2
\end{equation}
is used, while the cross-entropy loss 
\begin{equation}\label{eq:cost-ce}
\mathcal{L}_{\text{CE}}(\mathbf{x},\mathbf{\tilde{x}}) = -\frac1N\sum_{n=1}^N\ \bigg[x_n  \log_2 \tilde x_n + (1 - x_n)  \log_2 (1 - \tilde x_n)\bigg]
\end{equation}
is minimized for binary sequences.
A temporal context $n > 0$ is chosen to provide the GAE with more context to relate its input with the output. As shown in Figure \ref{fig:sensitivity_context}, besides focusing on $t$, the model will turn out to be specifically sensitive to onsets a quarter note ($t-3$) and a half note ($t-7$) before the prediction target at $t+1$.

\subsection{Data}\label{sec:data-relative}
We train the model both on symbolic music representations and on audio spectrograms.
For the symbolic data, the Mozart/Batik data set~\cite{widmer2003discovering} is used, consisting of 13 piano sonatas containing more than 106,000 notes.
The dataset is encoded as successive $60$ dimensional binary vectors (encoding MIDI note number $36$ to $96$), each representing a single time step of 1/16th note duration.
The pitch of an active note is encoded as a corresponding on-bit, and as multiple voices are encoded simultaneously, a vector may have multiple active bits.
The result is a piano roll-like representation.

The audio dataset consists of 100 random piano pieces of the MAPS dataset \cite{emiya2010multipitch} (subset MUS), at a sampling rate of 22.05 kHz. We choose a constant-Q transformed spectrogram using a hop size of $1984$, and Hann windows with different sizes depending on the frequency bin. The range comprises $120$ frequency bins (24 per octave), starting from a minimal frequency of $65.4$ Hz. Each time step is contrast-normalized to zero mean and unit variance.

\subsection{Training}\label{sec:training_inter}
The model is trained with stochastic gradient descent in order to minimize the cost function (cf. Equations~\ref{eq:cost-mse} and \ref{eq:cost-ce}) using the data described in Section~\ref{sec:data-relative}.
However, rather than using the data as is, we use data-augmentation in combination with an altered training procedure to explicitly aim at transposition invariance of the mapping codes.

\subsubsection{Enforcing Transposition-Invariance}\label{sec:enforc-transp-invar}
As described in Section~\ref{sec:model_interval} the classical GAE training procedure derives a mapping code from an input/target pair, and subsequently penalizes the reconstruction error of the target given the input and the derived mapping code.
Although this procedure naturally tends to lead to similar mapping codes for input target pairs that have the same interval relationships, the training does not explicitly enforce such similarities, and consequently, the mappings may not be maximally transposition invariant.

Under ideal transposition invariance, by definition the mappings would be identical across different pitch transpositions of an input/target pair. Suppose that a pair $(\mathbf{x}_{t-n}^{t}, \mathbf{x}_{t+1})$ leads to a mapping $\mathbf{m}$ (by Equation~\ref{eq:gamap}).
Transposition invariance implies that reconstructing a target $\mathbf{x}^{\prime}_{t+1}$ from the pair $({\mathbf{x}^{\prime}}_{t-n}^{t}, \mathbf{m})$ should be as successful as reconstructing $\mathbf{x}_{t+1}$ from the pair $(\mathbf{x}_{t-n}^{t}, \mathbf{m})$ when $({\mathbf{x}^{\prime}}_{t-n}^{t}, \mathbf{x}^{\prime}_{t+1})$ can be obtained from $(\mathbf{x}_{t-n}^{t}, \mathbf{x}_{t+1})$ by a single pitch transposition.

Our altered training procedure explicitly aims to achieve this characteristic of the mapping codes by penalizing the reconstruction error using mappings obtained from transposed input/target pairs.
More formally, we define a transposition function $\textit{shift}(\mathbf{x}, \delta)$, shifting the values of a vector $\mathbf{x}$ of length $M$ by $\delta$ steps (MIDI note numbers and CQT frequency bins for symbolic and audio data, respectively):
\begin{equation}\label{eq:shift}
\textit{shift}(\mathbf{x}, \delta) = (x_{(0+\delta)\inmod{M}}, \dots, x_{(M-1+\delta)\inmod{M}})^\top,
\end{equation}
and $\textit{shift}(\mathbf{x}_{t-n}^{t}, \delta)$ denotes the transposition of each single time step vector \emph{before} concatenation and linearization.


The training procedure is then as follows. First, the mapping code $\mathbf{m}_{t}$ of an input/target pair is inferred as shown in Equation \ref{eq:gamap}. Then, $\mathbf{m}_{t}$ is used to reconstruct a \emph{transposed} version of the target, from an equally \emph{transposed} input (modifying Equation \ref{recony}) as
\begin{equation}\label{reconyshift}
\mathbf{\tilde{x}}'_{t+1} = \sigma_g(\mathbf{V}^\top (\mathbf{W}_0^\top\mathbf{W}_1^\top \mathbf{m}_t \cdot \mathbf{U}\textit{shift}(\mathbf{x}_{t-n}^{t},\delta))),
\end{equation}
with $\delta \in [-30,30]$ for the symbolic, and  $\delta \in [-60,60]$ for the audio data.
Finally, we penalize the error between the reconstruction of the transposed target and the actual transposed target (i.e., employing Equations \ref{eq:cost-mse} and \ref{eq:cost-ce}) as
\begin{equation}
\mathcal{L}(\textit{shift}(\mathbf{x}_{t+1},\delta),\mathbf{\tilde{x}}'_{t+1}).
\end{equation}

\begin{figure}
\includegraphics[width=1.\linewidth]{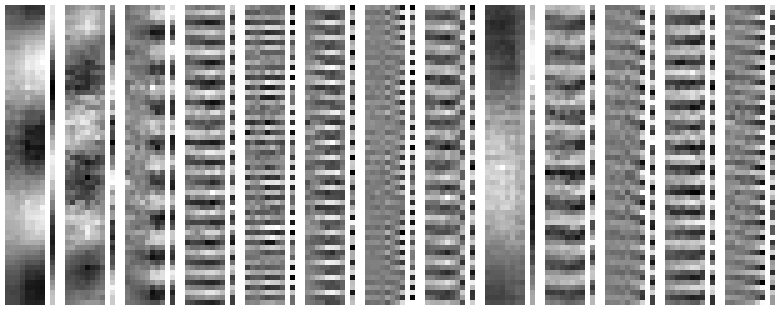} \\
\vspace{-8mm} \\
\caption{Some filter pairs $\in$ \{$\mathbf{U}, \mathbf{V}$\} of a GAE trained on polyphonic Mozart piano pieces.
}
\label{fig:filters}
\end{figure}

The transposition distance $\delta$ is randomly chosen for each training batch.
This method amounts to both, a form of guided training and data augmentation.
Some weights (i.e., filters) in $\mathbf{U}$ and $\mathbf{V}$ resulting from that training are depicted in Figure \ref{fig:filters}.

\subsubsection{Architecture and Training Details}
The architecture and training details of the GAE are as follows: A temporal context length of $n = 8$ is used (the choice of $n>1$ leads to higher robustness of the mapping codes to diatonic transposition). The factor layer has $1024$ units for the symbolic data, and $512$ units for the spectrogram data.
Furthermore, for all datasets, there are $128$ neurons in the first mapping layer and $64$ neurons in the second mapping layer (resulting in $\mathbf{m}_t \in \mathbb{R}^{64}$). 

L2 weight regularization for weights $\mathbf{U}$ and $\mathbf{V}$ is applied, as well as sparsity regularization \cite{Lee:2007uz} on the topmost mapping layer. 
The deviation of the norms of the columns of both weight matrices $\mathbf{U}$ and $\mathbf{V}$ from their average norm is penalized. 
Furthermore, we restrict these norms to a maximum value.
We apply $50\%$ dropout on the input and no dropout on the target, as proposed in \cite{memisevic2011gradient}. 
The learning rate (1e-3) is gradually decremented to zero throughout the training.


\subsection{Experiments}\label{sec:experimentsInverval}
In this Section we describe several experimental analyses to validate the proposed approach.
They are intended to test the degree of transposition-invariance of the learned mappings, and to assess their musical relevance (Sections~\ref{sec:class-clust-analys} and~\ref{sec:sensitivity-analysis}). Furthermore, we put the learned representations to practice in a repeated section discovery task for symbolic music and audio (Section~\ref{sec:disc-repe-them}).

\subsubsection{Classification and Cluster Analysis}\label{sec:class-clust-analys}
Our hypothesis is that the model learns relative pitch representations (i.e. intervals) from polyphonic absolute pitch sequences.
In order to test this hypothesis, we conduct two experiments using symbolic data.

In the first experiment a ten-fold k-nn classification of intervals is performed (where k = 10), where the task is to identify all pitch intervals between notes in the input and the target of an input/target pair.
If the learned mappings actually represent intervals, the classifier will perform substantially better on the mappings than on the input space.
As intervals in music are transposition-invariant, the interval labels do not change when performing transposition in the input space.
Thus, we perform the classification on the mappings of the original data and of randomly transposed data, to test if the mappings are indeed transposition-invariant.

We label the symbolic train data input/target pairs according to all intervals which occur between any two notes from the pair, independent of the temporal distance of the notes exhibiting the intervals. Thus, each pair can have multiple labels.
For each pair in the test set, the k-nn classifier predicts the set of interval labels that are present in the k neighbors of that pair.
Using the Euclidean distance, the classification is performed in the input space (i.e., on concatenated input / target pairs), and in the mapping space.
From the resulting predictions, we determine the precision, recall, and F-score over the test set (cf. Table~\ref{tab:knn}).
For example, when a pair contains 6 intervals and the classifier estimate yield 4 true-positive and 4 false-positive interval occurrences, that pair is assigned a precision of 0.5 and a recall of 0.67.

\begin{figure}
\begin{center}
\includegraphics[width=1.\linewidth]{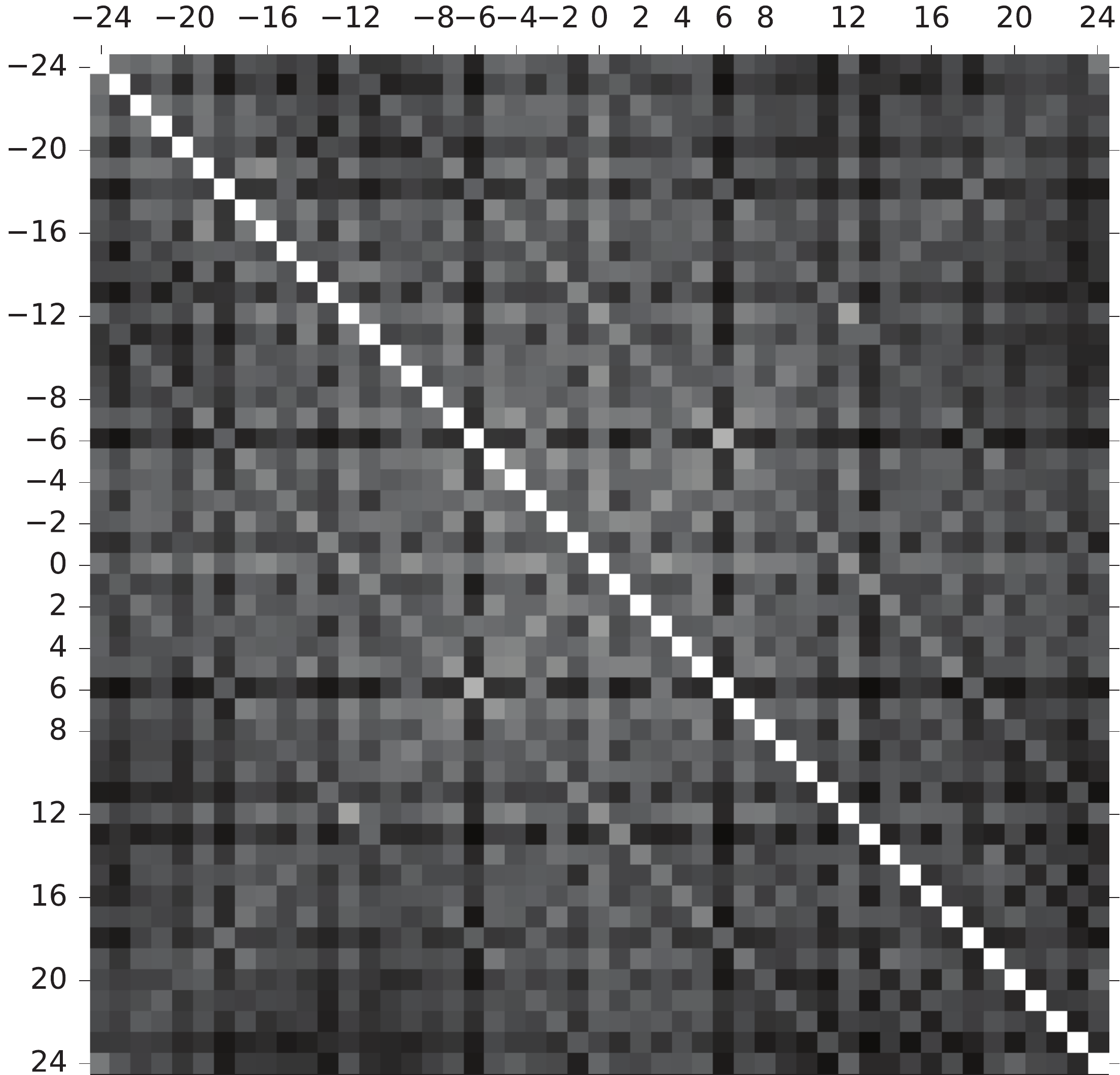}
\caption{Distance matrix of cluster centers of intervals represented in mapping space. Darker cells indicate higher distances between respective clusters, brighter cells indicate closeness.}
\label{fig:close_matrix}
\end{center}
\end{figure}

In the second part of the experiment, the cluster centers of all intervals in the mapping space are determined.
Again, each pair projected into the mapping space accounts for all intervals it exhibits and can, therefore, participate in more than one cluster.
The mutual Euclidean distances between all cluster centers are displayed as a matrix (cf. Figure~\ref{fig:close_matrix}).
An interpretation of the results follows in Section~\ref{sec:res_inter}.

\begin{table}[t]
\centering
\vspace{-7mm}
\begin{tabular}{llll}
\toprule
Data     & Precision & Recall & F1 \\
\midrule
\rule{-6pt}{0ex}
\textbf{Original input} & & & \\
\rule{-2pt}{2ex}
Mapping space & \textbf{91.27}     & 70.25  & 76.66   \\
Input space & 65.58    & 46.05  & 50.59   \\
\midrule
\rule{-6pt}{0ex}
\textbf{Transposed input} & & & \\
\rule{-2pt}{2ex}
Mapping space & 90.78     & \textbf{71.44}  & \textbf{77.31}   \\
Input space & 51.81    & 32.99  & 37.43   \\
\midrule
All & 26.40     & 100.0   & 40.05  \\
None & 0.0 & 0.0 & 0.0 \\
\bottomrule
\end{tabular}
\caption{Results of the k-nn classification in the mapping space and in the input space for the original symbolic data and data randomly transposed by $[-24,24]$ semitones. ``All'' is a lower bound (always predict all intervals), ``None'' returns the empty set.}
\label{tab:knn}
\end{table}

\subsubsection{Discovery of Repeated Themes and Sections}\label{sec:disc-repe-them}
\urldef\mirexurl\url{http://www.music-ir.org/mirex/wiki/2017:Discovery_of_Repeated_Themes_&_Sections}
The MIREX Task for Discovery of Repeated Themes and Sections for Symbolic Music and Audio\footnote{\mirexurl{}} 
tests algorithms for their ability to identify repeated patterns in music.
The commonly used JKUPDD dataset \cite{collins2017discovery} contains 26 motifs, themes, and repeated sections annotated in 5 pieces by J. S. Bach, L.~v.~Beethoven, F. Chopin, O. Gibbons, and W. A. Mozart.
We use the MIDI and the audio versions of the dataset and pre-process them as described in Section~\ref{sec:data-relative}.

\begin{figure}[t]
\begin{center}
\includegraphics[width=1.\linewidth]{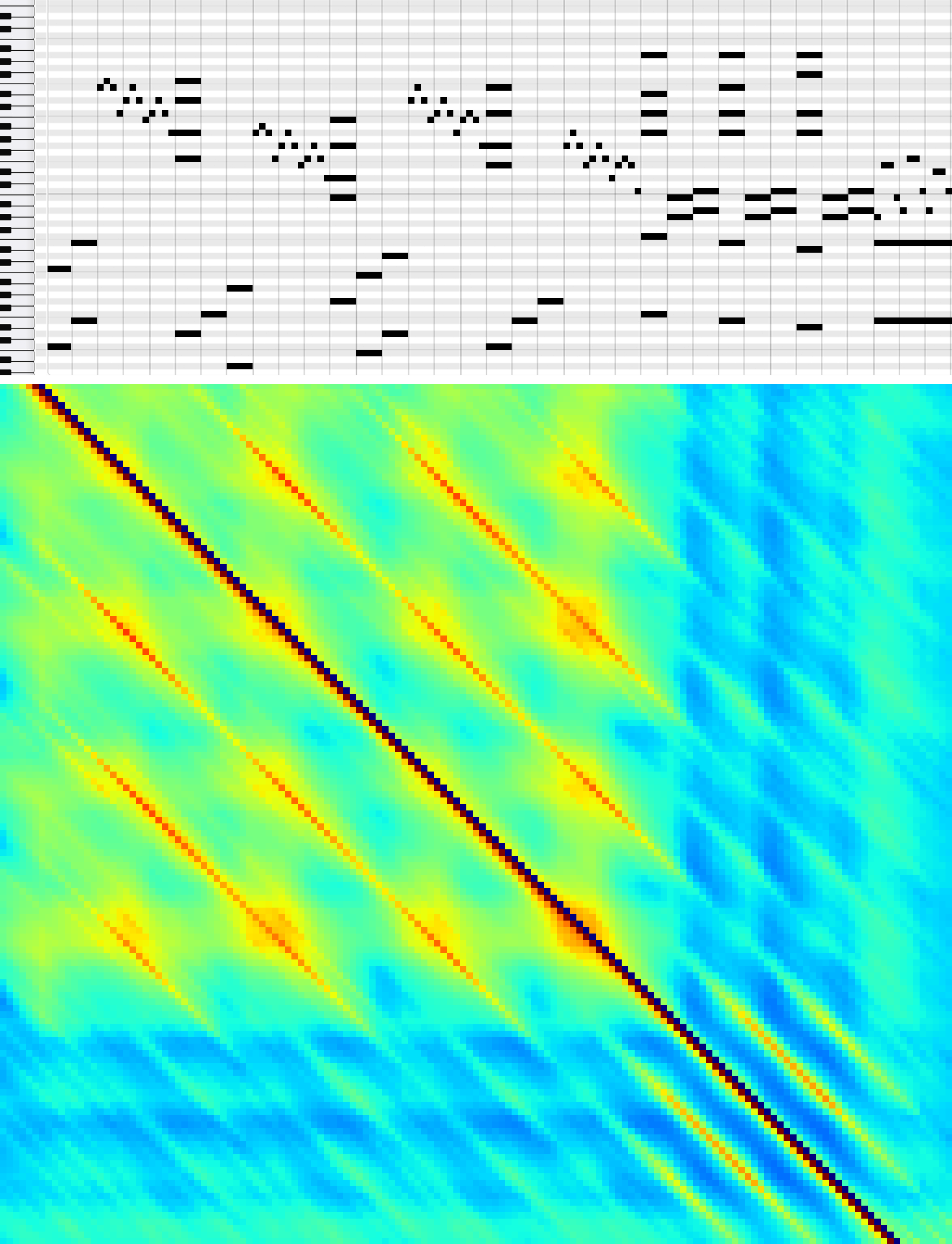}
\caption{Symbolic music and corresponding self-similarity matrix calculated from transposition-invariant mapping codes. Warmer colors indicate similarity, colder colors indicate dissimilarity.}
\label{fig:self-sim-invar}
\end{center}
\end{figure}

We calculate the reciprocal of the Euclidean distances between all representations $\mathbf{m}_t$ of a song, resulting in a transposition-invariant similarity matrix $X$.
Then, the values of the main diagonal are set to the minimum value of the matrix.
Subsequently, the matrix is normalized and convolved with an identity matrix of size $15 \times 15$ to emphasize and smooth diagonals (Figure~\ref{fig:self-sim-invar} shows a resulting matrix). 
The method used to determine repeated parts based on diagonals of high values in the self-similarity matrix is adopted from \cite{nieto2014identifying}, with a different method to identify diagonals, as described below.

The function 
\begin{equation}\label{eq:diag1}
s(i,j,N) = \sum_{k=N-m}^{N}{\frac{X(i+k,j+k) w_k}{m} }
\end{equation}
returns the score for a diagonal starting at $X(i,j)$ with length $N$, and diagonals with high score are considered to be repeated sections. For each $i,j$, we iteratively evaluate the score with $N$ increasing from $1$ in integer steps, until the score undercuts a threshold $\gamma$. Only the last $m$ values, $m = \min(10,N)$, of the diagonal are taken into account, because those values indicate when to stop tracing.
The factor
\begin{equation}\label{eq:diag2}
w_k = \frac{1+k+m-N}{m}
\end{equation}
linearly weights the last $m$ values of the diagonal so that later values have more impact on the overall score.

Three empirically determined parameters influence the functioning of the method: (1) from the diagonals found, we only keep those spanning more than $\mathit{2}$ \emph{whole notes}, (2) when aligning the repetitions of a section with the first occurrence of the section, their start and end may differ by not more than \emph{a half note} to be still considered repetitions of each other, (3) the thresholds $\gamma$ determining if a diagonal should be considered a repetition in the symbolic and the audio data are set to $0.9$ and $0.81$, respectively. The results are shown in Table~\ref{tab:secdiscovery} and are discussed in Section~\ref{sec:res_inter}.

\begin{landscape}
\begin{table*}
\hspace{-1cm}
\centering
\footnotesize
\begin{tabular}{llllllllllllll}
\toprule
Algorithm     & $F_{\text{est}}$  & $P_{\text{est}}$  & $R_{\text{est}}$  & $F_{\text{o(.5)}}$   & $P_{\text{o(.5)}}$   & $R_{\text{o(.5)}}$
 & $F_{\text{o(.75)}}$   & $P_{\text{o(.75)}}$   & $R_{\text{o(.75)}}$  & $\mathbf{F_3}$    & $P_3$    & $R_3$    & Time (s) \\
\midrule
\rule{-6pt}{0ex}
\textbf{Symbolic} & & & & & & & & & & & & & \\
\rule{-2pt}{2ex}
RelativePitch & 59.07 & \textbf{77.60} & 58.30 & 68.92 & \textbf{80.24} & 67.46 & \textbf{77.51} & \textbf{91.38} & 73.29 & 50.44 & 60.36 & 53.23 & \textbf{127}  \\
VMO symbolic*  & \textbf{60.79} & 74.57 & 56.94 & 71.92 & 79.54 & 68.78 & 75.98 & 75.98 & 75.99 & \textbf{56.68} & \textbf{68.98} & 53.56 & 4333 \\
SIARCT-CFP*  & 33.70  & 21.50  & \textbf{78.00}  & \textbf{76.50}  & 78.30  & \textbf{74.70}  & -     & -     & -     & -     & -     & -     & -    \\
COSIATEC*  & 50.20  & 43.60  & 63.80  & 63.20  & 57.00  & 71.60  & 68.40  & 65.40  & \textbf{76.40}  & 44.20  & 40.40  & \textbf{54.40}  & 7297 \\
\midrule
\rule{-6pt}{0ex}
\textbf{Audio} & & & & & & & & & & & & & \\
\rule{-2pt}{2ex}
RelativePitch & \textbf{57.67} & \textbf{67.46} & 59.52 & 58.85 & 61.89 & 56.54 & 68.44 & 72.62 & 64.86 & \textbf{51.61} & 59.60 & \textbf{55.13}  & 194 \\ 
VMO deadpan* & 56.15 & 66.80  & 57.83 & \textbf{67.78} & \textbf{72.93} & \textbf{64.30}  & \textbf{70.58} & \textbf{72.81} & \textbf{68.66} & 50.60  & \textbf{61.36} & 52.25 & \textbf{96}   \\
SIARCT-CFP*  & 23.94 & 14.90  & \textbf{60.90}  & 56.87 & 62.90 & 51.90  & -     & -     & -     & -     & -     & -     & -    \\
Nieto*  & 49.80  & 54.96 & 51.73 & 38.73 & 34.98 & 45.17 & 31.79 & 37.58 & 27.61 & 32.01 & 35.12 & 35.28 & 454 \\
\bottomrule
\end{tabular}
\hspace{-1cm}
\caption{Different precision, recall and f-scores (details on the measures are given in \protect\cite{collins2017discovery}) of different methods in the Discovery of Repeated Themes and Sections MIREX task, for symbolic music and audio. The $\mathbf{F_3}$ score constitutes a summarization of all measures. \\ \\
*Results of \\
VMO symbolic and VMP deadpan from \protect\cite{wang2015music}, \\
SIARCT-CFP from \protect\cite{collins2013siarct}, \\
COSIATEC from \protect\cite{meredith2013cosiatec} and \\
VMO deadpan from \protect\cite{nieto2014identifying}.
}
\label{tab:secdiscovery}
\end{table*}
\end{landscape}

\subsubsection{Sensitivity Analysis}\label{sec:sensitivity-analysis}
\begin{figure}
\begin{center}
\includegraphics[width=.7\linewidth]{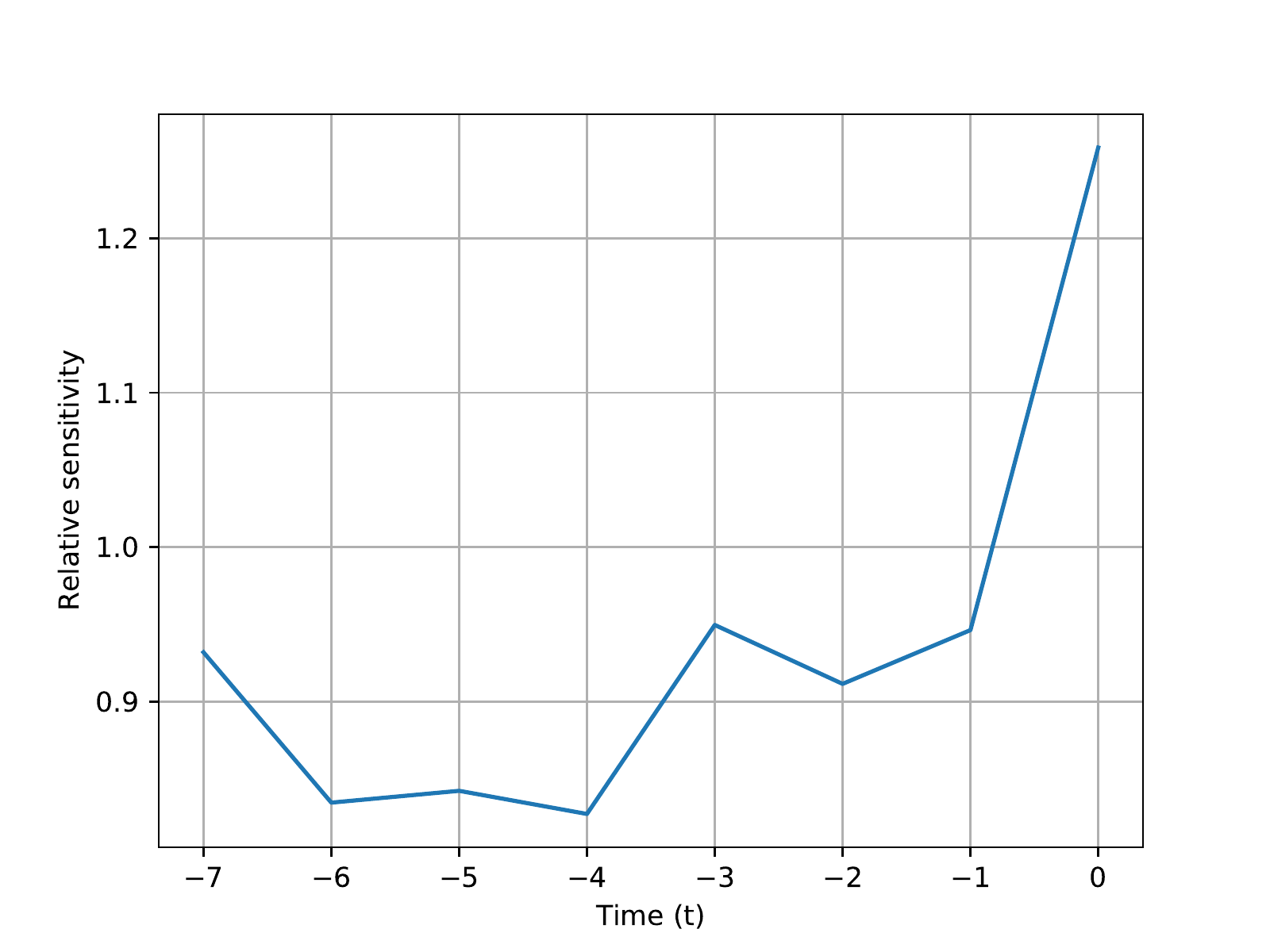}
\caption{Absolute sensitivity of the model when looking backwards on the temporal context, averaged over the whole dataset.}
\label{fig:sensitivity_context}
\end{center}
\end{figure}

The sensitivity of the model to specific context information provides important insights into the functioning of the model.
A common way of determining the sensitivity of a network is by calculating the absolute value of the gradients of the network's predictions with respect to the input, holding the network parameters fixed \cite{simonyan2013deep}.
Figure~\ref{fig:sensitivity_context} shows the sensitivity of the model with respect to the temporal context.
The model is particularly sensitive to note occurrences at $t \in \{0,-3,-7\}$.
This shows that the most informative notes for a prediction are direct predecessors ($t=0$), and notes which occur a quarter ($t=-3$) and a half note ($t=-7$, i.e., eight sixteenth notes) before the prediction.

\subsection{Results and Discussion}\label{sec:res_inter}
The results of the k-nn classification on the raw data and representations learned by the model are shown in Table~\ref{tab:knn}.
Classification in the mapping space appreciably outperforms classification in the input space and obtains similar values for mappings of the original data and the randomly transposed data.
In contrast, when performing classification in the input space, the results deteriorate for the randomly transposed input and do not exceed the theoretical lower bound (i.e., always predict all intervals).
As the register and keys of the original data are limited, correlations between absolute and relative pitch exist.
When transposing the input, the classifier cannot make use of these absolute cues for relative pitch anymore and performs weakly in the input space.

Figure~\ref{fig:close_matrix} indicates which intervals are close to each other in the mapping space.
An obvious regularity are the slightly brighter k-diagonals (i.e., parallels to the main diagonal) with $k \in \{-24,-12,12,24\}$,
showing that two pitch intervals lead to similar mapping codes when they result in the same pitch class, such as the intervals +8 and -4 semitones, or -7 and -19 semitones.
This is an indication that the model learns the phenomenon of octave equivalence, even if the input to the model represents only absolute pitch.
Another distinct feature is the stripe which is orthogonal to the main diagonal (i.e., where $y=-x$).
This indicates that the model develops some notion of relative distances, by positioning intervals of the same distance (but different signs) close to each other.

Note also that the mappings of certain intervals, notably $6$ and $-6$, are distant to those of most other intervals (dark horizontal and vertical lines).
This likely reflects the fact that tritone intervals are rare in diatonic music, and is further evidence of the musical significance of the learned mappings.

Table~\ref{tab:secdiscovery} shows results of the repeated themes and section discovery task, where the $F_3$ score is a good indicator for the overall performance of the models (see \cite{collins2017discovery} for a thorough explanation on the respective measures).
For the audio data, the current state-of-the-art $F_3$ score was raised from $50.60$ to $51.61$ by our proposed method.
The method performs slightly worse on the symbolic data, which is counterintuitive at first sight, given that results of other models suggest that this task is easier.
We hypothesize that for the discovery of repeated sections, approximate matching leads to better results than exact comparison, simply because musical variation goes beyond chromatic transposition (towards which our model is invariant).
For approximate matching, a spectrogram representation is better suited than symbolic vectors, as notes are blurred over more than one frequency bin, and harmonics may provide additional cues for similarity estimation.
The proposed approach is computationally efficient because the diagonal detector (cf. Equations~\ref{eq:diag1} and~\ref{eq:diag2}) is rather simple and the transposition-invariance of the representations does not require explicit comparison of mutually transposed musical textures.

\subsection{Conclusion and Future work}\label{sec:concl-future-work}
In this section, we have presented a computational approach to deriving (pitch) transposition-invariant vector space representations of music both in the symbolic and the audio domain.
The representations encode pitch intervals that occur in the music in a musically meaningful way,
with tritone intervals---a rare interval in diatonic music---leading to more distinct representations, and octaves leading to more similar representations.
Furthermore, the temporal sensitivity of the model reveals a beat pattern that shows increased sensitivity to pitch intervals occurring at beat multiples of each other.

The transposition-invariance of the representations makes it possible to detect transposed repetitions of musical sections in the symbolic and in the spectral domain of audio.
We have demonstrated that this is beneficial in tasks such as the MIREX task \emph{Discovery of Repeated Themes and Sections}.
A simple diagonal finding approach on a transposition-invariant self-similarity matrix produced by our model is sufficient to outperform the state-of-the-art in the audio version of the task.


We believe it is worthwhile to further explore the utility of transposition-invariant music representations for other applications, including speech recognition, music summarization, music classification, transposition-invariant music alignment (including a cappella voices with pitch drift), query by humming, fast melody-based retrieval in large audio collections, and music generation.
First results show that the proposed representations are useful for audio-to-score alignment \cite{arzt2018alignment} and for music prediction tasks (see Section \ref{sec:rgae}).

In the next section, we show that interval representations which take into account reference pitches at a constant time lag in the past (e.g., one measure) are a useful feature to encode (transposed) repetitions, which gives rise to modeling repetition structure in music.
Furthermore, by combining the proposed GAE architecture with an RNN, which operates on the interval representations provided by the GAE, generalization in sequence learning tasks can be improved by reducing sparsity in the input data.



%
%
%
%
%
\section{Learning Sequences of Intervals and Repetition Structure}\label{sec:rgae}

%
%

In this section, we further modify the GAE architecture proposed in Section \ref{sec:interval} by combining it with a Recurrent Neural Network (RNN), to obtain a ``relative pitch'' sequence model---the Recurrent Gated Autoencoder (RGAE).
The objective of sequence models for music prediction is to predict (the probability of) musical events at the next time step, given some prior musical context.
In the (most common) case of predicting note events, this task involves finding relationships between past and future occurrences of absolute pitch values.
However, many music theoretical constructs that might help to find such relationships are defined in relative terms, such as diatonic scale steps and cadences.
The discrepancy between the relative nature of musical constructs and the absolute pitch representation is problematic for modeling tasks, because it leads to high sparsity in the input data, bigger models, and altogether reduced generalization in music modeling.

To remedy these problems, musical input sequences can be transposed to a common key before training, augmented by random transpositions during training, or, in case of symbolic monophonic music, transformed into interval representations before training.
In this Section, we propose a sequence model which \emph{learns} both interval representations from absolute pitch sequences and temporal dependencies between such intervals.
By not only learning the intervals between two successive notes but all intervals within a window of $n$ pitches, the model is more robust to diatonic transposition and can also learn repetition structure.
To that end, a recurrent neural network (RNN) is employed on top of a gated autoencoder (GAE), which we refer to as Recurrent Gated Autoencoder (RGAE).
The GAE portion learns the intervals between its input and target pitches and represents them in its latent space.
The RNN portion operates on these interval representations, to learn their temporal dependencies.
The implicit transformation to intervals allows this architecture to operate directly on absolute musical textures, without the need for data pre-processing.
Besides, relative pitch modeling reduces the sparsity in the data, and the representations learned by the GAE are transposition-invariant.
Therefore, the RGAE requires less temporal connections than a common RNN while achieving higher prediction accuracy.

Also, operating on the intervals of input sequences brings added value to sequence modeling.
By relating its prediction with events using specific time lags, the RGAE can learn copy-and-transpose operations of short melodic fragments.
More precisely, a transposed copy of a fragment is generated by repeatedly applying a constant interval (i.e., the transposition distance) to events occurring a constant time lag (i.e., the length of the fragment) in the past.
Moreover, the RGAE can learn sequences of such copy-and-transpose operations (i.e., ``structure schemes'').
As a result, an arbitrary melodic fragment can be used as a building block from which transposed copies are generated over and over to create a new sequence obeying a learned scheme (see Section \ref{sec:copy-shift}).

Learning copy-and-transpose operations can be useful for music modeling, where repeated sections often occur as a transposed version of the initial section.
With common sequence models, like RNNs, it is challenging to learn such self-similarity relationships.
Common RNNs are specialized in learning the statistics of musical textures and are ``blind'' towards similarity and (transposed) repetition (i.e., there is no content-independent ``repetition neuron'').
As a result, when sampling music using such models, repeated fragments occur either due to chance or as a phenomenon of an entanglement with a learned texture.
In contrast, when RGAEs learn copy-and-transpose operations, they separate self-similarity structure from the actual content with which the structure is instantiated, leading to improvements in music prediction and music generation tasks.

In conclusion, we show in this Section that the RGAE is competitive in a music prediction task.
Furthermore, by combining the predictions of absolute pitch models with predictions of the (relative pitch) RGAE, we can further improve the prediction accuracy. This shows that the RGAE is complementary to common absolute pitch models.
Lastly, we show that the RGAE is particularly suited for learning sequences of copy-and-transpose operations.
It can learn to recognize and continue pre-defined ``structure schemes'', abstracted from the actual texture, with which the scheme is realized.

In Section \ref{sec:models}, the GAE and the proposed extensions yielding the RGAE are described, as well as the baseline RNN used for comparison and combined prediction.
General training details concerning the GAE are given in Section \ref{sec:training_rgae}.
The experiments conducted, including data pre-processing, training details and discussions, are introduced in Section \ref{sec:experimentsGAEpredict}.
A final discussion is given in Section \ref{sec:discussion} and Section \ref{sec:conclusion} concludes this section and provides further directions.

\subsection{Models}\label{sec:models}
\subsubsection{Gated Autoencoder}
For the GAE portion of our architecture, we largely adopt the GAE model described in Section \ref{sec:model_interval}, but we only use one mapping layer for simplicity (see Figure \ref{fig:rgae}, GAE portion), and different non-linearities.
Accordingly, the GAE learns to encode intervals in its latent mapping space
\begin{equation}\label{eq:gamap_rgae}
\mathbf{m}_{t+1} = \sigma_q(\mathbf{W}_m (\mathbf{Q}\mathbf{x}_{t-n}^{t} \cdot \mathbf{V}\mathbf{x}_{t+1})),
\end{equation}
where $\mathbf{Q}, \mathbf{V}$ and $\mathbf{W}_m$ are weight matrices, and $\sigma_q$ is the softplus non-linearity.
The operator~$\cdot$ (indicated as a triangle in Figure \ref{fig:rgae}) depicts the Hadamard product of the filter responses $\mathbf{Q}\mathbf{x}_{t-n}^{t}$ and $\mathbf{V}\mathbf{x}_{t+1}$, denoted as \emph{factors}.

The reconstruction of the target $\mathbf{x}_{t+1}$ is defined as a function of the input $\mathbf{x}_{t-n}^{t}$ and the mapping $\mathbf{m}_{t+1}$ as
\begin{equation}\label{recony_rgae}
\mathbf{\tilde{x}}_{t+1} = \sigma_g(\mathbf{V}^\top (\mathbf{W}_m^\top \mathbf{m}_{t+1} \cdot \mathbf{Q}\mathbf{x}_{t-n}^{t})),
\end{equation}
where $\sigma_g$ is the sigmoid non-linearity.
The GAE portion of the RGAE is pre-trained by minimizing the binary cross-entropy loss of the reconstruction (see Equation \ref{eq:cost-ce}).

\subsubsection{Recurrent Gated Autoencoder}\label{sec:model}
\begin{figure}
\begin{center}
\includegraphics[width=.7\linewidth]{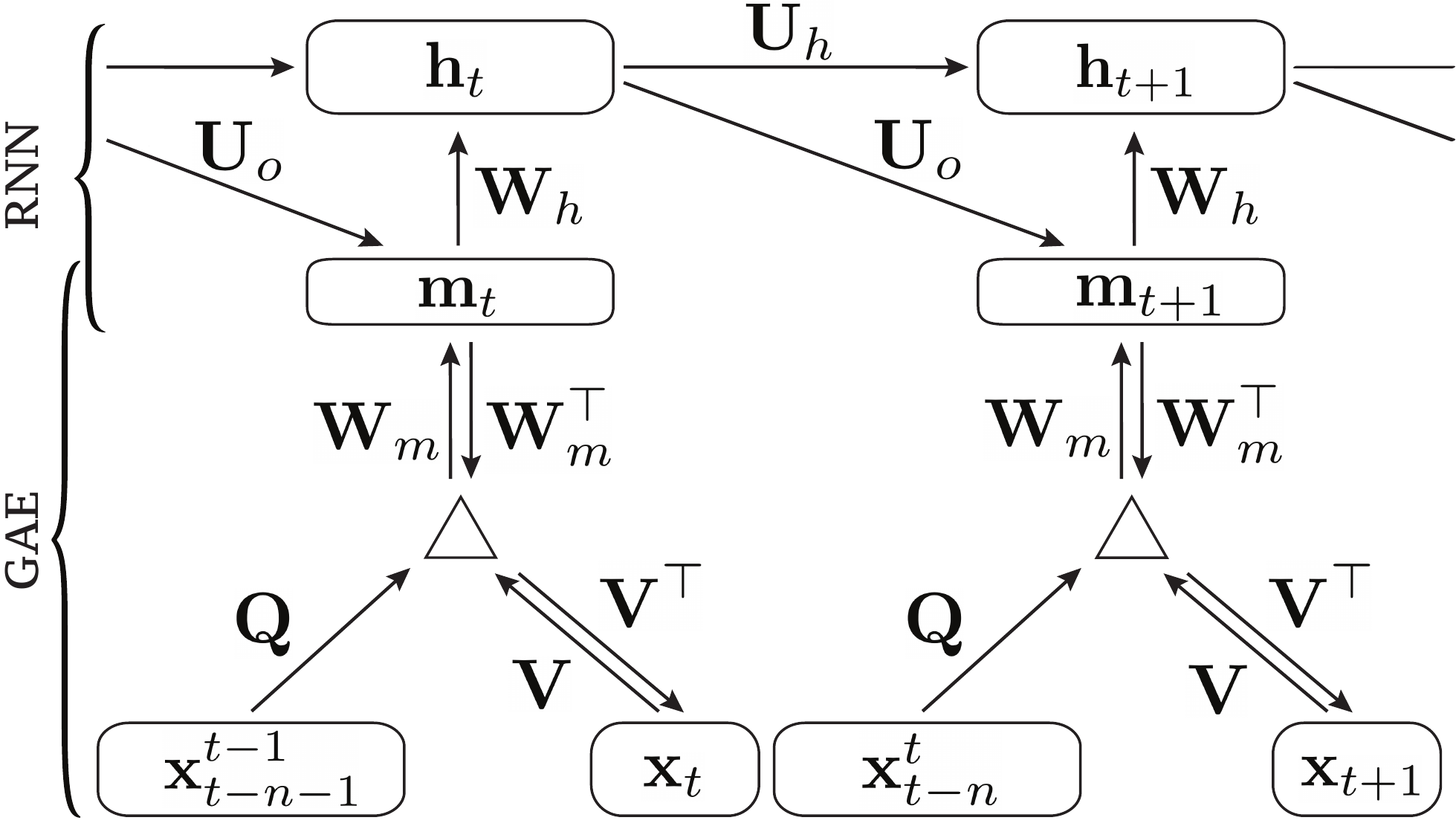}
\caption{Schematic illustration of the proposed Recurrent Gated Autoencoder architecture. Arrows represent weight matrices, rounded rectangles represent vectors. The triangles depict the Hadamard product. The specifics of the Gated Recurrent Unit are omitted for better clarity.}
\label{fig:rgae}
\end{center}
\end{figure}

The proposed model is a combination of a gated autoencoder (GAE) and a recurrent neural network (RNN) as depicted in Figure \ref{fig:rgae}.
The GAE learns relative pitch (i.e., interval) representations of the musical surface, and the RNN learns their temporal dependencies.

We use gated recurrent units (GRUs) \cite{cho2014properties} for the RNN portion of the RGAE. This type of units has been shown to be often as efficient as long short-term memory units (LSTMs, \cite{hochreiter1997long}) while being conceptually simpler \cite{chung2014empirical}.
It is intuitively clear that any RNN variant can be potentially attached on a GAE.
The input to the RNN at time t is the GAE's mapping $\mathbf{m}_{t}$, resulting in the following specification:
\begin{equation}\label{eq:z}
\mathbf{z}_t = \sigma_g(\mathbf{W}_{z} \mathbf{m}_{t} + \mathbf{U}_{z} \mathbf{h}_{t-1} + \mathbf{b}_z),
\end{equation}
\begin{equation}\label{eq:r}
\mathbf{r}_t = \sigma_g(\mathbf{W}_{r} \mathbf{m}_{t} + \mathbf{U}_{r} \mathbf{h}_{t-1} + \mathbf{b}_r),
\end{equation}
\begin{equation}\label{eq:h}
\mathbf{h}_t =  \mathbf{z}_t \cdot \mathbf{h}_{t-1} + (1-\mathbf{z}_t) \cdot \sigma_h(\mathbf{W}_{h} \mathbf{m}_{t} + \mathbf{U}_{h} (\mathbf{r}_t \cdot \mathbf{h}_{t-1}) + \mathbf{b}_h),
\end{equation}
where $\mathbf{h}_t$ is the hidden state at time $t$, $\mathbf{z}_t$ is the update gate vector, $\mathbf{r}_t$ is the reset gate vector, and $\mathbf{W}$, $\mathbf{U}$ and $\mathbf{b}$ are parameter matrices and vectors.
The RNN predicts the next mapping of the GAE as
\begin{equation}\label{eq:pred}
\mathbf{\widetilde{m}}_{t+1} = \sigma_q(\mathbf{U}_o \mathbf{h}_t),
\end{equation}
which is used to reconstruct the target configuration at $t+1$ as
\begin{equation}\label{recony_rnn}
\mathbf{\tilde{x}}_{t+1} = \sigma_s(\mathbf{V}^\top (\mathbf{W}_m^\top \mathbf{\widetilde{m}}_{t+1} \cdot \mathbf{Q}\mathbf{x}_{t-n}^{t}))\,.
\end{equation}

Here, we use the softmax non-linearity $\sigma_s$, as the data the RGAE is trained on is monophonic.
The full architecture is trained with Backpropagation through time (BPTT) to minimize the \emph{categorical} cross-entropy loss for the reconstructed target as
\begin{equation}\label{eq:cost_softmax}
\mathcal{L}(\mathbf{x},\mathbf{\tilde{x}}) = -\frac1N\sum_{n=1}^N\ x_n  \log_2 \tilde x_n\,.
\end{equation}

When the RGAE is applied to polyphonic music, in Equation~\ref{recony_rgae} the sigmoid non-linearity, together with the binary cross-entropy loss (see Equation \ref{eq:cost-ce}) has to be used.

\subsubsection{Baseline RNN}
As a baseline, we employ an RNN with GRUs to directly operate on the data.
Accordingly, Equations \ref{eq:z}, \ref{eq:r}, and \ref{eq:h} are adapted to consume $\mathbf{x}_t$ instead of $\mathbf{m}_{t}$ as input.
Consequently, the prediction of the baseline RNN amounts to
\begin{equation}\label{eq:pred_rnn}
\mathbf{\tilde{x}}_{t+1} = \sigma_s(\mathbf{U}_o \mathbf{h}_t),
\end{equation}
where the softmax non-linearity is applied, making the categorical cross-entropy loss (see Equation \ref{eq:cost_softmax}) applicable in training.


\subsection{Gated Autoencoder Pre-Training}\label{sec:training_rgae}


Due to the relatively high number of parameters in its GAE portion, the RGAE is prone to overfitting.
To circumvent this, and to establish robust interval representations, we pre-train the GAE first, using the cross-entropy of the reconstruction as the cost function (cf. Equation \ref{eq:cost-ce}).
In the second training iteration, we train the RNN portion of the GAE to minimize the cross-entropy error of the architecture's prediction (cf. Equation \ref{eq:cost_softmax}).
The datasets may differ between the training iterations as long as the included relations are identical (e.g., ``intervals of western tonal music'').
Consequently, the GAE parameters trained on one dataset can be used for prediction tasks on several datasets.
Fine-tuning the whole architecture in the last few epochs of predictive training can make up for possible bias.

In the following, we describe how the GAE is pre-trained in our experiments.
Details varying between the experiments are given later in the experiments section (see Section \ref{sec:experimentsGAEpredict}).

\subsubsubsection{Pre-training details}\label{sec;pretrain-details}
The GAE portion is pre-trained using the method to support the learning of intervals, as introduced in Section \ref{sec:enforc-transp-invar}.
The training details are as follows.
We use $512$ units in the factor layer and $64$ units in the mapping layer of the GAE.
On the latter, sparsity regularization \cite{Lee:2007uz} is applied.
The deviation of the norms of the columns of both weight matrices $\mathbf{U}$ and $\mathbf{V}$ from their average norm is penalized.
Furthermore, we restrict these norms to a maximum value.
The learning rate is reduced from 1e-3 to 0 throughout training, and RMSProp \cite{hinton2012neural} is used.

\subsection{Experiments}\label{sec:experimentsGAEpredict}
\subsubsection{Experiment 1: Folk Song Prediction}\label{sec:predict_folksongs}
We test the RGAE and RNN in a sequence learning task using the data described in Section \ref{sec:data-efsc}.
In order to make the results comparable, we use the same experimental setup as in \cite{pearce2004improved, cherla2016neural}.

\subsubsubsection{Data}\label{sec:data-efsc}
The EFSC subset (comprising a total of 54,308 note events) of the Essen Folk Song Collection (EFSC) \cite{TheEssenFolksongC:1995um} constitutes the data for the actual training and evaluation.
It consists of 119 Yugoslavian folk songs, 91 Alsatian folk songs, 93 Swiss folk songs, 104 Austrian folk songs, the German subset \emph{kinder} (213 songs), and 237 songs of the Chinese subset \emph{shanxi}.
The melodies are represented as a series of pitches ignoring note durations.

For pre-training the GAE portion of the RGAE, we again use the polyphonic Mozart piano music dataset described in Section \ref{sec:TrainingDataInterval} (\cite{widmer2003discovering}, comprising 13 piano sonatas with more than 106,000 notes) in piano-roll representation (i.e., using a regular time grid of 1/8th note resolution, and an active note can span several time steps).
We pre-train on that data because polyphonic music acts as a better regularizer for learning interval representations than monophonic music.

\subsubsubsection{Training and Architecture}\label{sec:train_rgae}
We use only $16$ hidden units in the RNN portion of the RGAE.
The lookback window of the GAE is $n = 8$ pitches, and we apply $50$\% dropout on the input in pre-training and when training the whole architecture.
We pre-train the GAE for 250 epochs on the Mozart piano pieces (cf. Section \ref{sec:data-efsc}).
Subsequently, the RNN portion is trained for 110 epochs on the interval representations (i.e., mappings provided by the GAE) of the EFSC datasets.
In the last $10$ epochs the whole architecture is fine-tuned.

The baseline RNN with $50$ hidden units is trained for 70 epochs on the EFSC data.
The learning rate scheme is adopted from that described in Section \ref{sec;pretrain-details} for all models.

\subsubsubsection{Combining Model Predictions}\label{sec:geom-mean}
We hypothesize that the RNN and the RGAE are complementary in how they process musical sequences.
For example, the RNN may have better stability in remembering absolute reference pitches, like the tonic of a piece, and is superior in modeling prior probabilities, to keep predictions in a plausible pitch range.
In contrast, the RGAE can make use of structural cues indicating repetitions and can generalize better due to relative pitch processing.
There are several possibilities to combine the predictions of statistical models.
Next to the ad-hoc approach of merely averaging their outputs, we can also use information about the certainty of the models and weight their outputs accordingly.
A measure for the certainty of a prediction is given by the Shannon entropy \cite{shannon2001mathematical}:
\begin{equation}
H(p) = -\sum_{a\in A}{p(a) \log_2 p(a)},
\end{equation}
where $p(a\in \mathcal{A}) = P(\mathcal{X} = a)$ is a probability mass function over a discrete alphabet $\mathcal{A}$.
The method which worked best in our experiments is calculating the entropy-weighted geometric mean of both predictions, as proposed in \cite{pearce2004methods}:
\begin{equation}
p(t) = \frac{1}{R} \prod_{m\in M}{p_m(t)^{w_m}},
\end{equation}
where $p_m(t)$ is the predicted distribution of model $m$ at time $t$, $w_m = H_{\text{relative}}(p_m)^{-b}$ is the weight of model $m$, non-linearly scaled using a bias $b$ (set to $0.5$ in our experiments), and $R$ is a normalization constant.
The relative entropy $H_{\text{relative}}(p_m)$ for model $m$ is given by
\begin{equation}
H_{\text{relative}}(p_m) = \frac{H(p_m)}{H_{\text{max}}(p_m)},
\end{equation}
where $H_{\text{max}}(p_m) > 0$ is the entropy of the probability mass uniformly distributed over the alphabet (indicating maximal uncertainty of the model).

\subsubsubsection{Evaluation}
Since the datasets are rather small, a fixed training/test set split would lead to a poor estimation of the performance of the models.
Therefore, and in accordance with \cite{pearce2004improved, cherla2016neural}, a 10-fold cross validation is performed for each dataset and the categorical cross-entropy loss (cf. Equation \ref{eq:cost_softmax}) is reported.

\begin{table*}[t]
\setlength{\tabcolsep}{5pt}
\centering
\footnotesize
\begin{tabular}{lll@{\hskip -0pt}llllll}
\toprule
    &  RNN & RTDRBM & RGAE & RNN + & RNN + & RTDRBM + \\
Data      & (GRU) & \cite{cherla2016neural} &  & RTDRBM & RGAE & RGAE \\
\midrule
Alsatian folk songs & $2.890$ & $2.897$ &  $2.872$  & $2.844$ & $2.788$ & $\mathbf{2.771}$\\
Yugoslavian folk songs  & $2.717$ & $2.655$ & $2.676$ & $2.617$ & $2.586$ & $\mathbf{2.530}$\\
Swiss folk songs  & $2.954$ & $2.932$ & $2.895$  & $2.851$& $2.831$ & $\mathbf{2.769}$\\
Austrian folk songs  & $3.185$ & $3.259$ &  $3.171$  & $3.163$ & $\mathbf{3.070}$ & $3.085$\ \\
German folk songs & $2.358$ & $2.301$ &  $2.305$  & $2.257$ & $2.233$ & $\mathbf{2.184}$\\
Chinese folk songs & $2.725$ & $2.685$ &  $2.752$  & $2.612$& $2.650$ & $\mathbf{2.595}$\\
\midrule
Average  & $2.805$ & $2.788$ & $2.779$ & $2.724$ & $2.693$ & $\mathbf{2.656}$ \\
\bottomrule
\end{tabular}
\caption{Cross-Entropies of the 10-fold cross validation in the prediction task for different data sets and different models. 
When combining the RGAE with an absolute pitch model (i.e., RNN, RTDRBM), results improve substantially.
The results suggest that absolute and relative pitch models are complementary in the aspects they learn about music and can be effectively used in an ensemble method.}
\label{tab:folksong-results}
\end{table*}
\newpage
\subsubsubsection{Results and Discussion}
The results are shown in Table \ref{tab:folksong-results}.
The current state-of-the-art results for general connectionist sequence models on the datasets are achieved by the RTDRBM model introduced in \cite{cherla2016neural}.
The results show that the RGAE slightly outperforms the RTDRBM and is clearly superior to the baseline RNN.
Note that, as stated in Section \ref{sec:train_rgae}, the RGAE needs only $16$ hidden units for learning temporal dependencies (the GAE portion mainly transforms absolute pitch input to relative pitch representations).
This compactness suggests that the relative processing of music indeed supports generalization by reducing the sparsity in the data.

When combining the predictions of the RGAE with an absolute pitch model (i.e., RNN or RTDRBM) based on the entropy-weighted geometric mean (cf. Section \ref{sec:geom-mean}), a more substantial improvement is achieved than when combining the two absolute pitch models.
This result shows that absolute and relative processing of music are complementary and can, therefore, be effectively used together in an ensemble method.

\subsubsection{Experiment 2: Copy-and-Transpose Operations}\label{sec:copy-shift}
This experiment is intended as a proof-of-concept for the RGAE's ability to learning sequences of copy-and-transpose operations (i.e., structure schemes).
We oppose our model to an RNN with GRUs, which is known to have difficulties in learning tasks in the form ``whatever has been generated before, now create a (transposed) copy of it''.
The hypothesis is that the RGAE, due to its modeling of intervals, is superior in solving this task.
It has shown in previous studies that it can learn content-invariant transformations between data instances \cite{memisevic2013aperture}, a necessary capability for learning content-invariant structure schemes.

\subsubsubsection{Data}
\begin{table}[]
\centering
\begin{tabular}{l}
\toprule
Transposition Schemes \\
\midrule
$\{+5,+5,+5,\dots \}$ \\
$\{+7,+7,+7,\dots \}$ \\
$\{-5,-5,-5,\dots \}$ \\
$\{-7,-7,-7,\dots \}$ \\
$\{+12,-12,+12,\dots \}$ \\
$\{+3,-3,+3,\dots \}$ \\
$\{+4,-4,+4,\dots \}$ \\
$\{+9,-9,+9,\dots \}$ \\
$\{+4,-8,+4,-8,\dots \}$ \\
$\{-4,+8,-4,+8,\dots \}$ \\
\bottomrule
\end{tabular}
\caption{The different relative transposition schemes used in the ``learning copy-and-transpose operations'' experiment.}
\label{tab:transpatterns}
\end{table}
In order to obtain a controlled setup for testing the model performances, we construct data obeying different recurring (chromatic) transposition patterns.
To this end, the EFSC dataset is transformed into a piano-roll representation with a resolution of 1/8th note.
From that, short fragments of length $4$, $8$, and $16$ ($\le$ the input length of the models) are randomly sampled (rests are omitted).
It is necessary that the RGAE has access to all past events with which the prediction should be related.
Choosing longer fragment lengths than the lengths of the receptive fields yields considerably worse results, also for the baseline RNN, which already performs weakly in this setup.
The fragments are copied and transposed according to some pre-defined transposition schemes (cf. Table \ref{tab:transpatterns}).
For each of the $10$ schemes and fragment lengths, 26 sequences (512 time steps each, resulting in $133\,120$ time steps) are generated, where 20 sequences are used for training, 5 sequences are used for testing and 1 for evaluation.
This results in a total of $600$ sequences for training, $150$ sequences for testing and 30 sequences for evaluation.

\subsubsubsection{Training and Architecture}
The lookback window of the RGAE is $n = 16$ time steps, the RNN portion has 64 units, and we do not use dropout on the input.
For the baseline RNN, we also input the 16 preceding time steps, as this supports copy operations by freeing up memory in the hidden units (i.e., that way, we obtain a direct path from the relevant information to the output).
The baseline RNN model size (512 units) is selected by starting from 64 units and always doubling that number until no substantial improvement occurs on the evaluation set.

The GAE portion of the RGAE is pre-trained for 50 epochs on the structured sequences described above.
Subsequently, the RGAE is trained for $50$ epochs, holding the parameters of the GAE fixed.
As the data of the pre-training does not differ from the sequences in the prediction task, fine-tuning is not necessary.

The baseline RNN is trained for 60 epochs.
Again, for both models, the learning rate scheme described in Section \ref{sec;pretrain-details} is employed.
Note that in this task, we always randomly transpose the input to the models in all training phases.
Therefore, we need no dropout on the input of the RGAE, and the baseline RNN does not overfit, despite its high number of parameters.

\subsubsubsection{Evaluation}
The models have to learn to continue sequences from the test set after exposition to the first $64$ time steps of each sequence.
The experiment is different from typical prediction tasks in that possibly incorrect predictions are fed back to the models, causing errors to accumulate.
To obtain more stable continuations, we do not sample from the predicted distributions of the models, but instead, treat the experiment as a classification task and choose the pitch with the highest predicted probability.
Accordingly, the precision is merely the percentage of correctly predicted pitches over time.
Also, we quantify how many sequences are correctly continued until the end by considering all sequences with an overall precision above $99\%$ as correctly continued.
Furthermore, like in Experiment 1, the categorical cross-entropy loss (cf. Equation \ref{eq:cost_softmax}) is computed.

\subsubsubsection{Results and Discussion}\label{sec:discussion}
\begin{table}
\centering
\begin{tabular}{lllll}
\toprule
Model & Pr (\%) & $> 99\%$ & CE & \# Params \\
\midrule
RNN & $41.38$ & $6.67$ & $0.287$ & $\sim 2\,300\,000$ \\
RGAE & $\mathbf{99.43}$ & $\mathbf{92.00}$ & $\mathbf{0.033}$ & $\sim \mathbf{600\,000}$ \\
\bottomrule
\end{tabular}
\caption{Results of the ``learning copy-and-transpose operations'' task. Average precision (Pr), percentage of continuations above $99\%$ precision, cross-entropy (CE) and number of parameters of the respective model.}
\label{tab:quant-res-struct}

\end{table}
\begin{figure}
\includegraphics[width=1.\linewidth]{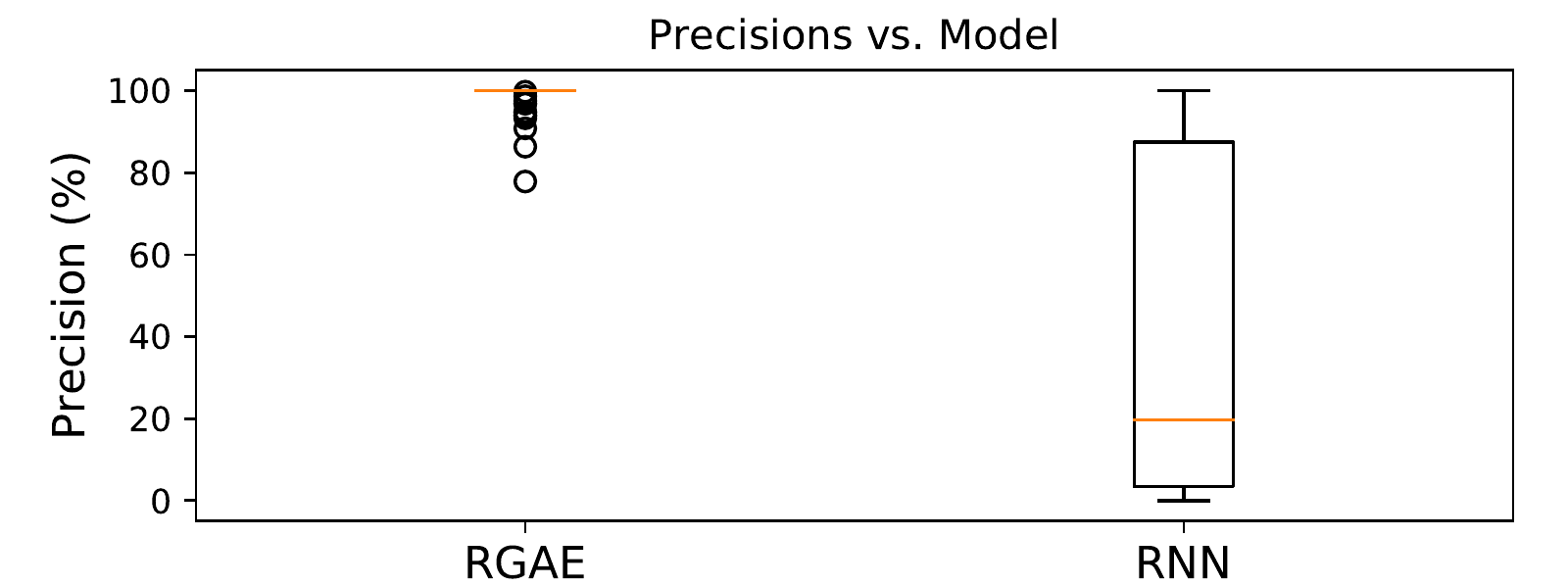} \\
\vspace{-0.3cm}
\caption{Distribution of precisions for continuation of highly structured sequences in the test set of size $150$. The median is marked with a orange line, the boxes indicate the interquartile range, and circles indicate outliers.}
\label{fig:prec-boxpl}
\end{figure}
\begin{figure}[t]
\includegraphics[width=1.\linewidth]{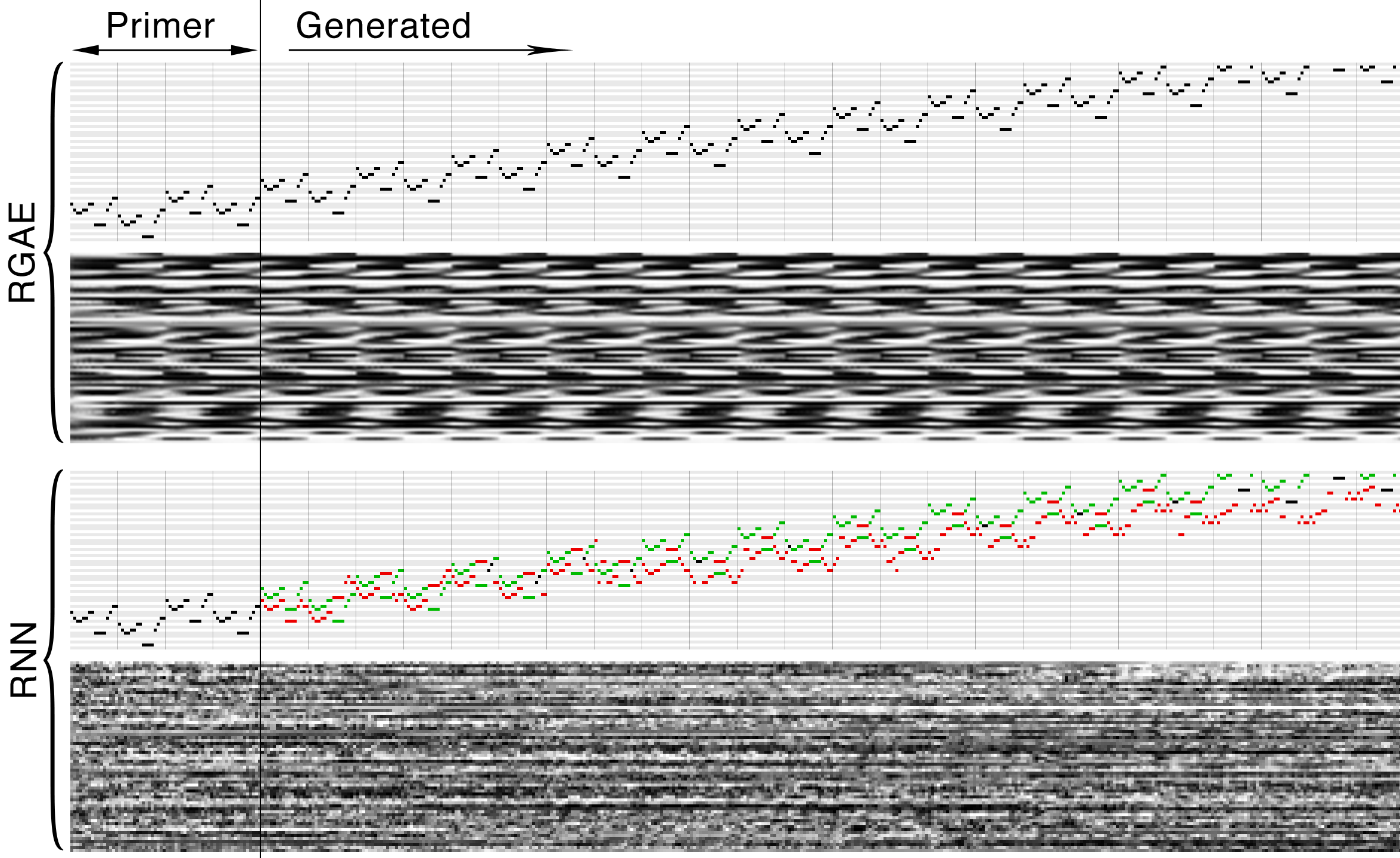} \\
\vspace{-1mm}
\caption{Generated structure schemes and hidden unit activations of the RGAE and the RNN models after input of a primer indicating the $\{-4,+8,-4,+8,\dots \}$ scheme, realized with melodies of length 16 not contained in the train set. Black notes indicate correct continuation, green notes indicate false negatives, red notes indicate false positives. Hidden units activations of the RNN are pruned due to space limitation.}
\label{fig:seq-continued}
\end{figure}
Table \ref{tab:quant-res-struct} shows the quantitative results of the experiment, and Figure \ref{fig:prec-boxpl} shows a box plot comparing the precisions of the two models.
With an average precision of $99.43\%$ percent, where $92\%$ of all examples are flawlessly continued, the RGAE shows remarkable stability in continuing the structure scheme realizations.
The cross-entropy of the RGAE is about two orders of magnitude lower than that of the RNN.
In Figure \ref{fig:seq-continued}, a specific example of this sequence continuation task is depicted.
Note that the hidden unit activations of the RGAE are more regular because they only represent copy-and-transpose operations instead of the musical texture itself (as it is the case for the RNN).
The most challenging part for the RGAE is counting, in order to change the copy operation (i.e., transposition distance) at the right time (in fact, at most of the incorrectly continued sequences, the RGAE miscounted by a time step).
It is important to note that the hidden unit activations of the RNN portion are identical for identical schemes because they operate on transformations between events, rather than on the events themselves (i.e., they are largely content-invariant).

\subsection{Conclusion and Future Work}\label{sec:conclusion}

The principle of modeling sequences of first-order derivatives in music is a compelling concept with the potential to solve two persistent problems in MIR: Learning transposition-invariant interval representations, and learning representations of (chromatically transposed) repetition structure.
The proposed model is conceptually simple and can be trained as a generative model in sequence learning tasks.

Moreover, the RGAE could act as a building block for more complex architectures, in order to extend its capabilities.
For example, the temporal lookback window could be greatly extended by employing the RGAE on top of a (dilated) convolutional network, enabling it to learn higher-level repetition structure.
In another variant, an RGAE could be employed on top of an RNN.
Applied to music, the RNN would provide the RGAE with representations of important, absolute reference pitches (e.g., the tonic of a scale, or the root note of a chord), and the RGAE could learn sequences of intervals in relation to them.
Another interesting architecture would involve stacking more than one RGAE on top of one another to learn higher-order derivatives, for example, variations between mutually transposed parts in music.

The RGAE, however, is not limited to the symbolic, monophonic, domain of music.
As shown in Section \ref{sec:interval}, a GAE can also operate in the spectral domain of audio and in polyphonic symbolic music.
Finally, we note that the RGAE is general enough to apply to other domains where the derivatives of functions are of higher importance than their absolute course.
Possible applications include modeling temporal progressions of changes in loudness, tempo, mood, information density curves, and other musical properties, modeling moving or rotating objects, camera movements in video recordings, and signals in the time domain.



\chapter{Conclusion and Future Work}
In this section, the chapters of this thesis are briefly reviewed, and an outlook on possible future research is given.
I will focus mainly on two research directions that I consider to be promising for follow-up in musical structure learning: Information Theory and Transformation Learning.
The conclusion starts by pointing out the potential of Information Theory in music structure analysis and structured generation, referring to some examples in linguistics.
Then, I will review my work on transformation and invariance learning, including a brief discussion on future work and on two of my very recent publications, which are related but not part of this thesis.

\section{Information Theory and Structure Analysis}
In Chapter \ref{sec:segment}, I perform segmentation of monophonic melodies using a probabilistic approach.
Probabilistic segmentation is based on the assumption that the probability of events within a segment is higher than that of events that mark segment boundaries.
The metric employed in practice is the \emph{information content} (IC), defined as the negative log probability of a note.
IC can be interpreted as the "level of surprise" a listener would experience when encountering a note event.
We confirm prior studies (see \cite{Pearce:2010ij}) that the IC, in combination with a threshold, provides a valid proxy for determining segment boundaries.

Other computational models of segmentation are based on Gestalt Principles.
We discuss in Chapter \ref{sec:segment} that while there is a direct link between IC and Gestalt Principles, it is unclear which is more parsimonious and might have given rise for the other to emerge as a perceptual mechanism.
In any case, the competitiveness of the probabilistic approach is evidence for the usefulness of Information Theory in musical structure analysis.
In light of this observation, it seems worthwhile to consider other metrics of Information Theory for music structure modeling.

In Chapter \ref{sec:constraints}, I evaluate the generated musical sequences $\mathbf{x}_{0} \dots \mathbf{x}_{T}$ of length $T$ based on Information Rate (IR).
IR, in sequences also known as the mutual information between the current event $\mathbf{x}_t$ and its predecessors $\mathbf{x}_{<t}$, is defined as $H(\mathbf{x}_t) - H(\mathbf{x}_t \mid \mathbf{x}_{<t})$, where $H(\cdot)$ is the entropy.
The equation yields high IR, when the conditional entropy of all events $H(\mathbf{x}_t \mid \mathbf{x}_{<t})$ is minimal (i.e., when all events in a sequence are highly expected given the past), and when the prior entropy of the events $H(\mathbf{x}_t)$ is maximal (i.e., when all events are equally likely to occur). The IR is therefore high, when a sequence is well predictable (highly structured), but not trivial (i.e., not composed of, for example, just a single, constantly repeating event).
IR can also be interpreted as the relative reduction of uncertainty of the present when the past is known \cite{DBLP:conf/semco/DubnovAC11}.
In music, IR is a measure for ``well-structuredness'' and has been applied, for instance, in parameter selection for musical pattern discovery \cite{wang2015pattern}.

Another Information Theory metric, which has been less studied in the musical domain, is Uniform Information Density (UID).
UID is based on the proposal of Shannon that for optimal data flow through a noisy channel, the information density (i.e., the IC over time) should be as uniform as possible \cite{shannon2001mathematical}.
This means that in order to minimize comprehension difficulty in language and music processing, the sensory information should be structured so as to avoid peaks and troughs in information density \cite{jaeger2007speakers}.
It has been shown that speakers intuitively construct utterances in order to satisfy UID (e.g., by modifying the talking speed \cite{bell2003effects, aylett2004smooth}, or by omitting words which carry little information \cite{jaeger2007speakers}).
In order to avoid troughs in information density, the expectedness of some phrases may be deliberately reduced.
For example, when a phrase is semantically expected due to the preceding context, it can have a more complex syntactic form \cite{genzel2002entropy}.

In music, a \emph{repetition} usually causes a trough in information density, because it is easy to predict the events of a sequence when the same sequence has already been encountered.
In order to balance the overall predictability of such a repeated musical phrase, some less predictable events (e.g., unlikely intervals) may be introduced.
Evidence for this is given by \citet{temperley2014information}, who shows that ``there is a tendency that when an intervallic pattern is repeated with alterations, the alterations tend to lower the probability of the pattern rather than raising it.''

As we have seen, Information Theory metrics can make predictions about language usage (UID), musical segmentation boundaries (IC), choices in music composition (UID), and parameters for musical pattern discovery systems (IR).
Considering the apparent links between Information Theory and structural properties in language and music, it seems obvious to further examine possible applications of Information Theory metrics to music composition and analysis.
As generative sequence models become ever more powerful \cite{musenet}, their precision in estimating Information Theory metrics is naturally increasing, too.
With the current state of the art, it would already be possible to perform large-scale corpus studies on audio data.
Future work could investigate the \emph{variance} of information in different genres and in different time periods (where higher variance suggests less UID, as proposed for language in \cite{collins2014information}).
Such a study, besides its potential to yield interesting insights on its own, could also help to quantify the degree of human-acceptable violations of UID, which in turn could be used for automatic music composition systems.
Analysis of musical works regarding UID could also be used in assisted composition systems to point out ``problematic'' sections (i.e., such with too high or too low variance in information).
Likewise, IR could inform automatic composition systems regarding the ``well-structuredness'' of their output.
Overall, it seems that the full potential of Information Theory in music generation and analysis has not yet been sufficiently exploited---a situation which should be improved in the future.

\section{Learning Transformations and Invariances in Music}

In Section \ref{sec:gae}, I explore a novel way of musical representation learning. To that end, I do not aim to learn the musical patterns themselves, but some ``rules'' defining how a given pattern can be \emph{transformed} into another pattern. Transformation Learning (TL) was initially proposed for image processing and had not yet been applied to music. In this section, I summarize our experiments in TL for music and suggest possible future research directions.

The models used throughout this thesis are based on Gated Autoencoders (GAE) which learn orthogonal transformations between data pairs.
We show that a GAE can learn \emph{chromatic transposition}, \emph{tempo-change}, and the \emph{retrograde movement} in music, but also more complex musical transformations, like \emph{diatonic transposition} (see Section \ref{sec:gae}).
Transformation Learning (TL) provides us with a different view on music data, and yields features complementary to other music descriptors (e.g., such as obtained by autoencoder learning or hand-crafted features). There are different possible research directions regarding TL in music. They involve using the transformation features themselves, using transformation-invariant features computed from TL models, and using TL models for music generation.

The most obvious usage of TL is to represent transformations between n-gram pairs which stem from a song or a corpus, as shown in Section \ref{sec:gae}. In the future, statistics of the thus obtained features could be used for \emph{comparison of musical works} (e.g., regarding the occurrence probability of certain transformations). In intra-opus analysis, \emph{all possible pairwise transformations} between the n-grams of a song could be computed. Such a representation would be informative for structure analysis, but also for music generation. Given a complete description of the transformations in a song, a new song could be found obeying these transformations but instantiated with different content. In Chapter \ref{sec:constraints}, I propose a constrained sampling approach to transfer structural properties from an existing piece to newly generated material. It would be straight-forward to use TL features as constraints in order to transfer the ``transformation structure'' between musical works.

A further application of Transformation Learning (TL) in music is to represent musical events as the result of transformation rules applied to \emph{preceding events}. In Section \ref{sec:interval}, I show that this approach yields features which behave like interval representations (i.e., their organization in the latent space is musically meaningful, and they are transposition-invariant). In such a setting, again, TL is useful for analysis \emph{and} generation. Due to the invariance property of the interval features, a self-similarity analysis in the feature space of musical works is sufficient to find \emph{mutually transposed sections}. Furthermore, in audio-to-score and audio-to-audio alignment, conventional dynamic time-warping approaches can be applied to \emph{mutually transposed tracks}, if the alignment is performed in the transposition-invariant feature space \cite{arzt2018alignment}. When applying sequence models to the invariant interval space, the accuracy in predicting musical events can be improved compared to using common input representations (see Section \ref{sec:rgae}). Furthermore, in music generation, using these features I show that it is easier for sequence models to generate copies of musical fragments, where a \emph{literal copy} is just a special case of the \emph{transposed copy} operation (see Section \ref{sec:copy-shift}).

Importantly, learning transformations and learning invariances are two sides of the same coin (as specific invariances are defined with respect to specific transformations). This becomes even clearer when considering the functioning of the Gated Autoencoder (GAE). When learning a particular orthogonal transformation between data pairs, a GAE internally performs an eigenvector-decomposition of that transformation. In general, the eigenvectors of orthogonal transformations are complex-valued. When projecting data points onto a complex eigenvector pair of a particular transformation, the norm of the resulting coefficients is invariant to that transformation. A well-known example of this property is the Fourier transform. As sine and cosine represent the complex eigenvectors of \emph{translation}, the amplitude spectrum of a signal is invariant to translation (i.e., phase shift) of the signal. Even though they do not emerge through conventional GAE training, it has been shown in \cite{memisevic2013aperture}, that a GAE can be constrained to obtain such complex eigenvector pairs. Very recently, I introduced the Complex Autoencoder \cite{lattner2019complex}, a model which explicitly learns such complex eigenvectors from data pairs and showed that the features thus obtained are competitive in alignment tasks and repeated section discovery.

A further example of the usefulness of transformation learning in music is another recent study on conditional rhythm generation \cite{DBLP:conf/waspaa/Lattner19}. Following a similar intuition as in Section \ref{sec:interval}, where a GAE learns the pitch intervals between temporally successive events, we now learn the \emph{inter-onset-intervals between different rhythm instruments} in a song. More precisely, a GAE is extended to a convolutional architecture and used for the generation of kick drum rhythms, conditioned on snare drum, bass, and estimated beat and downbeat information. Like in the experiments mentioned above, the model benefits from the invariance properties of Transformation Learning (TL). In fact, it is possible to represent a large part of the dynamic rhythmic relationships in a song by a single \emph{style vector}. This style vector is invariant to time (i.e., the model produces plausible output when the style vector is kept constant over time), and invariant to the tempo (i.e., the output tempo is determined by the input tempo and not the style vector). These properties render the approach very useful for high-level user control because a desired style vector can be found by manual exploration of a ``style space'' and can then be kept constant for the whole song or section while obtaining an output rhythm which dynamically adapts to the rhythmic context.

The idea of conditional generation of rhythmic structures using TL could be further extended to conditional generation of \emph{tonal structures} in the future. Instead of learning the pitch intervals between \emph{successive events} like in Section \ref{sec:interval}, one could learn the pitch intervals between \emph{concurrent events} of different instruments. In that case, a transposition-invariant style vector would, for example, yield the \emph{root note} with respect to the tonal context, while another style vector configuration would yield the \emph{third} or the \emph{fifth}. While keeping the style vector fixed, the output pitch would still change, when the tonal context changes. This would considerably reduce the amount of information in the representation space of a model and could lead to more compact music generation systems.

A disadvantage of Transformation Learning (TL) models like the GAE is their restriction to orthogonal transformations. Learning more general transformations in an unsupervised manner with neural network architectures could be relevant in music and other domains and should be investigated in the future. A further problem is the selection of data pairs to feed to the model. GAEs do not learn well if a considerable portion of the input pairs is not related. Preliminary experiments suggest that this problem can be overcome by using methods from curriculum learning, which can choose to ignore data pairs which cause high losses \cite{DBLP:conf/aaai/JiangMZSH15}.

The given examples show that TL can provide us with a different view on musical data, leading to improvements or advantageous behaviors in music analysis and generation systems. I also pointed out some promising future research directions and challenges. As TL in music is still a new field, it offers a high potential for novel findings and applications.

\cleardoublepage

\backmatter
\bibliographystyle{plainnat}
\bibliography{bib/bib_sl,bib/bib_mg,bib/bib_cc}

\cleardoublepage

\linespread{1.5}
\pagestyle{empty}


\chapter*{Curriculum Vitae of the Author}
\addcontentsline{toc}{chapter}{Curriculum Vitae of the Author}

\section*{Personal data}


\begin{table}[h]
\begin{tabular}{p{2.5cm} p{12cm}}
Name: & Stefan Lattner \\
Date of birth: & February 22nd, 1984 \\
Place of birth: & Kirchdorf a.\,d. Krems, Austria \\
Email: & \href{mailto:me@stefanlattner.at}{me@stefanlattner.at}  \\
Website: & \href{http://www.stefanlattner.at}{www.stefanlattner.at}  \\
\end{tabular}
\end{table}

\linespread{1.25}

\section*{Education}

\begin{table}[h]
\begin{tabular}{p{2.5cm} p{12cm}}
2014--2019 & PhD in Computer Science, Johannes-Kepler University Linz. \\
	& \emph{Thesis:} ``\mytitle'' \\
\addlinespace[0.2cm]
2010--2014 & Master of Science in Pervasive Computing, Johannes-Kepler University Linz, Austria. \emph{Thesis:} ``Hierarchical Temporal Memory - Investigations, Ideas, and Experiments''\\
\addlinespace[0.2cm]
2013 & Study abroad at the University of Tasmania, Hobart, Australia \\
\addlinespace[0.2cm]
2006--2009 & Bachelor of Science in Media Technology and -Design, University of Applied Sciences, Hagenberg Campus, Austria. \emph{Thesis:} ``Konzepte algorithmischer Komposition: Ein vergleichender Überblick''\\
\addlinespace[0.2cm]
2005--2006 & Bioinformatics, University of Applied Sciences, Hagenberg Campus, Austria \\
\end{tabular}
\end{table}

\section*{Experience}
\begin{table}[h]
\begin{tabular}{p{2.5cm} p{12cm}}
2018--present & Associate Researcher at the Sony Computer Science Laboratories, Paris, France \\
\addlinespace[0.2cm]
2017--2018 & Research Assistant at the Johannes-Kepler University, Linz, Austria \\
\addlinespace[0.2cm]
2016--2017 & Lectures in Generative Music (Markov Models and Neural Networks), Technical University, Vienna \\
\addlinespace[0.2cm]
2014--2017 & Research Assistant at the Austrian Research Institute for Artificial Intelligence (OFAI), Project \emph{Lrn2Cre8}, European Union Seventh Framework Programme \\
\addlinespace[0.2cm]
2009--2014 & Chief Developer and Project Manager for the Music AI Application \emph{Liquid Notes} at Re-Compose GmbH, Vienna, Austria \\
\addlinespace[0.2cm]
2008--2009 & Tutor in Algorithms and Data Structures, University of Applied Sciences, Hagenberg Campus, Austria \\
\end{tabular}
\end{table}

\section*{Scientific Services}
\begin{table}[h]
\begin{tabular}{p{2.5cm} p{12cm}}
Reviewer & Society for Music Information Retrieval Conference (2015--2019) \\
\addlinespace[0.2cm]
	& Special Issue on Deep Learning for Music and Audio in Springer’s Neural Computing and Applications (2018) \\
\addlinespace[0.2cm]
	& Association for the Advancement of Artificial Intelligence (2017) \\
\addlinespace[0.1cm]
\end{tabular}
\end{table}

\end{document}